\documentclass{aa} 

\usepackage[colorlinks=true, linkcolor=blue, citecolor=blue, urlcolor=blue]{hyperref}
\makeatletter
\renewcommand*\aa@pageof{, page \thepage{} of \pageref*{LastPage}}
\makeatother
\usepackage{graphicx}
\usepackage{txfonts}
\usepackage{siunitx}
\usepackage{booktabs}
\usepackage{subcaption}
\usepackage{gensymb}
\usepackage{bm}

\newcommand{\Vaisala}{V\"{a}is\"{a}l\"{a}~}
\usepackage{ulem}
\usepackage[dvipsnames]{xcolor}
\definecolor{vi}{HTML}{5E1111}

\begin{document}

   \title{Non-linear saturation of gravito-inertial modes excited by tidal resonances in binary neutron stars}%in neutron-star binaries?

   \author{Alexis Reboul-Salze
          \inst{1}, Aur{\'e}lie Astoul \inst{2}, Hao-Jui Kuan \inst{1}, Arthur G. Suvorov \inst{3,4}}
   \institute{\inst{1} Max Planck Institute for Gravitational Physics (Albert Einstein Institute), D-14476 Potsdam, Germany\\
            \email{alexis.reboul-salze@aei.mpg.de}  \\\inst{2} School of Mathematics, University of Leeds, Leeds LS2 9JT, UK \\
            %\email{a.a.v.astoul@leeds.ac.uk}\\
              \inst{3}Departament de F{\'i}sica Aplicada, Universitat d'Alacant, Ap. Correus 99, E-03080 Alacant, Spain\\
              \inst{4}{Theoretical Astrophysics, IAAT, University of T{\"u}bingen, T{\"u}bingen, D-72076, Germany}
             }

   \date{Received ??; ??}

% \abstract{}{}{}{}{} 
% 5 {} token are mandatory
 
  \abstract
   {During the last seconds of a binary neutron-star merger, the tidal force 
   can excite stellar oscillation modes to large amplitudes. From the perspective of premerger electromagnetic emissions and next-generation gravitational-wave detectors, gravity ($g-$) modes constitute a propitious class. However, existing estimates for their impact employ linear schemes which may be inaccurate for large amplitudes, as achieved by tidal resonances. With rotation, inertial modes can be excited as well and while their non-linear saturation has been studied, an extension to fully-consistent gravito-inertial modes, especially in the neutron-star context, is an open problem.}
   {We study the linear and non-linear saturation of gravito-inertial modes and investigate the astrophysical consequences for binary neutron-star mergers, including the possibility of resonance-induced dynamo activity.}
   {A new (non-)linear formulation based on the separation of equilibrium and dynamical tides is developed. Implementing this into the 3D pseudo-spectral code \texttt{MagIC}, a suite of non-linear simulations of tidally-excited flows with an entropy/composition gradient in a stably-stratified Boussinesq spherical-shell  
   are carried out.}
   {The new formulation accurately reproduces results of linear calculations for gravito-inertial modes with a free surface for low frequencies. 
   For a constant-density cavity, we show that the axisymmetric differential rotation induced by nonlinear $_2g$ and $_1g$ modes may theoretically be large enough to amplify an ambient magnetic field to $\gtrsim 10^{14}$~G. In addition, rich 
   non-linear dynamics are observed in the form of a parametric instability for the $_1g$ mode. The stars are also spun-up, 
   which extends the resonance window for any given mode.}
   {This study provides non-linear numerical support for a recently-proposed scenario where, to accommodate the non-thermal precursor flares seen in some short gamma-ray bursts, the magnetic field of a premerger star is amplified by resonant $g$-modes. It further suggests that $g$-mode resonances may have a stronger impact on gravitational-wave signals than previously estimated.}

   \titlerunning{Tidal resonances of non-linear gravito-inertial modes}
   \authorrunning{A. Reboul-Salze, A. Astoul, H.-J. Kuan, A. G. Suvorov}
   \maketitle
%
%-------------------------------------------------------------------

\section{Introduction}

Binaries are common in Nature, with some estimates of stellar multiplicity being close to unity based on modern astrometric surveys \citep{duch13,bad24}. Tidal forces play an important role in the rotational and orbital evolution of close systems, such as binary stars \citep[e.g.][and references therein]{zahn2013,barker22}, %zahn66
exoplanetary hot-Jupiter systems \citep[e.g.][]{jack08,BM2016,laz24}, planet-satellite systems \citep[for a review]{counsel73,fuller24}, and merging neutron stars \cite[NSs;][]{dam09,tani10}. Their impact can generally be quantified in terms of two components, being the equilibrium and dynamical tides. The former exist even in the zero-frequency limit and describe bulk geometric large-scale deformations, while the latter relate to the excitation of internal (damped) waves \citep[e.g.][]{og14,hind16,ande20}.

A perturbative, and often even linear, treatment of tides provides an excellent approximation in many astrophysical systems of interest. There are cases however where the validity of a linear scheme is less certain \cite[see section 4 of][for a review]{og14}. One such scenario -- central to this paper -- involves resonances in NS mergers. As the binary compactifies due to gravitational-wave (GW) radiation-reaction, the orbital frequency `chirps' upwards towards coalescence \cite[e.g.][]{peters64}. When the tidal driving frequency approaches the oscillation frequency of a mode inside one of the stars, that mode's amplitude grows rapidly until saturating at a value that depends on the degree of orthogonality between its eigenfunction and the appropriate multipole component of the tidal field \cite[`overlap';][]{pres77,alex87}. For some mode classes, the saturation amplitude could theoretically reach fractions of unity where non-linear corrections need to be taken into account to more accurately assess how tides affect GW or electromagnetic observables; see \cite{skk24} for a recent review.

Numerical relativity simulations have highlighted the impact of non-linearities on GW dephasing for the fundamental ($f$-) mode \citep{kuan24}. Though subdominant, gravity ($g$-) modes, supported by internal composition and/or entropy gradients, may also non-negligibly accelerate inspiral in mergers as their eigenfrequencies are sufficiently low ($\sim10^{2}$~Hz) that they can be excited $\sim$~seconds prior to coalescence while maintaining sizeable overlaps \citep{ks95,kuan21a,pass21,hega24}. Since these modes offer valuable microphysical information and it is hoped that templates will be matched to GW data in the future, it is important that their non-linear saturation be quantified. For stars retaining a high angular frequency into old age ($\Omega \gtrsim 10^{2}$~Hz; `recycled pulsars') the inertial class of modes (restored by Coriolis acceleration) may also be relevant \citep{lw06,sk20}. With both buoyancy and rotation present, the two branches merge to some degree and go under a `gravito-inertial' label \cite[see, e.g.,][]{aerts23}. 
Here we present a method to model the non-linear saturation of such hybrid modes. Note in particular that nonlinearities are treated at the level of the magnetohydrodynamic (MHD) equations of motion, rather than via secondary couplings between linear modes as studied previously \cite[e.g.][]{wein13,pk15,zz17}. This marks the first time therefore that this class and their saturation have been modeled consistently at a non-linear level for rotating, tidally-forced NSs.

Non-linear, tidally-forced inertial modes have been studied in  numerical simulations modeling convective envelopes of giant gaseous planets and low-mass stars in close binaries systems \cite[][]{2014FavierNLIWtidal,2016BarkerNLTides,2022Astoul}. Due to the generation of differential rotation, transfer rates of tidal energy (i.e. tidal dissipation) can be either larger or smaller by up to $\sim 3$ orders of magnitude relative to linear-theory predictions at some frequencies \citep[especially for thin shells, high tidal amplitudes, and low viscosities;][]{2023Astoul}. 
In radiative regions, differential rotation can also arise from the deposition of gravity wave angular momentum at corotation resonances  -- when the phase velocity of the wave matches the local speed of the fluid -- as predicted by \citet{G1989} in early-type stars envelopes. This result has been confirmed for the radiative cores of solar-like stars in 2D and 3D spherical nonlinear simulations, where the core is progressively spun up from inside out by tidal gravity wave breaking (or weakly non-linear damping) near the center of the star \citep{BO2010,B2011,GO2023}.
Here, we also study such non-linear effects but in the context of gravito-inertial modes in NSs by building on the formulation used in \citet{2023Astoul} to split tides into equilibrium and dynamical components while adding buoyancy and thermal diffusion.

Based on this `split formulation', accounting for the fact that the Brunt-\Vaisala (BV) frequency of a cold NS is typically much smaller than its dynamical (hydrostatic) response frequency, we use the pseudo-spectral \texttt{MagIC}\footnote{\url{https://magic-sph.github.io}} code \citep{2002WichtMagic,2012GastineCode,2013SHTNS} to solve the hydrodynamical equations in a spherical shell under the Boussinesq approximation \citep{spi60}. The equations are coupled to the leading-order ($\ell = m = 2$) tidal potential for frequencies relevant to the late-stages of inspiral, with (effective) viscosity included in anticipation of damping and turbulence. Magnetic effects may be especially relevant in some mergers, as there is evidence to suggest that those systems which release nonthermal gamma-ray burst (GRB) precursor flares \cite[$\lesssim 5\%$ of mergers;][]{wang20,cop20} contain magnetars with near-surface fields of $B \gtrsim 10^{13}$~G \cite[e.g.][]{xiao24}. Fields of this strength can influence mode spectra \cite[e.g.][]{skk22,kuan21a} or merger dynamics directly \cite[e.g.][]{cio20}. A thorough comparison with a linear scheme, using the \texttt{Dedalus3} code \citep{burns20}, is also provided to explore the direct impact of non-linearities.

Although this is primarily a `methods' paper where the above-described strategy is presented in detail and then integrated numerically for a variety of setups, we also explore some astrophysical applications. For example, incorporating magnetic fields naturally allows for the investigation of dynamo action, namely the magnetorotational instability \cite[MRI;][]{1998ReviewMRIBalbus, astoulbark25}, sourced by the differential rotation induced by non-axisymmetric mode activity. The non-linear scheme allows, in principle, for us to compare with the idea suggested by \cite{2024Suvorov} that magnetar-level fields are generated on the verge of coalescence rather than preserved over cosmological timescales, as such longevity appears difficult to reconcile with magnetothermal evolutions \citep{pv19}. %We also explore the theoretical strain placed on a crust by comparing the non-linear and linear saturation amplitudes as relevant for GRB precursors, and damping in the convective envelopes of solar-type stars \as{[A: make sure we actually do this or kill it here.]}. 

This paper is organized as follows. Sec.~\ref{sec:model} introduces the neutron-star hybrid modes and the equations governing both the linear (Sec.~\ref{sec:lineqns}) and new non-linear (Sec.~\ref{sec:Nonlinear}) schemes used throughout, with details on numerical implementation provided in Sec.~\ref{sec:numerics}. Results for both $g$- and more general gravito-inertial modes in cases with rotation are given in Sec.~\ref{sec:Linear} for the linear and Sec.~\ref{sec:Sim} for the non-linear cases, respectively. Astrophysical implications are briefly explored in Sec.~\ref{sec:astrophysics}, with discussion provided in Sec.~\ref{sec:Disc}. Key results are summarized in Sec.~\ref{sec:Conclusion}.

%--------------------------------------------------------------------
\section{Modeling of the tidal resonance of gravito-inertial in a neutron star binary}
\label{sec:model}
\subsection{Simplified neutron star model}
\label{3Ddata}

In this article, we study the tidal forcing of gravito-inertial modes in the late stages of a binary neutron-star inspiral. To do this, we use methods and formulations from the stellar and planetary community, where linear codes and more recently non-linear simulations are developed following the split methodology mentioned earlier. 
As a first study in the context of binary neutron stars, we adopt the idealized Boussinesq approximation that allows us to study the impact of stable entropy/composition stratification while keeping the simulations computationally efficient so that the parameter space can be efficiently explored. This approximation corresponds to assuming that variations in mass density can be neglected except as they contribute to buoyancy, i.e., the fluid is otherwise incompressible and soundwaves are filtered \citep{spi60}. This approximation is justified because the velocities of resonant modes, as found in previous studies \citep{kuan21a, pass21,2024Suvorov}, are strictly subsonic.
We use the Newtonian Navier-Stokes equations throughout and neglect the effects of general relativity, as the Boussinesq approximation is already a more restrictive approximation. Note that we also assume the neutron-star interior to contain only a normal fluid, ignoring superfluidity and superconductivity.

The fluid stratification is characterized by the BV frequency,
\begin{equation}
    N^2 \equiv - \frac{g}{\rho_0} \left(\left. \frac{\partial \rho}{\partial S}\right|_{P,Y_e} \frac{d S}{d r} + \left. \frac{\partial \rho}{\partial Y_e}\right|_{P,S} \frac{d Y_e}{d r} \right)\,,
\end{equation}
where $g,\, \rho_0,\, \rho, S,\, P$ and $Y_e$ are, respectively, the gravitational acceleration, uniform density (see below), density, entropy, pressure, and electron fraction. 
As a simplification, we assume that the BV frequency depends only on one dimensionless buoyancy variable, chosen here as $b = -\frac{\rho}{\rho_0}$. In this work, we refer to the diffusion process associated with this buoyancy variable as the thermal diffusivity, $\kappa_{th}$.
This does not reduce the (numerical) generality of the problem as the buoyancy variable could be entropy or composition and would play exactly the same role.
For dissipation processes, we adopt uniform profiles of viscosity $\nu$ and  thermal $\kappa_{th}$ (or compositional $\kappa_{comp}$) diffusivities with $\nu=\kappa_{th}$ for the sake of simplicity\footnote{Note that a realistic estimate of the compositional Prandtl number for mature NSs can be found in \citet{2024Suvorov}, viz. $Pr\equiv\nu/\kappa_{comp} \sim 10^3-10^8$. The influence of $Pr$ on mode dynamics is left to future studies.}. 

The model for the NS we use is that of a spherical shell with a uniform density $\rho_0$. The latter quantity is determined by the mass of the primary, given as $M_{1} = \frac{4}{3} \pi \rho_0 R_o^3$, where $R_o$ is the stellar radius of the primary and depends on the equation of state (EOS).
To avoid modeling the stellar core, due to both sizable uncertainties (which only increase during evolution) in microphysical inputs we require and numerical stability reasons, we suppose the shell has an inner radius $R_i$ whose ratio to the stellar radius ($R_o$) is set as $\alpha \equiv \frac{R_i}{R_o} =0.5$. This value has been used in previous studies for planets and stars as a standard \citep[e.g.][]{2014FavierNLIWtidal,2022Astoul,2023Pontin}. For astrophysical applications to NSs, we take instead $\alpha = 0.1$ to approximate a whole NS without increasing numerical costs (see Sec.~\ref{sec:Astro_consequences}).

\subsection{Governing equations}
 We place ourselves in a frame rotating with the star and assume an initial hydrostatic balance, i.e. $\rho_0 g = - \frac{d\tilde{P}}{dr}$ where $g$ is the gravitational acceleration and $\tilde{P}$ is the background pressure. 
 The equations for the perturbed velocity components can then be written as 
\begin{align}
    \bm \nabla \cdot \bm u &=0, \label{eq:grad_u}\\
    \frac{\partial \bm u}{\partial t} + (\bm u \cdot \bm\nabla)\bm u & = -\frac{\bm\nabla p}{\rho_0} - 2\bm{\Omega_s} \times \bm u - b \bm g  - \bm\nabla \Psi + \bm\nabla \cdot \bm \tau_\nu, 
\end{align}
where $\boldsymbol{u},\ p$ and $b$ are velocity, pressure and buoyancy perturbations,  $\boldsymbol{\Omega_s}$ is the (constant) stellar angular frequency of the background star, $\Psi$ is the tidal potential, and \(\bm \tau_\nu\) is the viscous stress tensor. Note that the components of velocity are designated by $u_{i}$ with $i \in [r,\theta,\phi]$ instead of the contravariant equivalent ($u^{i}$) in GR calculations. 
Centrifugal and other forces which may lead to a non-spherical cavity are neglected. 
We choose to neglect perturbations to the gravitational potential for the sake of simplicity and because they are expected to be quantitatively small in the context of our Boussinesq system \citep{2004OgilvieL}. This hypothesis is discussed in detail in Sec.~\ref{disc:self_gravity}.
This makes it easier to compare between formulations as we will neglect the contributions of the gravitational potential perturbations due to the equilibrium tides characterized by the Love number $k_2$ (See Sec.~\ref{disc:self_gravity}). \\

The heat equation thus takes the form
\begin{equation}
    \frac{\partial b}{\partial t} + (\bm u \cdot \bm\nabla) b + u_r \frac{d \tilde{b}}{dr} = \kappa_{th} \bm\nabla^2 b, 
    \label{eq:dsdt}
\end{equation}
where the global gradient of the buoyancy variable $\frac{d \tilde{b}}{dr}$ is linked to the radially dependent BV frequency through 
\begin{equation}
 N^2 = - \frac{g}{\rho_0}  \frac{d \tilde{b}}{d r}\,.
    \label{e:brunt}
\end{equation}
In the case of entropy or composition stratification, we would have $\frac{d \tilde{b}}{d r}=\left. \frac{\partial \rho}{\partial S}\right|_{P,Y_e} \frac{d S}{d r} $ or $\frac{d \tilde{b}}{d r}=\left. \frac{\partial \rho}{\partial Y_e}\right|_{P,S} \frac{d Y_e}{d r}$, respectively.

Leaving aside events involving dynamical capture, orbital eccentricity is expected to be erased long before coalescence in a neutron-star binary merger \cite[e.g.][]{peters64}. Because of this, we consider a circular but asynchronous orbit with no spin-orbit misalignment \cite[though cf.][]{kuan23b}. The dominant (quadrupolar) component of the tidal potential can be written as \cite[e.g.][]{zahn66} 
\begin{equation}
    \Psi = \Psi_0 r^2 Y_2^2(\theta,\phi) e^{-i \hat{\omega} t},
    \label{eq:tidal potential}
\end{equation}
where $Y_l^m$ is a spherical harmonic of degree $\ell$ and order $m$, and the strength of the potential is \citep[see, for instance, equation 3 and table 1 in][]{og14}
\begin{align}
    \Psi_0 = \sqrt{\frac{6 \pi}{5}} \frac{M_2}{M_1} \omega_d^2 \left(\frac{R_o}{a}\right)^3,
\end{align}
where $M_2$ is the companion mass, $a$ is the orbital semi-major axis, and $\omega_d = \sqrt{G M_1/R_o^3}$ is the dynamical frequency with $G$ being the gravitational constant. 
In the rotating frame introduced earlier, the tidal forcing frequency reads $\hat{\omega} = 2 \left(\Omega_o - \Omega_s \right)$, where $\Omega_o$ is the orbital frequency, and we note that $\hat{\omega}$ can be \emph{negative}. 
Equations \eqref{eq:grad_u}--\eqref{eq:dsdt} constitute the full non-linear equations considered here. 

\subsubsection{Linear equations} \label{sec:lineqns}
To have a point of comparison with previous literature we first derive the equations describing linear gravito-inertial modes. 
These are similar to those from \cite{2024Pontin}, which read 
\begin{align}
    \bm\nabla \cdot \bm u &=0, \\
    \frac{\partial \bm u}{\partial t} & = -\frac{\bm\nabla p}{\rho_0} - 2\bm{\Omega_s} \times \bm u - b \bm g  - \bm\nabla \Psi + \nu\bm\nabla^2\bm u, \\ 
    \frac{\partial b}{\partial t} & + u_r \frac{d \tilde{b}}{dr}  = \kappa_{th} \nabla^2 b.
\end{align}
In this study, we consider three distinct BV profiles: a uniform BV frequency in Sec.~\ref{sec:Linear}, a uniform buoyancy variable gradient, which leads to a BV frequency varying radially and linearly with the gravitational acceleration of a homogeneous body, $\bm{g} = -g \bm e_r=- g_0 r \bm{e_r}$ in Sec.~\ref{sec:Sim}, and a realistic profile from \cite{2024Suvorov} in Sec.~\ref{sec:astrophysics}.
To solve the linear system, we take a non-penetrating boundary condition for the radial velocity at the inner boundary and a free surface where normal stress vanishes at the outer boundary following \cite{2023Pontin,2024Pontin}.
We further impose stress-free conditions (no tangential stress) at both boundaries,
with fixed buoyancy variable $(b=0)$ or its gradient $(\frac{\partial b}{\partial r}=0)$ at both boundaries as it plays little impact on the resonances \citep{2023Pontin}.
This setup will be referred to as a ``full formulation'' hereafter as it encompasses both the equilibrium and dynamical tides.
However, the implementation of a free surface in a fully non-linear code is both prohibitively costly as surface oscillations would occur at frequencies of $ \sim \omega_d$, while the rotation rate and (gravito-inertial or BV frequency) mode frequencies are much lower for NSs.

\subsubsection{A new formulation to excite gravito-inertial modes } \label{sec:Nonlinear}

A derivation for separating equilibrium and dynamical tides, especially for inertial modes, can be found in numerous studies \citep[see, for example,][and references therein]{2004OgilvieL,zahn2013,2013Ogilvie,2022Astoul}.
For the treatment of the gravito-inertial modes, we take a slightly different approach as the one used in \cite{2024DhouibIWinstratified}:
as the equilibrium tide corresponds to the instantaneous hydrostatic response to the tidal force, we assume that reequilibriation between pressure and gravity is reached instantaneously compared to the buoyancy variables (i.e., $b$ in the current context).
Another way of describing this treatment is to assume that the advection of the background gradient of the buoyancy variables by the equilibrium tide is part of the heat equation \eqref{eq:dsdt} for the dynamical tides. Then the heat equation for the equilibrium buoyancy variable (i.e. $b_e$) would be
\begin{equation}
    \frac{\partial b_e}{\partial t} = \kappa_{th} \bm \nabla^2 b_e,
\end{equation}
which gives $b_e=0$ as there are no sources terms.
This corresponds to assuming that the equilibrium tide flow $\bm{u_e}$ is fully incompressible, and thus we adopt the expression valid for $\ell = m =2$ inertial modes \citep{2013Ogilvie,LO2018,2022Astoul,2023Astoul}
\begin{equation}\label{eq:u_e}
    \vec{u_e} = \text{Re}[i\hat{\omega}\bm\nabla X e^{-i\hat{\omega} t}]
\end{equation}
with 
\begin{equation}
    X(r,\theta,\phi) = \frac{C_t}{2(1-\alpha^5)}\left[r^2+\frac{2}{3}\alpha^5 r^{-3}\right] Y_2^2(\theta,\phi),
\end{equation}
where $\bm{u_\mathrm{e}}$ is satisfying the free-surface boundary condition at the top and impenetrability at the bottom.
Here $C_t = \left(1+\text{Re}[k_2^2]\right) \epsilon$ is the tidal force amplitude related to the real part of the quadrupolar Love number $\text{Re}[k_2^2]$ and the tidal amplitude parameter $\epsilon = \left(M_{2}/M_1\right)\left(R_o/a\right)^3$. 

Equation \eqref{eq:u_e} leads to a static pressure perturbation $p_e$ that, in addition to $\bm u_e$, compensates the tidal force in the Navier-Stokes equations.
With this equilibrium tide, the linear equations for the leftover velocity $\bm u'=\bm u-\bm{u_e}$, pressure $p'=p-p_e$ and $b'=b-b_e$ become
\begin{align}
    \bm\nabla \cdot \bm u' &=0, \\
    \frac{\partial \bm u'}{\partial t} & = -\frac{\bm\nabla p'}{\rho_0} - 2\bm\Omega_s \times \bm u' - 2\bm{\Omega_s}\times \bm{u_e} - s \bm g + \nu\bm\nabla^2 u', \\ 
    \frac{\partial b'}{\partial t} & + u'_r \frac{d \tilde{b}}{dr}  = - u_{e,r} \frac{d \tilde{b}}{dr} +  \kappa_{th} \bm\nabla^2 b'.
\end{align}
This system has two effective tidal forcing: $\bm{f_t}=- 2\bm{\Omega_s} \times \bm{u_e}$, the Coriolis acceleration on the equilibrium tidal flow $\bm{u_e}$ in the momentum equation ,and $f_N=- \bm{u_e} \cdot \bm \nabla \tilde{b}$, the advection of the background entropy gradient by the equilibrium tide in the heat equation.
This second term will allow us to excite g-modes even in a non-rotating case.
It is usually not included in more general split formulations, where background density is not uniform \citep{2004OgilvieL,2024DhouibIWinstratified}, as it is considered as part of a equilibrium tidal buoyancy variable (i.e., entropy/temperature) term. 
This approach may lead to issues for g-modes in a uniform density star that may not be present with a non-uniform density, depending on the definition of the equilibrium tide. Indeed, in the Boussinesq approximation, we consider that density perturbations are only due to entropy/composition perturbations but, as described by \citet{og14}, such a density perturbation corresponds to a delta spike at the surface with radial displacement. The heat equation would then force the radial displacement to be zero, yielding a contradiction. Our formalism is a workaround to avoid this issue.
A comparison between formalisms with a non-uniform density $\rho_0$ is planned in a further study.
The use of a fixed surface also means we cannot have surface modes and, therefore, our formalism is valid for $\Omega_s^2,N^2 \ll \omega_d^2$, which is the case for NSs and main-sequence stars \citep{2013Ogilvie,2024Suvorov}.

In the following, we omit primes when using the above-introduced formulation, hereafter called the "split formulation", where the non-linear equations read
\begin{align}
    \label{eq:split_non_lin}
    \bm\nabla \cdot \bm u &=0, \\
    \frac{\partial \bm u}{\partial t} + (\bm u \cdot \bm\nabla)\bm u & = -\frac{\bm\nabla p}{\rho_0} - 2\bm{\Omega_s} \times \bm u +\bm{f_t} - b \bm g  + \bm\nabla \cdot \bm \tau_\nu, \\ 
    \frac{\partial b}{\partial t} & + (\bm{u} \cdot \bm\nabla) b + u_r \frac{d \tilde{b}}{dr} = - u_{e,r}\frac{d \tilde{b}}{dr} +  \kappa_{th} \bm\nabla^2 b.
    \label{eq:split_non_lin_end}
\end{align}
With this split formulation, we use non-penetrating boundary conditions for the velocity at both outer and inner boundaries. We also keep the stress-free boundary conditions as in the previous formulations.
For both formulations, we tested that the buoyancy variable boundary conditions did not impact significantly the results and choose therefore the one that gave the strongest amplitude: setting the gradient of the buoyancy variable to zero, i.e. $\frac{d b}{d r} =0$.

\subsubsection{Energy equations}

As we have a new forcing term in the equations, we focus on deriving the new energy balance to understand the dissipation and saturation of tidally-forced modes.
The dissipation terms correspond to the volume integrated viscous $D_{\nu}$ and thermal $D_{\kappa}$ dissipation rates given by
\begin{equation}
    D_{\nu} = - \int_V \rho_0 \nu \bm u\cdot\bm \nabla^2 \bm u\, \mathrm{d}V,\\
    D_{\kappa} =  - \int_V \rho_0 \kappa_{th} \frac{g^2}{N^2} b \bm\nabla^2 b\, \mathrm{d}V.\\
\end{equation}
\\
By the standard definition of work, the power injected by the tidal force in the full formulation is 
\begin{equation}
 P_\Psi= \int_V \rho_0 \bm u \cdot \left(-\bm\nabla \Psi\right)\, \mathrm{d}V.
\end{equation}
In this formulation, the energy balance has already been derived by \cite{2023Pontin,2024Pontin} 
and is given by
\begin{equation}
    P_\Psi = \frac{d E_K}{dt} + \frac{d E_{PE}}{dt} + D_{\nu} + D_{\kappa},
\end{equation}
where $E_K$ and $E_{PE}$ are, respectively, the kinetic and potential energy of the star. Their respective rates of change are
\begin{equation}
    \frac{d E_K}{dt} = \int_V\frac{\rho_0}{2}\frac{\partial |u|^2}{\partial t} dV, 
 \ \frac{d E_{PE}}{dt} = \int_V\frac{g^2}{2N^2}\frac{\partial b^2}{\partial t} dV,
\end{equation}
except for $N^2 = 0$ where one has $\frac{d E_{PE}}{dt}=0$.
At saturation and averaged over a forcing period of $2\pi/\hat{\omega}$, energy conservation implies
\begin{equation}
    P_\Psi = D_{\nu} + D_{\kappa}.
\end{equation}
For the split formulation, the power injected by the effective body force $\bm{f_t}$ due to Coriolis force of the equilibrium tide is given by 
\begin{equation}
   P_{Cor} = \int_V \rho_0 \bm u \cdot \bm{f_t}\, \mathrm{d}V.
\end{equation}
To obtain the power for the effective forcing $f_N$ in the heat equation, it is easier to use the buoyancy power in the Navier-Stokes equation that is defined by
\begin{equation}
    P_{buoy} = - \int_V \rho_0 u_r b g\, \mathrm{d}V.
\end{equation}
By substituting $u_r$ using the heat equation, we obtain the buoyancy power due to the tidal entropy forcing  
\begin{equation}
    P_{buoy,e} = \int_V \rho_0 u_{e,r} b g \,\mathrm{d}V.
\end{equation}
This gives the global balance of energy in the split formulation
\begin{equation}
    P_{Cor} + P_{buoy,e} = D_\nu + D_\kappa.
    \label{eq:balance_th}
\end{equation}
Note that this final balance is obtained by substituting the buoyancy power, $P_{buoy}$, by its saturation balance in the heat equation, $P_{buoy}= P_{buoy,e} - D_\kappa$, which accounts for the impact of thermal dissipation on the available power for the tidal flows. 

\subsection{Parameter space corresponding to resonant g-modes for a neutron star binary}

Previous linear studies have shown that g-modes could be resonantly-excited in a neutron star binary.
We build on the study by \cite{2024Suvorov} 
for our NS models constructed with the APR4 EOS \citep{1998APR4} coupled to a Douchin and Haensel (DH) crust \citep{2001DHcrust}. 
For our tests of both linear (Section \ref{sec:Linear}) and non-linear calculations (Section \ref{sec:Sim}), we study the resonance of g-modes in a $2.0 M_\odot$ NS, where the frequency of the first g-mode is $N=102.89$~Hz. We also consider an equal mass binary, so that the characteristic frequency is $\omega_d = 2288$~Hz for both stars.
We vary the rotation rate $\Omega_s$ in this study with $\Omega_s=\frac{100}{\pi}$~Hz being a ``standard'' value. The resonant gravity, inertial, or mixed gravito-inertial modes are well-separated from  $\omega_d$ and lie in the low-frequency regime.
In addition, the tidal amplitude is as small as $\epsilon \sim 10^{-3}$ near
the g-mode resonance when $2 \Omega_{o} \sim N $.
According to previous studies on inertial modes \citep{2022Astoul,2023Astoul}, this places us firmly in the linear regime but this could change for a lower viscosity, higher aspect ratio ($\alpha > 0.5$), or for the resonant gravito-inertial modes with the highest kinetic energy.
 
With respect to dissipation, we choose our standard viscosity and thermal diffusivity to be $\nu=\kappa= 10^{-6} R_o^2 \omega_d \approx 1.7 \times 10^{10}$~cm$^2$ s$^{-1}$, similar to \cite{2023Pontin,2024Pontin}. This value is many orders of magnitude higher than the microphysical viscosity expected in NSs \citep{1993ThompsonPNS} but is chosen to prevent numerical issues. These choices give an Ekman number of 
\begin{equation}
    Ek \equiv \frac{\nu}{R_o^2 \Omega_s} \approx 7.2 \times 10^{-5}.
\end{equation}
The control parameter for the BV frequency in the simulations is the Rayleigh number, defined as 
\begin{equation}
    Ra \equiv - \frac{N^2 R_o^4}{\kappa \nu} \approx -2.02 \times 10^{9}. 
\end{equation}
As an astrophysical application, we study a more standard equal-mass neutron star binary of $1.6 M_\odot$ in Section \ref{sec:astrophysics}. The characteristic frequency is then $\omega_d=1941$ Hz. We also use a ``realistic'' BV frequency and its value at the outer boundary to define the the Rayleigh number. The dimensionless numbers are $Ra=-5.1 \times 10^{10}$ and $Ek =6.2\times 10^{-5}-1.99\times 10^{-5} $ with a rotation rate of $\Omega_s =\frac{100}{\pi}-100$ Hz, respectively.

\subsection{Numerical methods} \label{sec:numerics}

To solve the linear equations, we use the spectral MHD code \texttt{Dedalus3} \citep{burns20}. 
\texttt{Dedalus} has been used to handle different types of hydrodynamical and MHD problems, including convection, waves, and magnetic fields in an unstratified or stratified medium in many studies \citep{2017LecoanetgravityMHD,2018CoustonDedalus,2023JiFullerTayler}. Using the full formulation, we benchmarked the code results against the study by \citet{2023Pontin} and are therefore confident of its results for when we will use the split formulation. The resolution used for the linear problem was  $(n_r,n_{\theta})=(200-400,200-400)$ depending on the parameters.

In order to integrate the \emph{non-linear equations} in time, Eqs.~\eqref{eq:split_non_lin}--\eqref{eq:split_non_lin_end}, we use the benchmarked pseudo-spectral code \texttt{MagIC} \citep{2002WichtMagic,2012GastineCode,2013SHTNS}.
\texttt{MagIC} solves the 3D MHD equations in a spherical shell using a poloidal-toroidal decomposition for the velocity and the magnetic field (made possible by the solenoidal assumption on $\vec{u}$),
\begin{align}
\rho_0 \vec{u} &= \vec{\nabla} \times \vec{\nabla} \times \left(W \ \vec{e_r}\right) + \vec{\nabla} \times \left(Z \ \vec{e_r}\right), \\
\vec{B} &= \vec{\nabla} \times \vec{\nabla} \times \left(b \ \vec{e_r}\right) + \vec{\nabla} \times \left(a_j \ \vec{e_r}\right),
\end{align}
where $W$ and $Z$ are the poloidal and toroidal kinetic scalar potentials, respectively, while $b$ and $a_j$ are the magnetic potentials. 
The scalar potentials and the pressure $P$ are decomposed on spherical harmonics for the colatitude ($\theta$) and the azimuthal ($\phi$) angles, together with Chebyshev polynomials in the radial direction.
For more detailed descriptions of the associated spectral transforms and the numerical method, we refer to \citet{1981GilmanAnel}, \citet{1997TilgnerJFM}, and \citet{2015CHRISTENSEN245}.

The simulations presented in this paper were performed using a standard grid resolution of $(n_r,n_{\theta},n_{\phi})=(257,256,512)$.
Some of the simulations have a cutoff $m_{max}$ in the spherical harmonics to speed up the simulations; for simulations in Section \ref{sec:Sim}, $m_{max}=85$ is enough to ensure convergence as the dominant mode corresponds to $m=2$. 

\section{Linear tests of the formulation}
\label{sec:Linear}

In this section, we test the new split formulation. As a first step, we show that we are able to reproduce the properties of g-modes obtained with the ``full'' formulation (Sec.~\ref{sec:puregmodes}). This sets us up to study gravito-inertial modes to check whether the two effective tidal forcings give the correct geometry and amplitude of the gravito-inertial modes (Sec.~\ref{sec:gimodestest}).  
In order to compare more easily with \citet{2023Pontin}, we adopt in this section the same units as that study: $R_o= 1$, $\omega_d=1$ and $\rho_0=1$.

\subsection{Pure g-modes} \label{sec:puregmodes}

\begin{figure}%[ht] 
    \centering \includegraphics[width=0.45\textwidth]{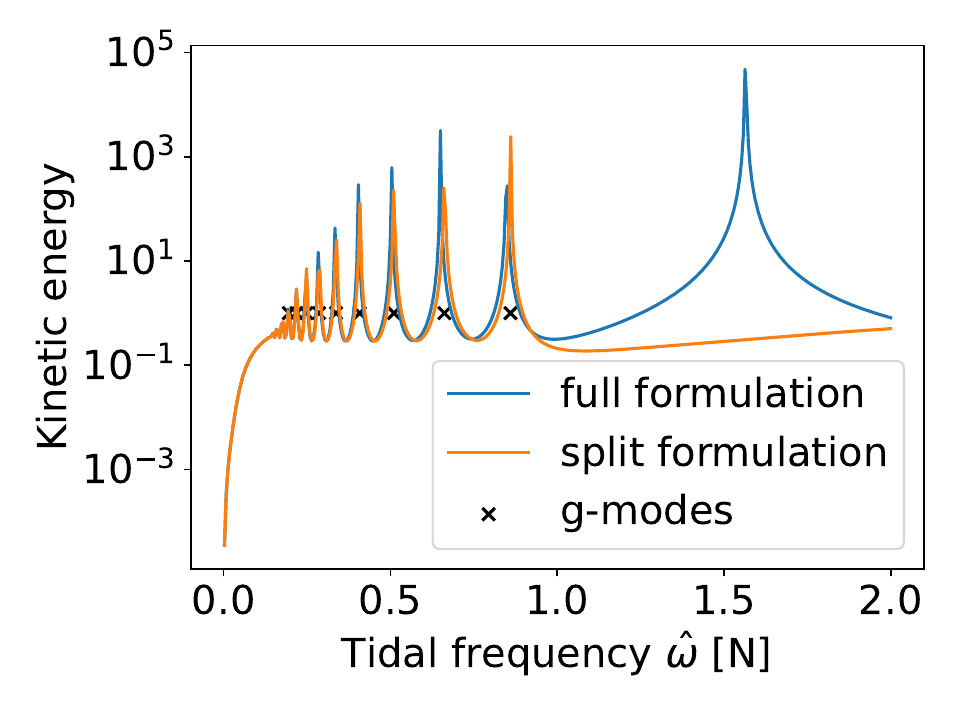}
\caption{Kinetic energy as a function of the tidal frequency $\hat{\omega}$ depending on the formulation used (see plot legends). Crosses correspond to the g-mode frequencies in a cavity \citep[Equation (B9) of][]{2023Pontin}. Code units are used such that $R_o= 1$, $\omega_d=1$, and $\rho_0=1$.}
\label{fig:diss_g_test}
\end{figure}

As a benchmark of the \texttt{Dedalus} code and the full and split formulations, 
we first consider non-rotating stars ($\Omega_{s} = 0$) and also fix $N^2=1$ and $\Psi_0 =1$.
For the split formulation, we further use a dimensionless tidal amplitude $C_t=\Psi_0/\omega_d^2= \sqrt{6\pi/5}\epsilon=1$ in this section.
We then solve the linear system for both the full formulation and split formulation. In order to verify that mode resonances behave the same, we first plot the kinetic energy as a function of the tidal forcing frequency $\hat{\omega}= 2 \Omega_o$.
Fig. \ref{fig:diss_g_test} shows that the dissipation peaks reside at the same frequencies until the tidal forcing frequency becomes close to the f-mode frequency, $\omega_d=1$. This shift was found in \citet{2023Pontin}, where the pure g-mode frequencies computed were slightly off the resonant frequencies, with disagreement growing as one approaches $\omega_d$. The split formulation recovers, with high accuracy (<$0.4\%$), the theoretical frequencies of $\ell = m =2$ g-modes ($_n g_2^2$ for radial node-number $n$); in a cavity for a constant BV frequency $N$, these are given by~\citet{2023Pontin}
\begin{equation}
\omega_{_n g_2^\ell}^2 = \frac{4l(\ell + 1)N^2\left(\ln(\alpha R_o)\right)^2}{(2\ell + 1)^2\left(\ln(\alpha R_o)\right)^2 + 4\pi^2n^2}.
\label{eq:gmodes}
\end{equation}

\begin{figure}[ht]
    \centering
\begin{subfigure}[b]{0.45\textwidth}
\includegraphics[width=0.49\textwidth]{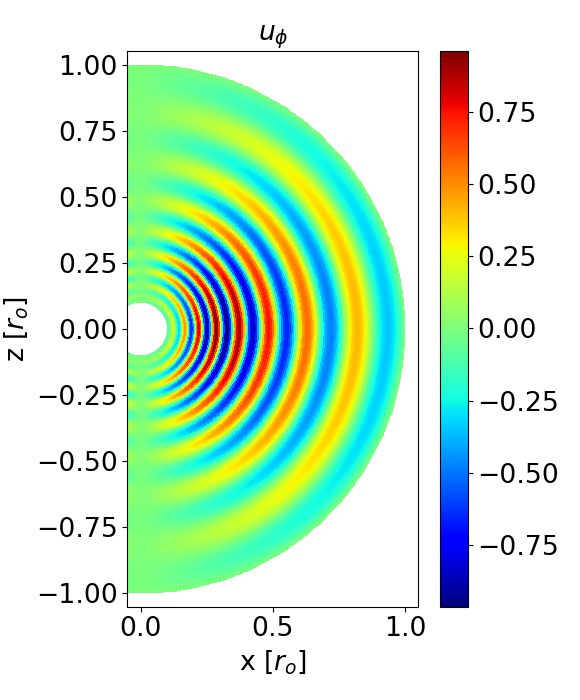}
\includegraphics[width=0.49 \textwidth]{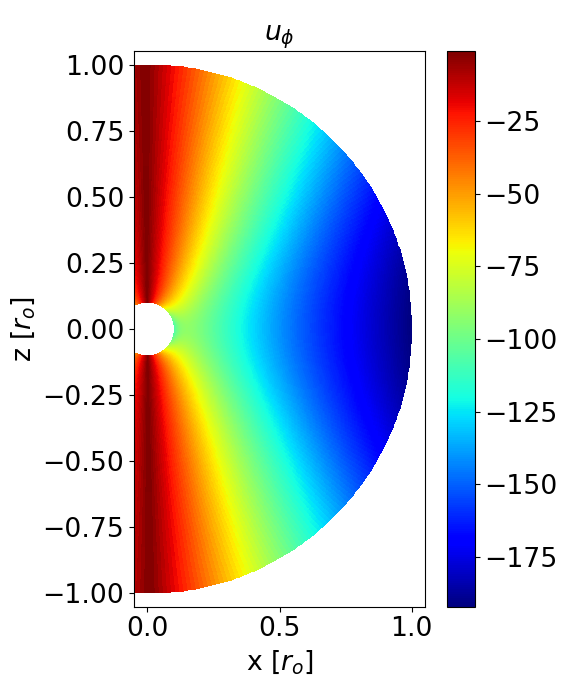}
     \caption{Toroidal velocity $u_\phi$ in the full formulation for $\hat{\omega}=0.1$ (left) and $\hat{\omega}=1.56$ (right) at $\phi=0$.}
\end{subfigure}
\begin{subfigure}[b]{0.45\textwidth}
\includegraphics[width=0.49\textwidth]{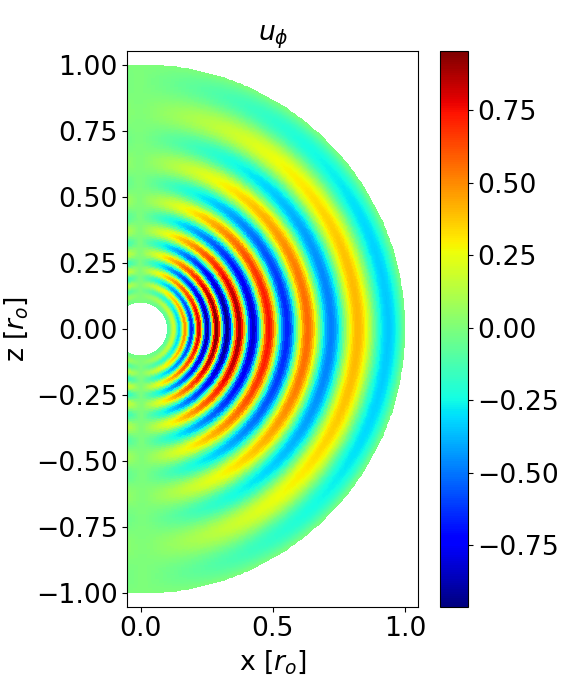}
\includegraphics[width=0.49\textwidth]{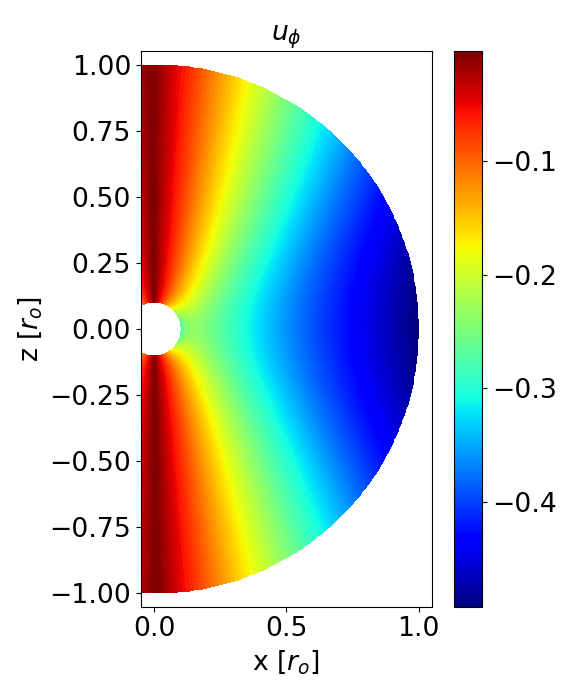}
    \caption{Toroidal velocity $u_\phi$ in the split formulation for $\hat{\omega}=0.1$ (left) and $\hat{\omega}=1.56$ (right) at $\phi=0$.}
\end{subfigure}
\caption{Comparison of the azimuthal eigenfunctions for resonant modes in the full formulation (top) and the split formulation (bottom) for two representative $\hat{\omega}$ (see subcaptions). 
}
\label{fig:uphi_g_test}
\end{figure}

In addition to the characteristic frequencies of g-modes, the split formulation in the regime where $\hat{\omega}$ is far from $\omega_d$ is similarly able to reproduce the eigenfunction of the resonant mode obtained by the full formulation to very high accuracy, within $<0.01\%$ (left panels of Figure \ref{fig:uphi_g_test}). 
However, when the tidal forcing frequency $\hat{\omega}$ is of the same order of magnitude $\omega_d$, the split formulation cannot reproduce the resonance as it is expected.
Indeed, the split formulation cannot reproduce f-modes as the surface needs to be free. 

\subsection{Gravito-inertial modes} \label{sec:gimodestest}

\begin{figure}[ht]
    \centering
    \includegraphics[width=0.45\textwidth]{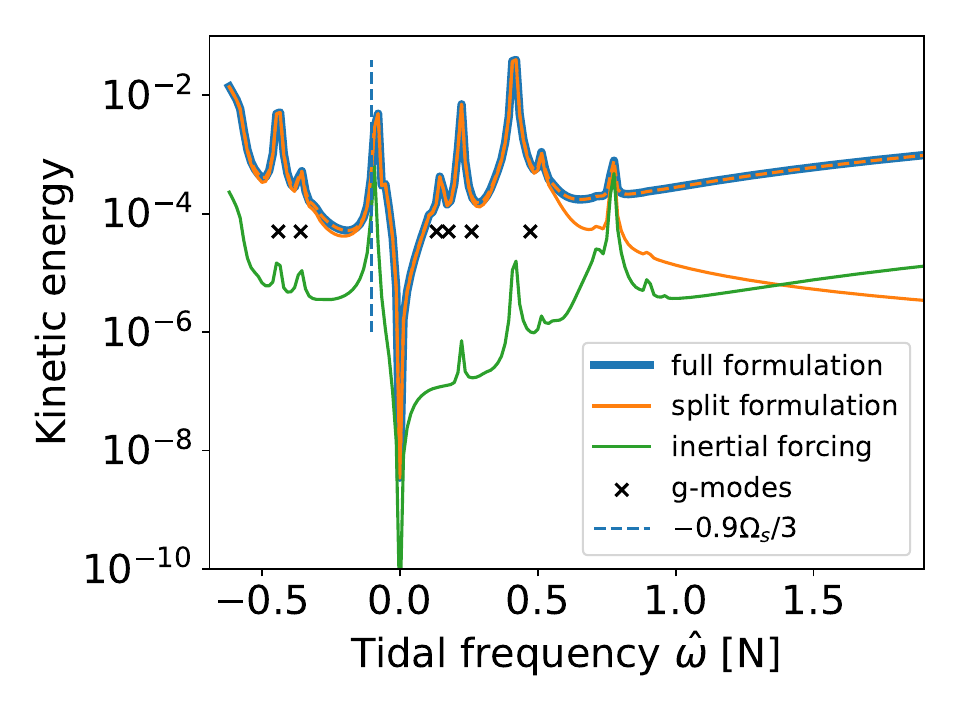}
    \caption{Kinetic energies of tidally-excited fluid motions as functions of the tidal forcing frequency $\hat{\omega}$ when the full tidal potential is taken into account (blue) and when the tidal forcing is effectively described using the equilibrium tidal flow (orange). The orange dashed line also corresponds to the split formulation but with different boundary conditions (see text), while the solid, green line corresponds to the modes obtained only with the effective tidal forcing $\bm{f_t}$ due to Coriolis force. Pure g-modes frequencies (black, Eq. \ref{eq:gmodes}) and negative g-modes (red) are designated by a cross. A Rossby mode resonance is highlighted at $\hat{\omega} = -0.9 \Omega_s/3$.} 
\label{fig:diss_g_iner_test}
\end{figure}

We apply the split formulation to tidal resonance problem for low-frequency modes in rotating neutron stars, where inertial modes, and gravito-inertial modes also come into play.
We are therefore in the regime where $\hat{\omega} \ll \omega_d$ and $\Omega_s, N \ll \omega_d$.
For our second test, we want to probe the impact of the rotation on the gravity modes (i.e. to compare g-modes with gravito-inertial modes) in the right parameter regime. We set $\Omega_s=0.0139\,\omega_d$ and $N=0.0450\,\omega_d$, respectively. 

\begin{figure}[ht] 
    \centering
\begin{subfigure}[b]{0.45\textwidth}
\includegraphics[width=0.45\textwidth]{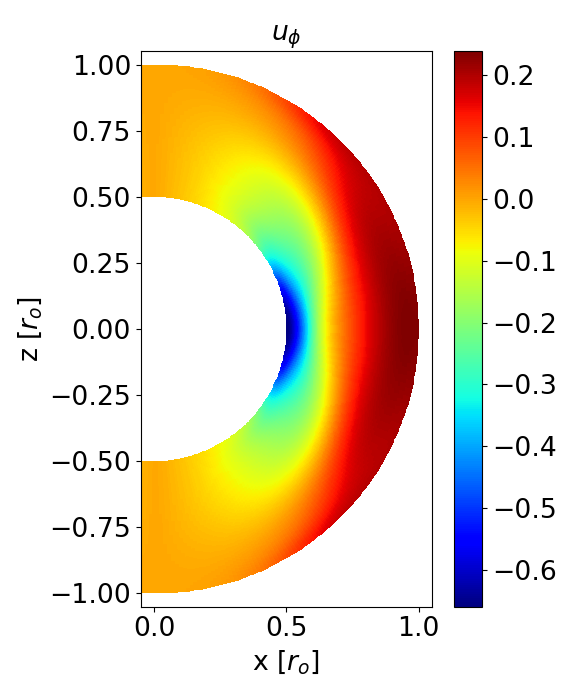}
\includegraphics[width=0.45 \textwidth]{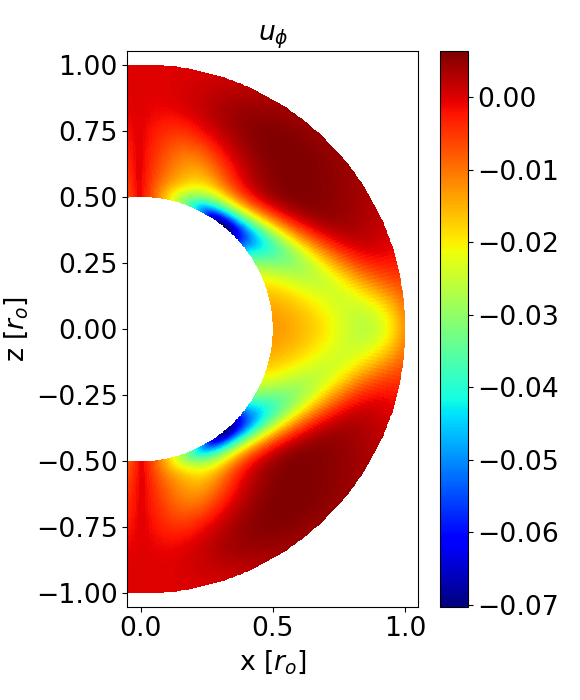}
\caption{Toroidal velocity $u_\phi$ at $\phi=0$ in the full formulation for $\hat{\omega}=0.421 N$ (left) and $\hat{\omega}=0.777 N$ (right).}
\end{subfigure}
\begin{subfigure}[b]{0.45\textwidth}
\includegraphics[width=0.45\textwidth]{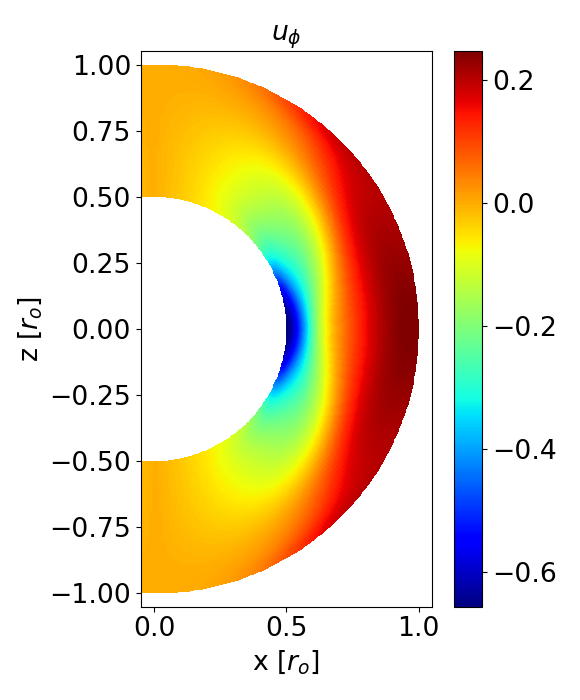} 
\includegraphics[width=0.45\textwidth]{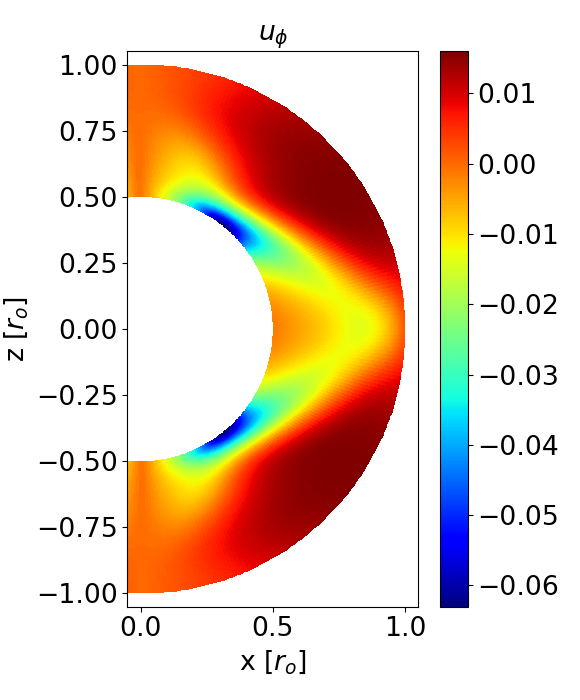}
    \caption{Toroidal velocity $u_\phi$ at $\phi=0$ in the split formulation for $\hat{\omega}=0.421 N$ (left) and $\hat{\omega}=0.777 N$ (right).}
\end{subfigure}
\begin{subfigure}[b]{0.45\textwidth}
\includegraphics[width=0.45\textwidth]{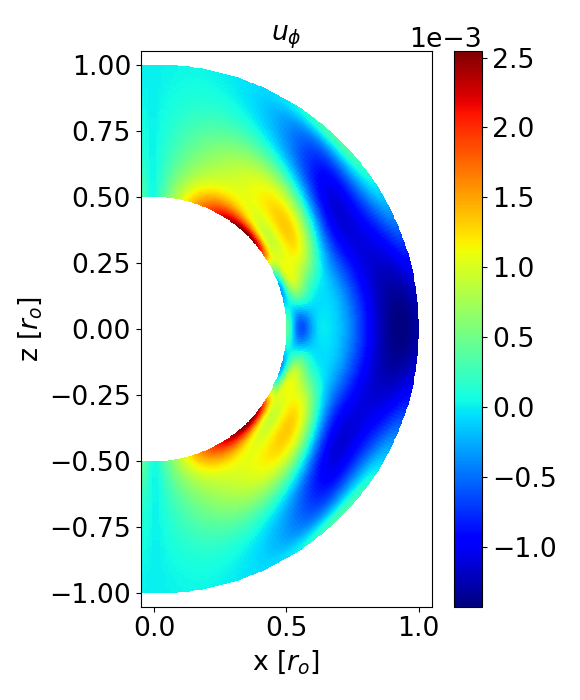}
\includegraphics[width=0.45\textwidth]{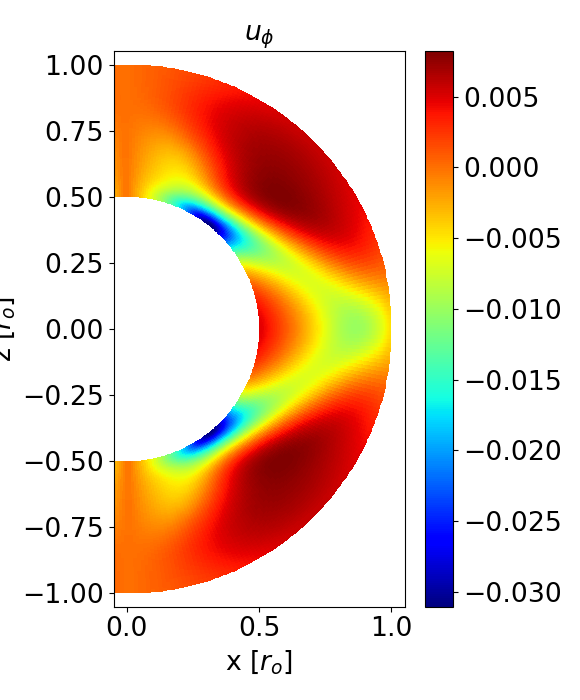}
    \caption{Toroidal velocity $u_\phi$ at $\phi=0$ with only the effective Coriolis forcing $\bm f_t$ for $\hat{\omega}=0.421 N$ (left) and $\hat{\omega}=0.777 N$ (right). 
    }
\end{subfigure}
\caption{Comparison of two gravito-inertial modes along $\phi=0$ for the different formulations. The first mode, with $\hat{\omega}=0.421 N$, corresponds to the $_1g^2_2$ mode impacted by the rotation, while the second one is closer to an inertial or gravito-inertial mode, $\hat{\omega}=0.777 N$. Velocities are given in units such that $R_o \omega_d=1.56 \times 10^{10}$ cm s$^{-1}$.}
\label{fig:uphi_g_iner_test}
\end{figure}
With these parameters, we compute the kinetic energy of the resonant modes depending on the tidal frequency for the two formulations. Since we are far away from the respective f-mode frequencies, we find that the peak kinetic energy frequencies are identical for both formulations (Figure \ref{fig:diss_g_iner_test}). 
We also find that the range of gravito-inertial mode is roughly  $[-2\Omega_s,N]$, which is consistent with the limit $\hat{\omega}_{max}=N\sqrt{1+4\Omega_s^2/N^2}\approx 1.18 N$ found in previous studies \citep{R2009,2024Pontin}. 
By looking at their radial wavelengths, we identify the gravito-inertial modes corresponding to the pure g-modes $_4g$ to $_1g$ with slightly smaller frequencies compared to Eq. (\ref{eq:gmodes}) in the positive frequencies and from $_3g$ to $_2g$ modes in the negative frequencies. We find that these negative modes corresponding to $n=2$ and $n=3$  seems to be roughly at frequencies corresponding to $\hat{\omega}_{_{n} g^2_{-2}}  \approx -\omega_{_n g^2_2}- \frac{3}{5}\Omega_s$ \citep[i.e. rotation breaks symmetry; in the non-rotating case, negative g-mode are perfectly symmetrical to the positive as in][]{2023Pontin}. 
The peak at the lower limit of frequencies $\hat{\omega}=-2\Omega_s$ is related to the negative $_1g$ mode but the resonance peak seems not to be reached (not shown here). 
We also identify a resonant mode close to the frequency of a resonant Rossby mode at $\hat{\omega} = -0.9 \frac{\Omega_s}{3}$. The usual frequency for this resonance is $\hat{\omega} = -\Omega_s/3$ but it is expected to decrease with $\alpha$ \citep{O2009,2024Pontin}.
The identification of the resonant modes with $\hat\omega>\omega_{_1g^2_2}\approx0.47$ is not straightforward since they cannot by predicted from Eq. (\ref{eq:gmodes}) and thus are not slightly shifted 
g-modes. 
For these other modes, the presence of stratification, which has the effect of shifting resonant peaks (compared to neutrally-stratified studies), % from Gordon13, Riccardo+ work
does not allow clear comparison with any of the regular or singular inertial modes of \citet{2013Ogilvie} or \citet{2023Astoul}.
The next resonant peak corresponding to $\hat\omega\approx0.51$ seems to be a  gravito-inertial mode as it has some mixed g-mode and inertial mode features and the $n=1$ radial wavelength. 
This mode is still in the sub-inertial range $\hat\omega<2\Omega_s$ where we expect to find equatorially trapped hyperbolic gravito-inertial modes \citep[as described in][]{DR1999,DR2000,M2009}.
The higher frequency peaks with $\hat\omega>2\Omega_s$ (in the super-inertial regime) presumably belong to the elliptic gravito-inertial mode family, as described in the aforementioned studies.

We also compare the kinetic energy to the split formulation but without the forcing term $f_N$ in the heat equation, so just with $\bm{f_t}$ which we call inertial forcing (green line on Figure \ref{fig:diss_g_iner_test}). Some of the peaks linked to the stable stratification and g-modes cannot be seen anymore, are strongly mitigated (especially for low frequency), or change geometry completely as shown on the $_1g$ mode (left panel of Figure \ref{fig:uphi_g_iner_test}). Without it, it corresponds to having gravito-inertial modes with the gravity component only secondarily excited by the effective Coriolis forcing. 
For example, the mode at $\hat{\omega} = 0.777 N$ is well recovered with only the effective Coriolis forcing although the amplitude is divided by two (right panel of Figure \ref{fig:uphi_g_iner_test} but at the frequency of the $_1g$ mode the mode is completely different (left panel of Figure \ref{fig:uphi_g_iner_test}). 

\begin{figure}[ht] 
    \centering
\includegraphics[width=0.24\textwidth]{Figures/pontin_up_eval_om107.png}
\includegraphics[width=0.24\textwidth]{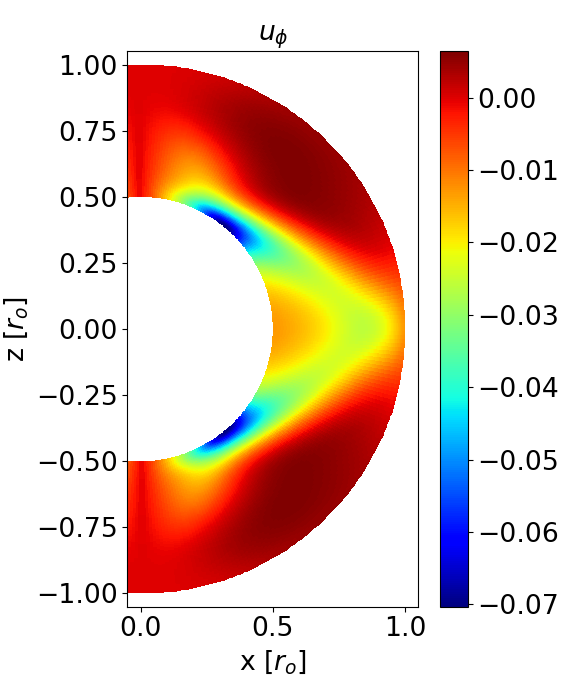}
    \caption{Toroidal velocity $u_\phi$ with $\hat{\omega}=0.777 N$ in the full formulation (left), and in the split formulation (right) with modified non stress-free boundary conditions (see text). %applied to $u+u_e$ instead of $u_e$ 
    }
\label{fig:stress_free_comp}
\end{figure}

For all the peaks below $\hat{\omega}<0.5 N$ , the kinetic energy for both formulations agrees within $<1\%$ but there are some differences with the full formulation as the amplitude of the velocity becomes lower in the split formulation for the tidal frequencies $\hat{\omega}$ larger than the BV frequency.
This means that some tidal response seems to not be captured by the split formalism when further away from the g-mode or gravito-inertial mode frequencies. 
In order to investigate this effect, we look at the resonant mode for two different tidal forcing frequencies. For $\hat{\omega}<N$, the gravito-inertial mode velocity is  reproduced within $0.5\%$ by the split formulation (two upper left panels of Figure \ref{fig:uphi_g_iner_test}). We can also see the impact of the rotation, which makes the mode more cylindrical than the spherical symmetry of pure g-modes. 
For $\hat{\omega} \approx N$, the toroidal velocity is not exactly reproduced by the split formalism, and the kinetic energy seems to decrease with increasing tidal frequency since we leave the allowed frequency range for gravito-inertial waves to be excited and equilibrium tide is not negligible anymore. 
This behavior can be explained by differences in the boundary conditions. Indeed, when outside of the  resonant peaks, the amplitude of the equilibrium tide $\bm{u_e}$ cannot be neglected compared to those of the dynamical tide $\bm u$. 
This leads to a difference in boundary conditions as the full formulation solve the equations for both tides $\bm{u}+\bm{u_e}$, while the split formulation solve the equation for the dynamical tide $\bm u$. In order to compare the two formalisms, we artificially take into account the impact of the equilibrium tide by 
changing our boundary conditions for the viscous rate tensor to $\tau_{\nu,r\theta} (\bm u) = - \tau_{\nu, r\theta} (\bm u_e) $ and $\tau_{\nu,r\phi} (\bm u) = - \tau_{\nu, r\phi} (\bm u_e)$.
Applying these changes, we recover the results from the full formulation within $<0.05\%$ (Figure \ref{fig:stress_free_comp}).

\begin{figure}[ht]
    \centering    \includegraphics[width=0.45\textwidth]{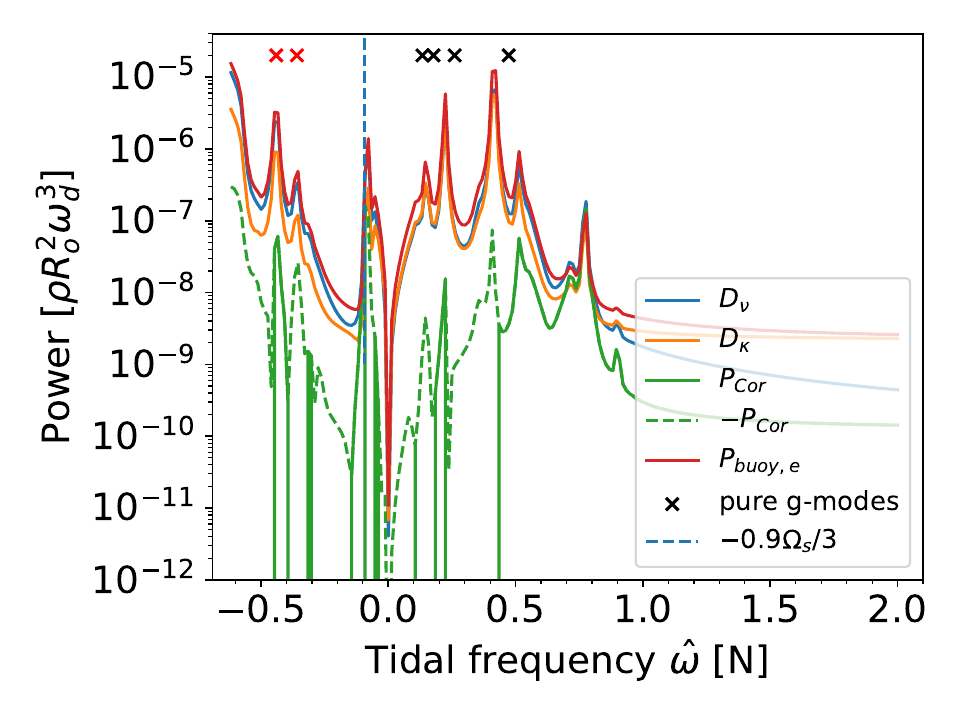}
    \caption{ Power/dissipation balance of the tidally-excited fluid motions as function of the tidal forcing frequency $\hat{\omega}$ with the split formulation of the tidal forcing with $N/\Omega_s=3.24$. Similar to Figure \ref{fig:diss_g_iner_test}, the modes are designated by crosses. 
    }
\label{fig:power_iner_test}
\end{figure}

Since the split formulation has two effective forcing terms, $\bm f_{t}$ and $f_{N}$, we can compute the power associated with both to assess their importance.
Figure \ref{fig:power_iner_test} shows the different components of the power of the forcings and the thermal and viscous dissipation due to the tidal flows. We first verify that the balance of Equation (\ref{eq:balance_th}) is verified and the error that depends on resolution is around of $10^{-7} \%$.
For almost all the modes, the power of the forcing $f_N$ due to the advection of the buoyancy variable gradient dominates, which is expected since we are in the regime $N > \Omega_s$. It even dominates at a resonant Rossby frequency $\hat{\omega}= - \Omega_s/3$ where the contribution of the effective Coriolis forcing is the strongest for negative frequencies and is around $P_{Cor} \approx 0.4 P_{buoy}$. 
The only exception is of the gravito-inertial modes found for $\hat{\omega} = 0.777 N$, described before.
We also find that the power due to the Coriolis effective forcing $\bm f_t$ alternates between negative and positive values. This may be linked to the splitting due to rotation of the gravity modes $m=2$ and $m=-2$. 

The case of pure inertial modes with the effective body force $\bm{f_t}$ has been well studied and therefore needs no testing.
Overall, these tests show that the split formulation with the two forcing terms is able to obtain well the resonant gravito-inertial modes without having a free surface implemented.

\section{Non-linear simulations of gravito-inertial modes}
\label{sec:Sim}

As the orbital separation $a$ shrinks with time due to the emission of GWs, the tidal forcing frequency and amplitude 
increase, and non-linear effects can start to be important for the dynamics of the resonance. In this section, we wish to test and study the importance of non-linear effects for binary NSs. 
In particular, we now need to look at the non-linear saturation of the gravito-inertial modes (i.e. when the non-linear simulation reach an average steady state).
 
We implement the split formulation in the code \texttt{MagIC} with the stress-free conditions applied %only
to the dynamical tide $\bm u$, since it does not change results for resonant modes. In a similar fashion to \citet{2023Astoul}, we also choose to neglect 
some non-linear interactions between the equilibrium and dynamical tides. Indeed, our model uses a non-deformed spherical shell which, in the presence of mixed (equilibrium-dynamical) nonlinear terms, would lead to a non-vanishing equilibrium tide at the surface $\bm{u_e} \cdot \vec{n} \neq 0$. In this case, the advection of the dynamical flow by the equilibrium flow artificially generates angular momentum and spin-up the star. We, therefore, neglect this advection term to avoid this setup artifact and because it is justified in the astrophysical regime \citep[see discussion of][]{2022Astoul}. 
For the saturation of the flow, we only consider the non-linear effects of the dynamical tide on itself $(\bm u\cdot\bm\nabla)\bm u$.

We use this implementation to run several simulations for a NS binary of 2 solar masses adopting an initial rotation of $\Omega_s=100/\pi$ Hz to continue the comparison as in the previous section. One notable difference is that, in this section, the gradient of buoyancy variable $\frac{d \tilde{b}}{dr} \propto \frac{N}{g}$ is constant instead of the BV frequency for the sake of simplicity. The BV frequency now depends on the radius and we have $N(r)=N r$ and $g = g_o r$. 

For the amplitude of the tidal force, we choose the tidal parameter to reflect the situation of having $\hat{\omega} \approx N$ so as to study the resonance of gravito-inertial modes, which gives a tidal amplitude of $C_t \approx 10^{-3}$ for $N = 102.87$ Hz, $\Omega_s = 100/\pi$ Hz, and $\omega_d=N/0.0450$. 
In order to investigate differences in non-linear effects between the different modes, we keep the tidal amplitude constant for all simulations in this section. Note that we would have $C_t=\sqrt{6\pi/5} \epsilon \propto (\Omega_o/\omega_d)^2$ (neglecting the real part of the Love number) in the realistic case (see Section~\ref{sec:astrophysics}). 

\subsection{Linear and non-linear comparison}

\begin{figure}[ht]
    \centering
    \includegraphics[width=0.49\textwidth]{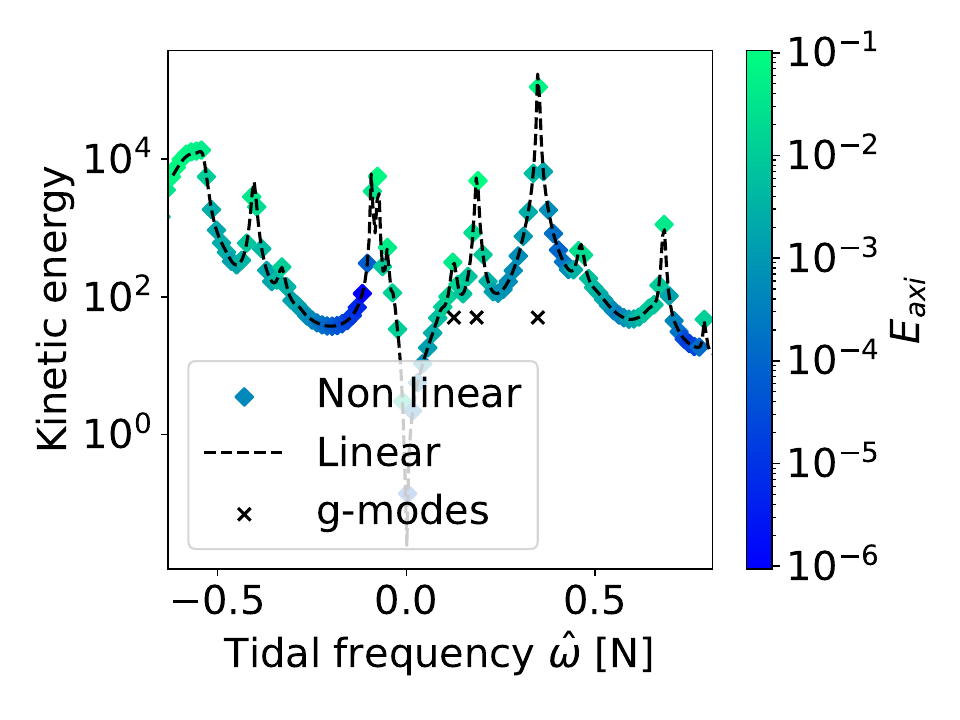}
\caption{Kinetic energy along the tidal frequency $\hat{\omega}$ for the linear code (dashed lines) or the non-linear simulations (scatter points). The colour scale of the non-linear simulations corresponds to the ratio of the axisymmetric kinetic to the total kinetic energy. }
\label{fig:non_lin_vs_lin}
\end{figure}

\begin{figure}[ht]
    \centering
    \includegraphics[width=0.23\textwidth]{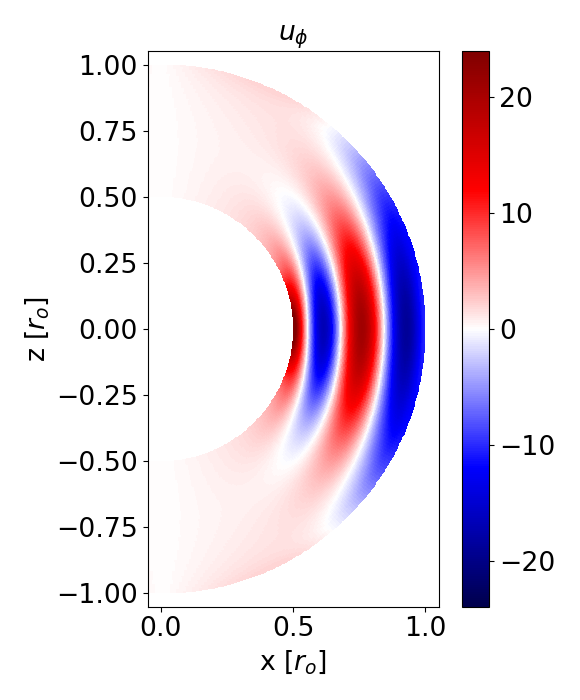}    
    \includegraphics[width=0.23\textwidth]{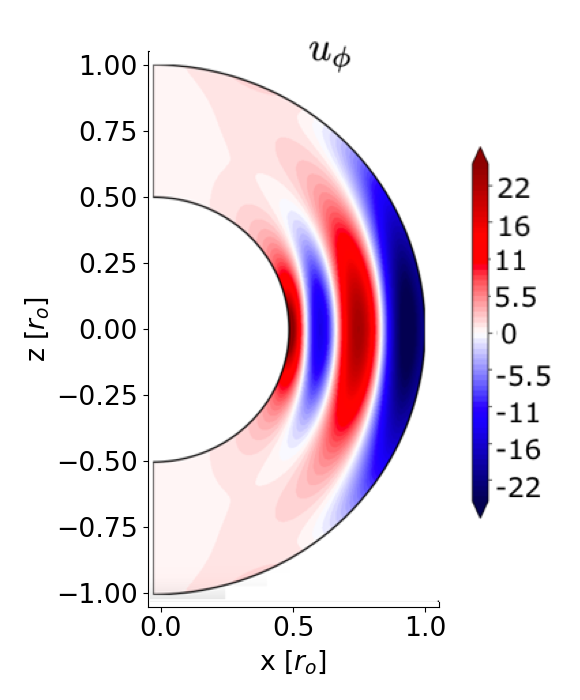}
    \caption{$_3$g-mode resonance at $\phi=0$ in the case of small rotation ($N/\Omega_s=3.24$). 
    Results of the linear code are shown on the left, while that of the non-linear code are on the right; notice the difference in the color bar for each panel.} 
    \label{fig:comp_mode_non_lin_lin}
\end{figure}

To be confident in our non-linear simulations, we first compare the \texttt{MagIC} simulation results to the linear results. We run the linear code for $
\hat{\omega} \in [-2\Omega_s,2 N]$ with a frequency resolution $N_{freq}= 500$, while $N_{freq}= 100$ for the non-linear simulations.
Figure \ref{fig:non_lin_vs_lin} shows the total kinetic energy of the different modes for both the linear code and the non-linear code \texttt{MagIC}. Overall, we find a good agreement between the two codes for many resonant modes, and the comparison of one of these cases can be seen for the $_3g$ mode (first mark furthest to the left on Figure \ref{fig:non_lin_vs_lin}) in Figure \ref{fig:comp_mode_non_lin_lin}.  
The mode is deformed by rotation and becomes more cylindrical compared to a pure g-mode (see left panels of Figure \ref{fig:uphi_g_test} for a pure g-mode at a different frequency).
We still notice small differences for some resonance modes, but such effects may stem from the two different frequency resolutions or due to small non-linear effects. 

To quantify the importance of non-linear effects, Figure \ref{fig:non_lin_vs_lin} shows the ratio of the axisymmetric-to-total kinetic energy in these simulations. It clearly shows that with the strongest resonance modes, a $m=0$ zonal flow component emerges in addition to the forced $m=2$ component. This means that non-linear effects are stronger for these modes. 
It is not the only criterion though, as the $_2g^2_2$ mode has the highest percentage of axisymmetric energy, while the $_1g^2_2$ mode has the highest mode energy.  

\subsection{Non-linear saturation}
\label{non_lin_sat_tests}

\begin{figure}[ht]
    \centering
    \includegraphics[width=0.49\textwidth]{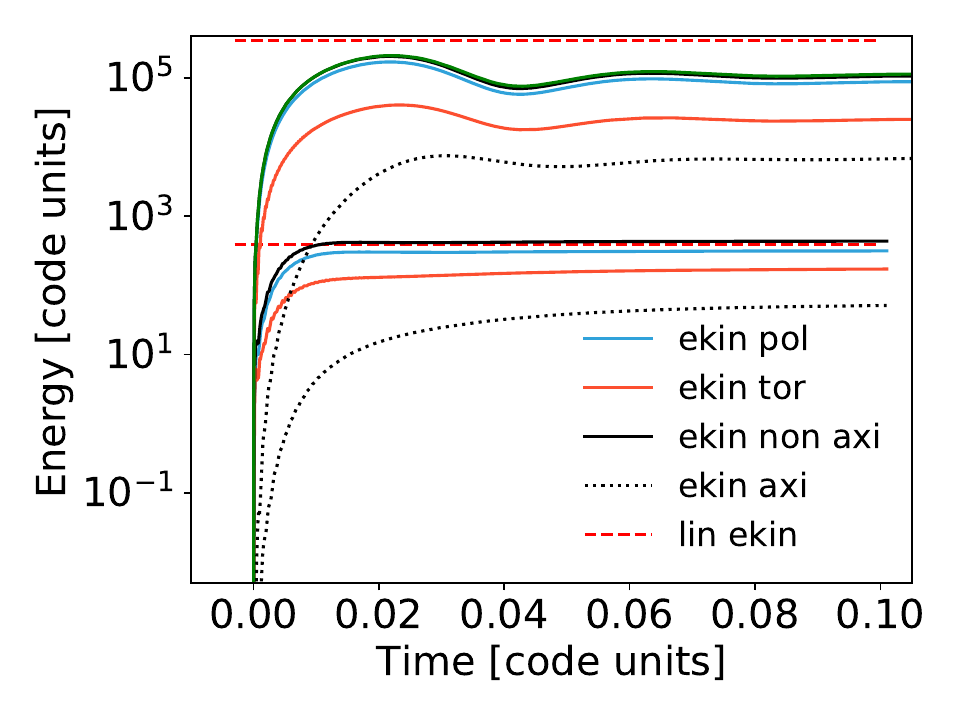}
\caption{Time evolution of the kinetic energy for the non-linear simulations of $_1g$ and $_2g$ modes compared to the linear saturations (dashed red lines). The above lines correspond to the $_1g$ mode and below lines $_2g-$ mode, which have been rescaled by a factor $0.1$ for improved visibility. 
}
\label{fig:non_lin_time_evolution}
\end{figure}

The previous section showed that for $C_t=10^{-3}$ the non-linear saturation of the kinetic energy matches the linear one. In this section, we examine this assumption to see if it still holds for the strongest resonances. Since non-linear effects could become important, we also compare the linear and non-linear simulations for cases with the highest percentage of axisymmetric energy and the highest absolute axisymmetric energy. They correspond to the $_2g-$ and $_1g$ modes, as discussed above. 
We first check that in the case of $C_t=10^{-9}$, the kinetic energy of both non-linear simulations agree with the associated linear simulations within a relative error $<10^{-6}$. 
Figure \ref{fig:non_lin_time_evolution} compares
the time evolution of the non-linear simulations and the saturated energies predicted by the linear calculations for both modes. 
We find that the non-axisymmetric energy of the $_2g$-mode agrees well with the linear kinetic energy. We note 
that its total saturation is slightly increased by the axisymmetric energy, which is $10\%$ of the total energy.
In the case of the $_1g$-mode, the non-linear saturation is lower by a factor $3$. This means that the saturation mechanism differs from the linear theory for this mode. It is possible that we only see it for this mode due to it having the strongest amplitude. 

\begin{figure}[ht]
    \centering
    \begin{subfigure}[b]{0.45\textwidth}
    \includegraphics[width=0.45\textwidth]{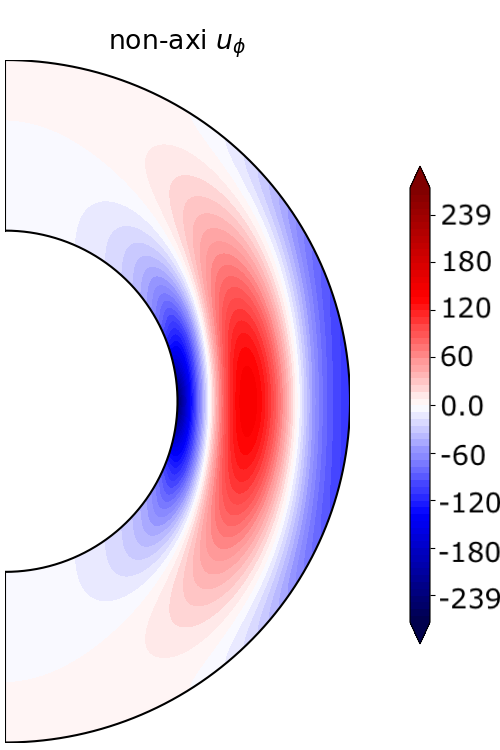}
    \includegraphics[width=0.45\textwidth]{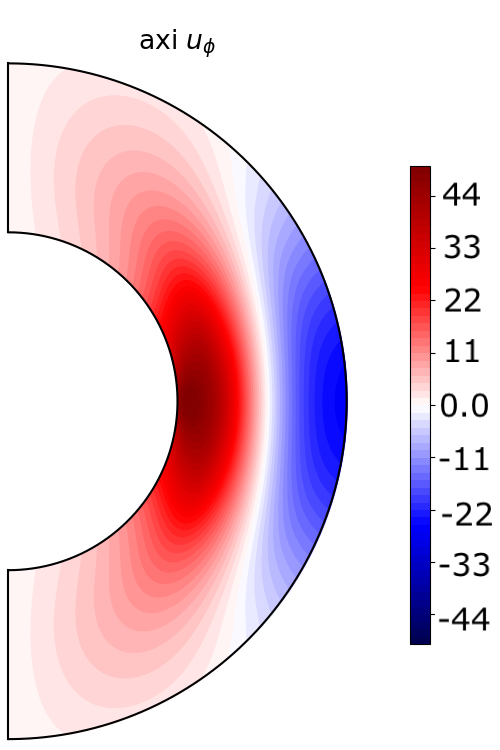}
    \caption{Non-axisymmetric (left) and axisymmetric (right) toroidal velocity $u_\phi$ for the $_2g$ mode.}
\end{subfigure}
\begin{subfigure}[b]{0.45\textwidth}
    \includegraphics[width=0.45\textwidth]{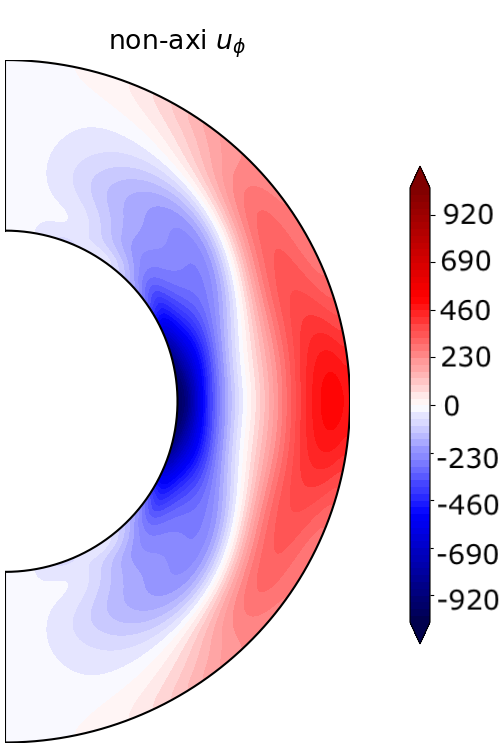}
    \includegraphics[width=0.45\textwidth]{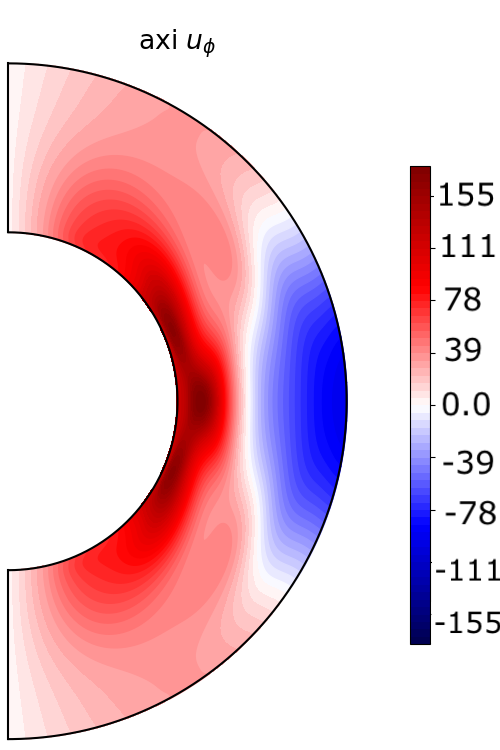}
    \caption{As above but for the $_1g$ mode.}
\end{subfigure}
    \caption{Strongest two g-mode resonances and their non-linear saturation in the case of small rotation ($N/\Omega_s=3.24$).}
    \label{fig:non_lin_saturation}
\end{figure}

Figure \ref{fig:non_lin_saturation} compares the non-axisymmetric and axisymmetric toroidal velocity of the two modes to see the impact of the non-linear saturation.
We find that axisymmetric rotation, decreasing with cylindrical radius, is generated for both modes.
We expect that the differential rotation emerges due to non-linear processes such as tidal wave-wave interactions \citep[$m=2$ and $m=-2$ non-linear coupling to form an $m=0$ mode, as explained in][for inertial waves]{BT2018,2022Astoul} and/or wave breaking as in \citet{BO2010} in a 2D cylindrical geometry, \citet{B2011} in a 3D box, and \citet{GO2023} in a 2D circular cavity \citep[see also e.g.][for experimental generation of a mean flow by internal gravity waves]{SF2016}. 
In the latter, the geometrical focusing of gravity waves launched from the surface towards the center generates differential rotation from inside out by wave breaking (or weak linear damping) of (subcritical) waves and their progressive deposition of angular momentum. 
From the simulations presented here, the presence of both Coriolis and buoyancy forces promotes a mix between cylindrical and spherical geometry of the mean flow \citep[which is reminiscent of latitudinal differential rotation produced in Fig. 6 of][]{B2011}.
In proportion to the amplitude, the differential rotation is stronger for the $_2g$-mode. 

The emergence of differential rotation supports the ability of the resonant mixed gravito-inertial modes to be able to amplify some ambient magnetic field through winding in a way that is qualitatively similar to that studied by \citet{astoulbark25}.
In future work, these results should be checked for different BV frequencies that are more realistic for astrophysical sources to assess the impact of the differential rotation amplitude.

\section{An astrophysical application: gravito-inertial modes in a neutron star binary} \label{sec:astrophysics}

Depending on the stellar spin, the eigenfrequencies of gravito-inertial modes imply that they may become resonant some seconds prior to merger in a binary neutron-star inspiral. Because of their polar character, the leading-order overlap integrals and hence amplitudes may also be sizeable. These aspects together indicate that these modes may be important in the description for mergers for a few reasons:
\begin{itemize}
\item[$\boldsymbol{\star}$]{A leading theory for explaining the ignition of GRB precursor flares, seen in $\lesssim 5\%$ of mergers \citep{cop20,wang20}, is that of crustal failure: resonant modes exert strain on the ion lattice defining the crust, which may exceed its elastic maximum if the mode amplitudes are large enough. Such an overstraining relieves the star of some magnetoelastic energy, which can fuel a gamma-ray flash \citep{tsang12}. Previous estimates using a linear, general-relativistic (GR) framework \citep{pass21,kuan21b,skk22} found that some $g$-modes may exceed modern estimates for polycrystal strain maxima \citep{bc18,baiko24}; a direct comparison is made with our nonlinear models in Sec.~\ref{sec:grlin}.}
\item[$\boldsymbol{\star}$]{Energy may be siphoned from the orbit into the kinetic energy of the modes, leading to GW \emph{dephasing}. Linear theory predicts that while the dephasing due to gravity(-inertial) modes is unlikely to be visible to current interferometers, it may be relevant for next-generation ones \citep[see also \citealt{skk24}]{2023Ho}. In general, however, we expect that treating mode excitations via non-linear perturbation theory would \emph{favour} measurability since amplitudes in post-resonance regimes do not remain constant as in linear theory \citep{yu24,2024Kwon,2025KwonNL}. In addition, the onset of GW dephasing depends on the resonance timing, which will differ between linear and nonlinear models, as explored in Sec.~\ref{sec:grlin}. Deducing their non-linear saturation amplitudes is a necessary step towards providing realistic assessments of dephasing; such a comparison is made in Sec.~\ref{sec:nlinsat}.}
\item[$\boldsymbol{\star}$]{Observations of nonthermal GRB precursors and other aspects \cite[see, e.g.][]{xiao24} hint that some magnetar-like objects participate in a rare subclass of merger. This is difficult to reconcile with the fact that Ohmic decay times are thought to be much shorter than a characteristic inspiral time \cite[e.g.][]{pv19}. One solution is that actually strong fields are not preserved but rather generated prior to coalescence by a mode-driven dynamo, as suggested by \cite{2024Suvorov}. The validity of this scenario depends on microphysical specifics as well as the mode properties; some MHD aspects of this in the context of the simulations presented here are discussed in Sec.~\ref{sec:Astro_consequences}.}
\end{itemize}

\begin{figure}[ht]
    \centering
    \includegraphics[width=0.45 \textwidth]{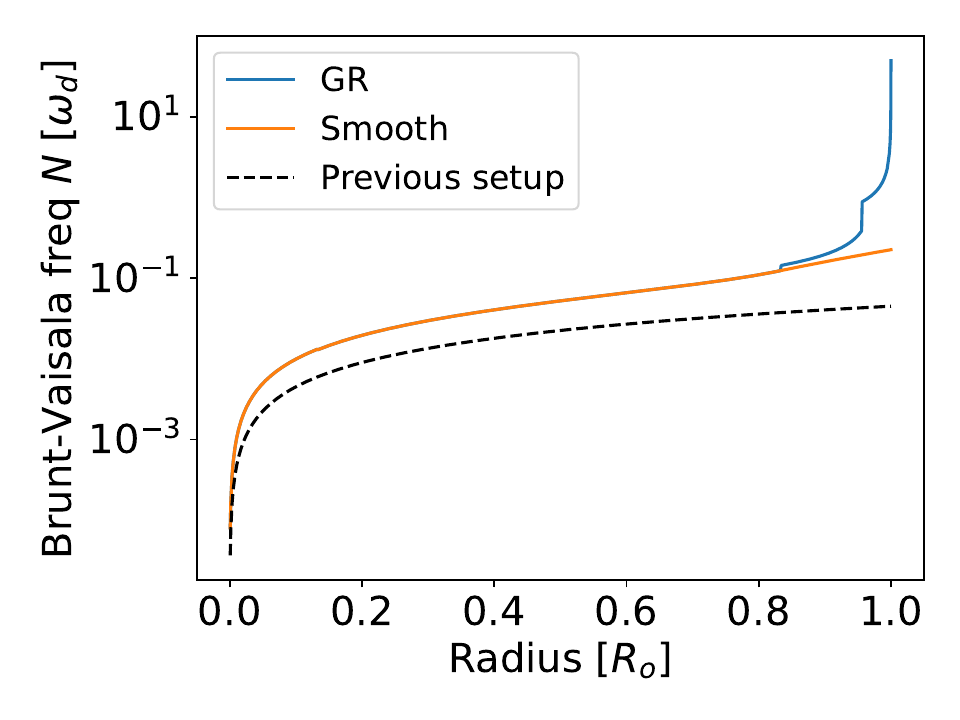}
    \caption{BV frequency for a 1.6 $M_\odot$ NS with the APR4 EOS, giving a radius of $R_o=11$ km from \citet{2024Suvorov}. The orange line is the profile that is used in the Newtonian simulations presented here (see main text for details). For reference, we overlay the BV frequency of the setup in Sec.~\ref{sec:Sim} (dashed line).} 
    \label{fig:GR_N}
\end{figure}

In this section, we wish to come back closer to the astrophysical objects of a standard NS binary. For that, we change our model and consider a more standard NS of $1.6 M_\odot$, anticipating that for objects with non-negligible spin some prior epoch of accretion may have increased the birth value. We now take a radial profile for the BV frequency inspired by a realistic model of the NS from \citet{2024Suvorov}.
We neglect (physical) discontinuities arising in the BV frequency in our ``crustal cavity'' due to different layers of nuclear composition, as results are very likely to change with density variations between the core and the crust. The study of core-crust interfacial modes \cite[i.e. $i$-modes;][]{tsang12} due to these discontinuities is left to future studies.
In Figure \ref{fig:GR_N}, we show the smoothed BV frequency: it is visually identical to that predicted by the GR calculations except in the crust. 
The other main difference in our non-linear simulations from the linear calculations in \citet{2024Suvorov} would be that the density is still kept constant, as we first aim to understand non-linear effects in a simpler setup. 

\subsection{Comparison to GR linear calculations} \label{sec:grlin}

First, before treating rotation self-consistently, we can directly compare  GR linear calculations to our linear ones to estimate the impact of our numerical hypotheses. To compute these modes, we use a radius ratio of $\alpha=0.1$ and the smooth BV profile from Fig.~\ref{fig:GR_N}.
As it turns out, the agreemnent is encouraging: the resonant frequencies are only slightly shifted to lower or higher values by $3.5 \%$ and $2.3 \%$  for the $_2g$ and $_1g$ modes, respectively.

\begin{figure}[ht]
    \centering
 \includegraphics[width=0.45 \textwidth]{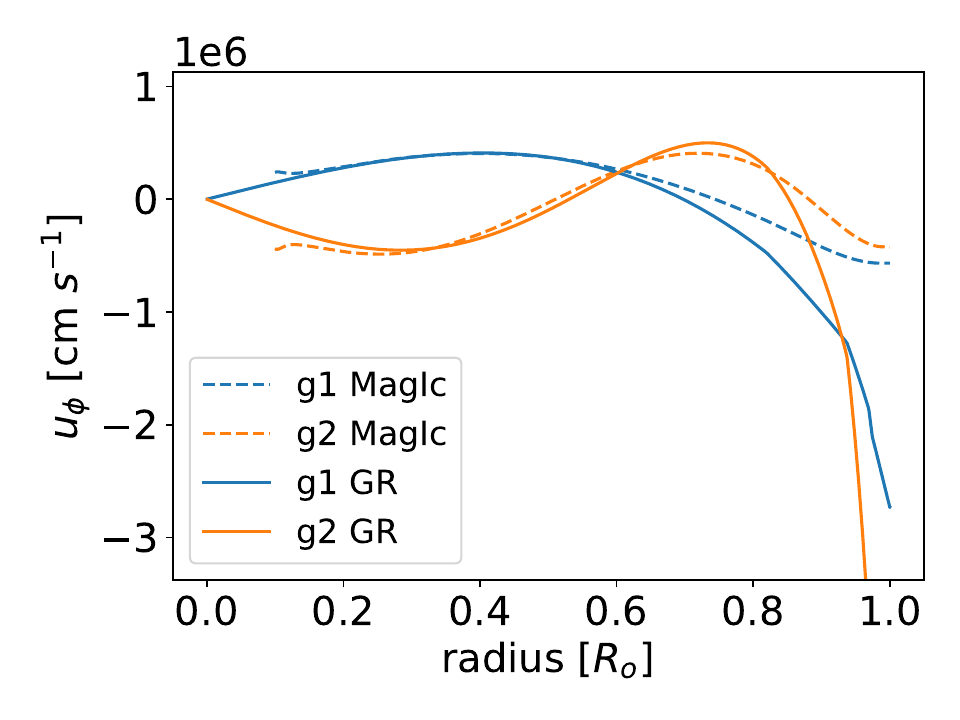}
    \caption{Toroidal velocity for $_1g_{22}$ and $_2g_{22}$ modes. Solid lines represent the cases as those in Fig.~1 of \cite{2024Suvorov}. 
    }
    \label{fig:GR_profile}
\end{figure}

The radial eigenfunctions between the two also agree well until a radius of $\approx 0.8 R_o$, where the impact of the density gradient and, subsequently, the discontinuities become more important (see Figure \ref{fig:GR_profile} for the case of $u_{\phi}$, which corresponds to $u^\phi$ in GR calculations). In terms of amplitudes, the Newtonian $_1g$ and $_2g$ modes are rescaled by a factor of $\sim 0.02$ and $\sim 0.0175$ for the radial profiles to match, respectively. This rescaling is due to the way the linear modes are computed: in the GR case, the saturation is determined by the resonance time window, while, in the Newtonian case, the saturation is determined by the diffusivities. 
The rough agreement between the rescaling factors shows that the results from the Newtonian case in the core are consistent with the GR linear calculations. We can therefore use the smooth BV frequency to study the resonant modes and their non-linear saturation.
However, the results will be mostly relevant to core-like regions of the NS, and a proper treatment of the density and BV discontinuities is required to study quantitatively the modes in the crust.

%\subsection{Linear calculations of the resonant frequencies} \label{sec:linfreq}

\subsection{Non-linear saturation of the strongest resonance} \label{sec:nlinsat}

In this section, we test the impact of rotation on the $_1g$ and $_2g$  modes that generate the strongest non-linear differential rotation in Section \ref{non_lin_sat_tests}. We first start with the $_2g$ mode as it has the lowest tidal amplitude so non-linear effects are expected to be smaller.
We therefore consider three different rotation rates $\Omega_s \in [0,100/\pi,100]$ Hz (i.e., astrophysically corresponding to irrotational, moderate, or ``recycled'' objects). To keep the viscosity the same as previous simulations, the same Rayleigh number $Ra$ is maintained but the Ekman number now, respectively, varies from $Ek=6.2\times 10^{-5}$ to $Ek=1.99\times 10^{-5}$ in the slow and fast rotating cases (note that the Ekman number is not defined for $\Omega_s=0.0$).  
With the new BV profile, which is normalized at the outer boundary, the Rayleigh number is now $Ra=-5.1\times 10^{10}$.
For the same tidal forcing frequency, the orbital frequency $\Omega_o$ is effectively higher since the spin of the NS is increased, and thus we have a larger tidal amplitude parameter $\epsilon$.
Our strategy is to first use the linear code to compute the resonant frequencies 
which is fed into the non-linear code to compute the saturation amplitudes. As one of our motivations is to investigate the strength of the magnetic field that could be theoretically generated by differential rotation, energies are shown in ergs and compared to the strength of a magnetic field that would have the same energy as the flow kinetic energy. This comparison is done by using the energy of uniform $10^{14}$ G magnetic field, $E_{B14}= \frac{1}{8\pi} \left[(10^{14} \rm G)^2 V\right] = 2.27 \times 10^{45}$ erg, where $V= \frac{4\pi}{3} (1-\alpha^3) R_o^3$ is the volume of the domain. 
%Ek=1.98561707

\subsubsection{$_2g$ mode}

\begin{figure}[ht]
    \centering
    \includegraphics[width=0.45\textwidth]{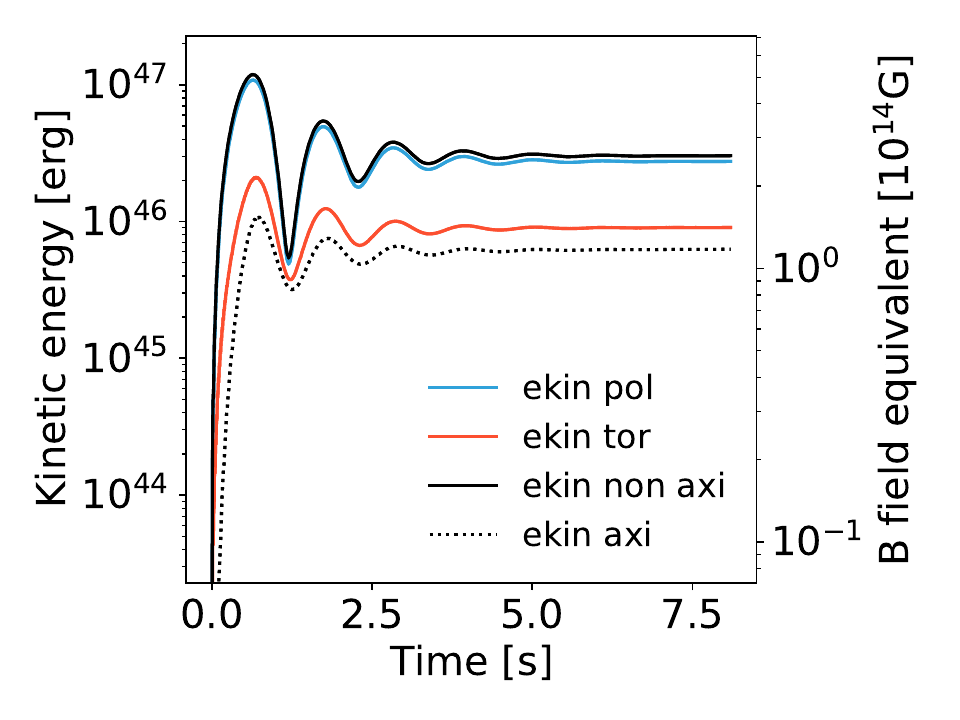}
        \includegraphics[width=0.45\textwidth]  {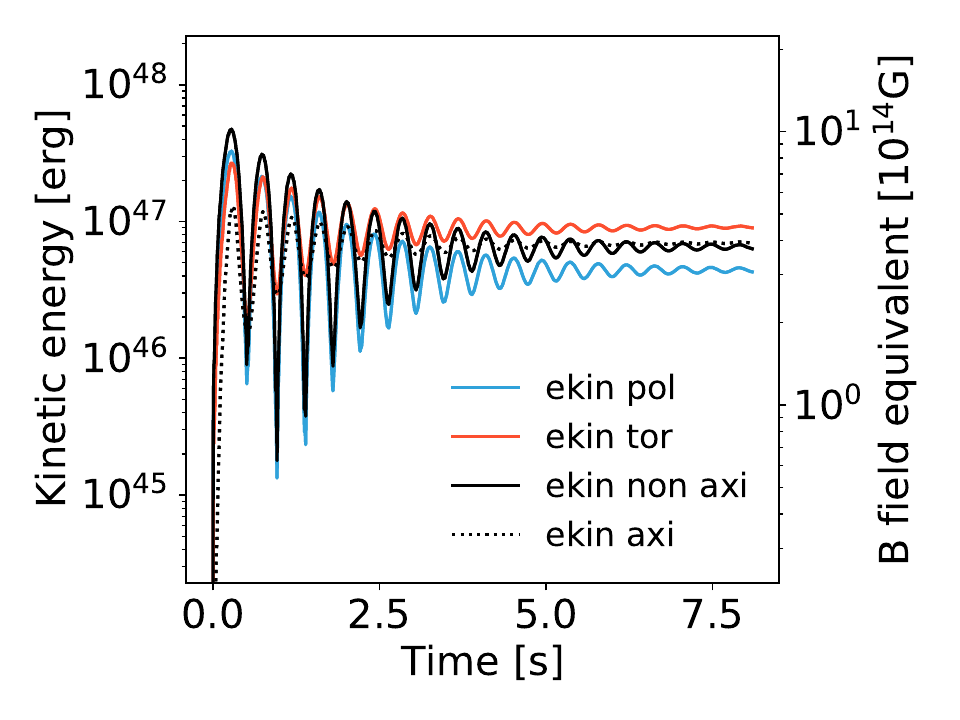}
        \includegraphics[width=0.45\textwidth]      {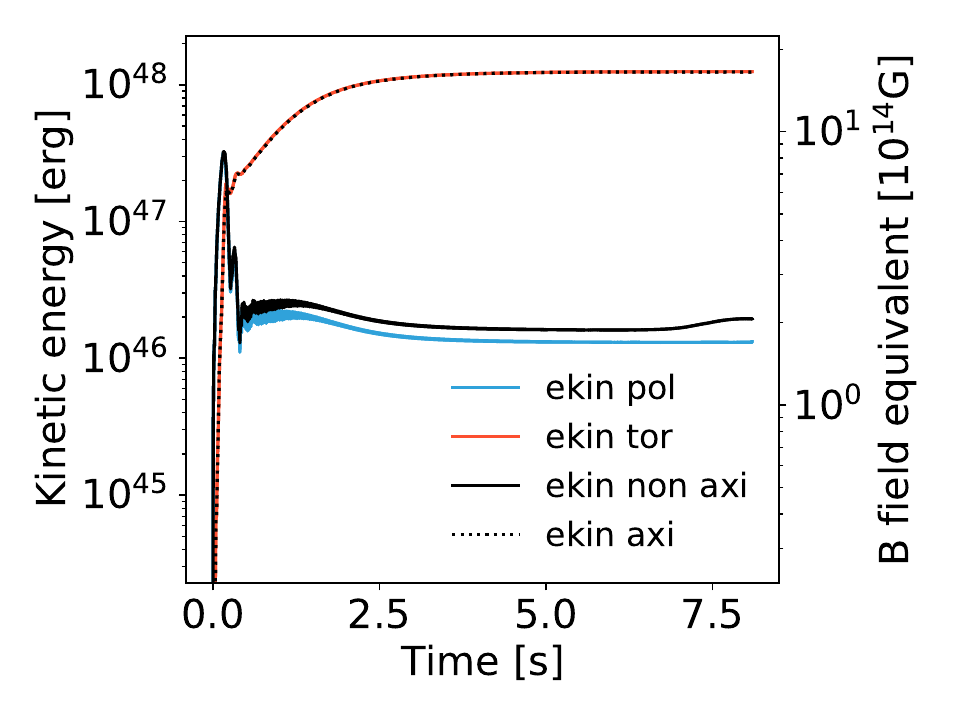}
\caption{Time evolution of the kinetic energy for the $_2g$ mode for  $\Omega_s = 100/\pi$~Hz, (top), $\Omega_s = 100$~Hz (middle), and $\Omega_s=0$ with an increased tidal amplitude (bottom). 
}
\label{fig:astro_g2_non_lin}
\end{figure}

%9.281
We first start with the description of the $_2g$ mode for different rotation rates, $\Omega_s \in [0,100/\pi,100]$ Hz.
For these spins, 
the tidal forcing frequencies of the $_2g$  mode are respectively $\hat{\omega}/N \in [0.146, 0.132, 0.127]$  while the tidal dimensionless amplitudes are $C_t = 9.3 \times 10^{-4}$ and $C_t = 4.1 \times 10^{-3}$ for the two rotating cases. In order to compare the efficiency of the non-linearities in the non-rotating case, we manually adjust the tidal amplitude in the non-rotating case to the fast rotating case, namely $C_t = 4.1 \times 10^{-3}$ instead of the realistic one $C_t = 5.29 \times 10^{-4}$.
This allows us to effectively mimic non-linear effects at lower viscosities since the amplitude of the resonant modes increases with decreasing viscosity \citep[in the linear regime, e.g.][]{O2009,2023Pontin}.
Figure \ref{fig:astro_g2_non_lin} shows the time evolution of the different kinetic energies, which for all simulations would be enough to generate a  magnetic field stronger than %>
$10^{14}$~G if we assume that this kinetic energy would be subdivided through dynamical means into equal parts kinetic and magnetic energy (equipartition). 
We see that non-linearities become relevant for both modes after less than $1$ s of evolution. The non-linearities for the $_2g$ mode and their impact are mostly seen in the toroidal kinetic energy which is expected physically as zonal flows mainly develop in the azimuthal direction. 
For $\Omega_s=100/\pi$~Hz, the poloidal kinetic energy dominates and non-linear effects are weaker than the other case due to the lower tidal amplitude. 
We find that the axisymmetric energy represents $\sim 20\%$ of the total kinetic energy. This is higher than but still consistent with the results of the previous section. 
For a faster star with $\Omega_s=100 \mathrm{Hz}$, the toroidal component becomes dominant and the axisymmetric kinetic energy is comparable to the non-axisymmetric energy, representing $\sim 50\%$ of the total energy.

To understand whether this increase is due to rotation or the tidal amplitude, we compare it to the non-rotating case with the same tidal amplitude.
We find a striking difference for $\Omega_s=0$: the non-linear effects become important much faster, and the axisymmetric toroidal velocity becomes dominant after $1$ s. The toroidal kinetic energy saturates at a higher level compared to the other two rotating cases. 
This different evolution may be due to the fact that the amplitude of linear g-modes also saturate at greater values with lower rotation because of overlap integral properties \citep[see Figure 6 of][]{2024Pontin}. For a given tidal amplitude, the non-linear effects are therefore more important with lower rotation, which is what we observe in our simulations.

\begin{figure}[ht]
    \centering
\begin{subfigure}[b]{0.47\textwidth}    \includegraphics[width=0.48\textwidth]{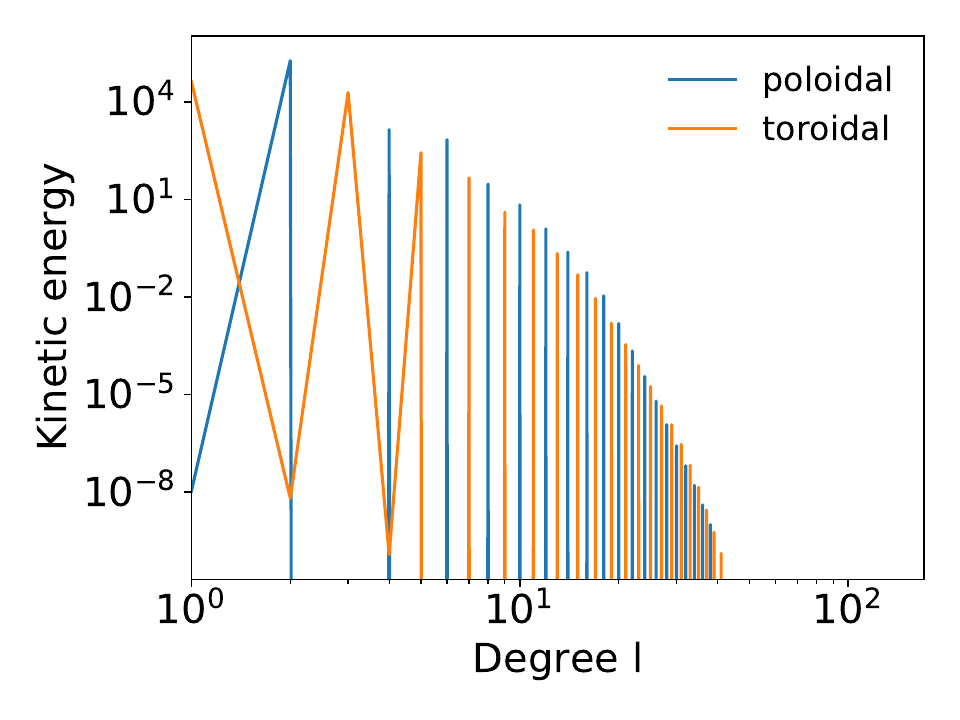}
    \includegraphics[width=0.48\textwidth]{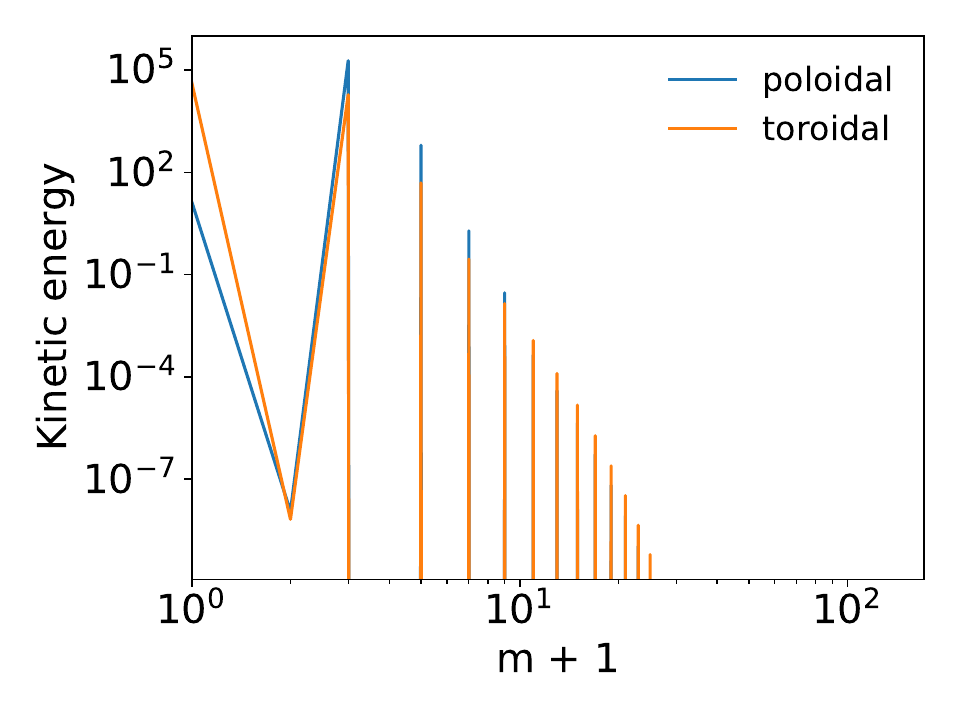}
    \caption{Kinetic energy spectra of the $_2g$ mode for $\Omega_s = 100/\pi$ Hz and $C_t = 9.3 \times 10^{-4}$.}
\end{subfigure}
\begin{subfigure}[b]{0.47\textwidth}    \includegraphics[width=0.48\textwidth]{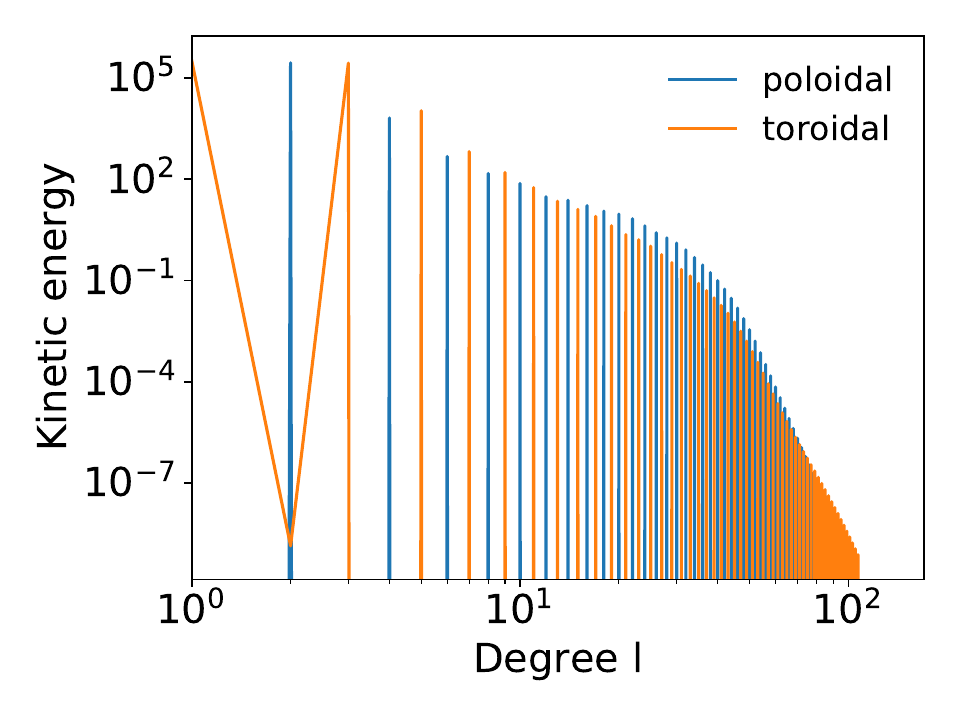}
   \includegraphics[width=0.48\textwidth]{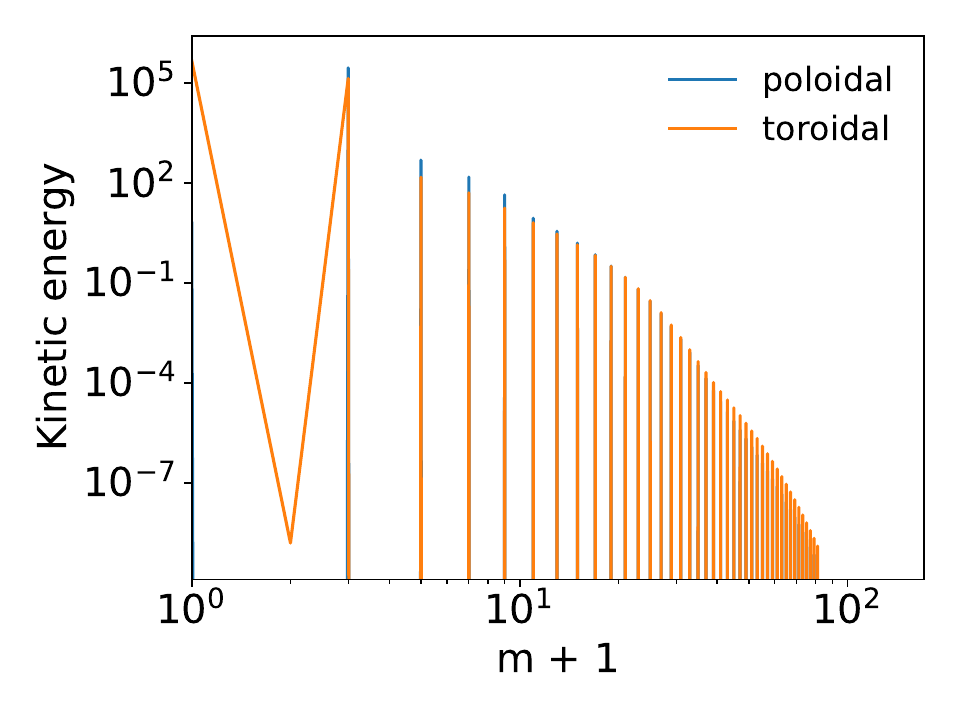}
    \caption{As above but for $\Omega_s=100$ Hz and $C_t = 4.1 \times 10^{-3}$.}
\end{subfigure}
\begin{subfigure}[b]{0.47\textwidth}
    \includegraphics[width=0.48\textwidth]{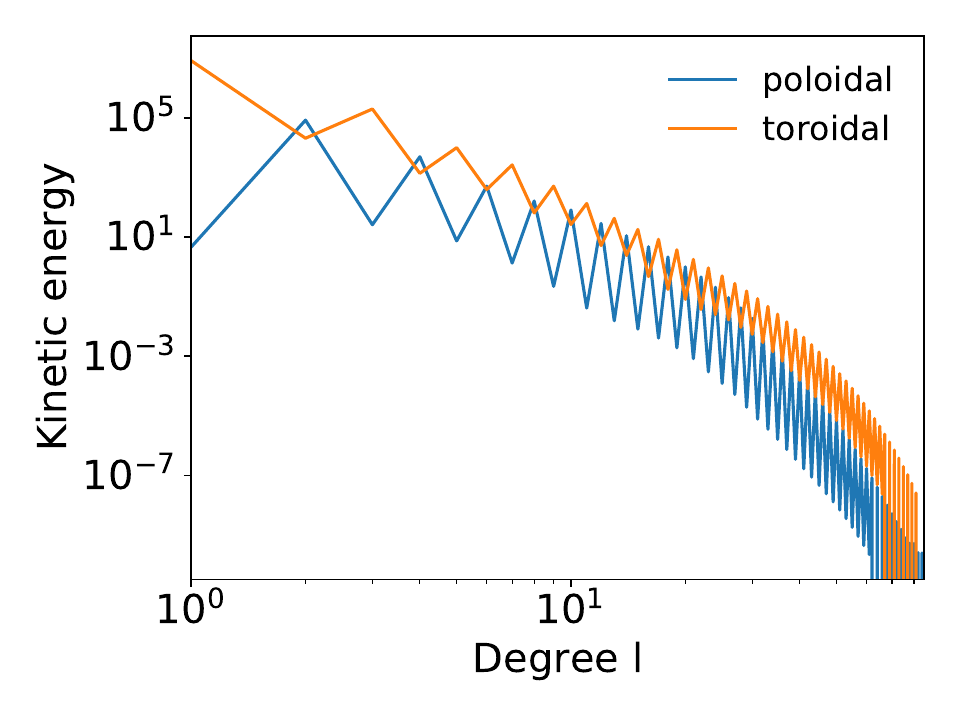}
    \includegraphics[width=0.48\textwidth]{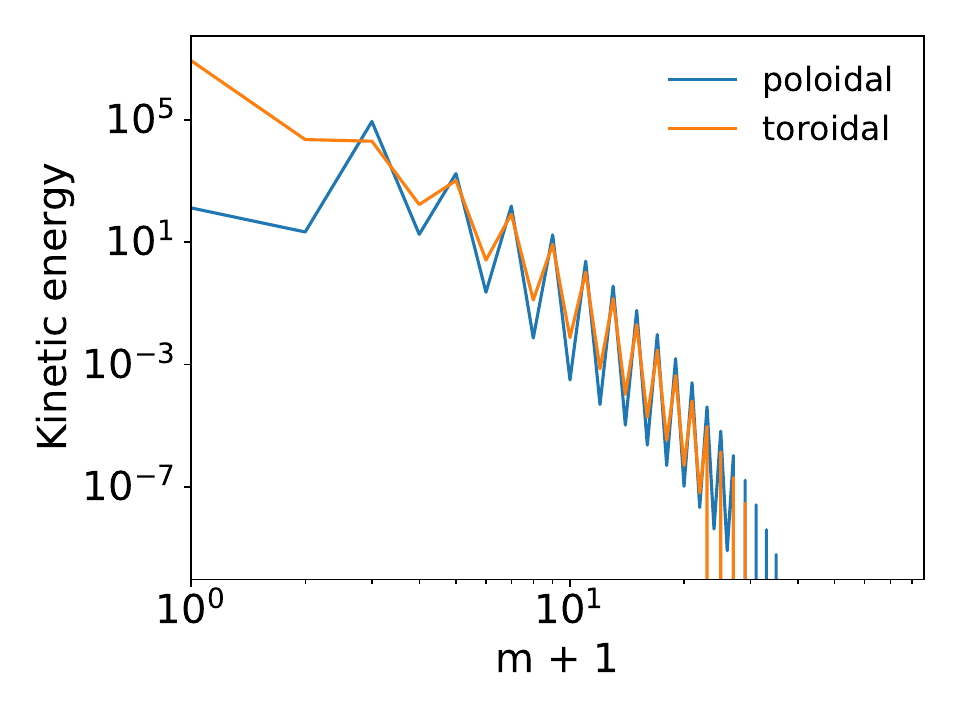}
    \caption{As above but for $\Omega_s=0$ with tidal amplitude $C_t = 4.1 \times 10^{-3}$.}
\end{subfigure}
    \caption{Kinetic energy spectra of the $_2g$-mode resonance for $\Omega_s = 100/\pi$~Hz, (top), $\Omega_s = 100$~Hz (middle), and $\Omega_s=0$ with an increased tidal amplitude (bottom). Left panels are kinetic energy integrated over $r$ and $m$, while varying $\ell$; and right panels are integrated over $r$ and $\ell$, while varying $m$. 
    }
    \label{fig:non_lin_g2_spec_astro}
\end{figure}

To compare the non-linear effects for the three cases in more detail, we look at the kinetic energy spectra along degrees $\ell$ and orders $m$. The $m=2$ mode would dominate for purely linear solutions, and the $\ell=2$ mode should dominate for pure g-modes as the different $\ell$ modes are only coupled by the Coriolis force. Note that $\ell=2$ is also dominant for linear inertial waves due to the tidal potential have a $\ell=2$ geometry.
In addition, we expect to have even $\ell$ modes dominating for the poloidal field and odd $\ell$ modes for the toroidal field due to the equatorial plane symmetry of the equilibrium tide used for the effective forcing. 
We see these behaviors in the spectrum depending on the $\ell$ modes of the two rotating cases and, to a lesser extent, in the non-rotating case too (left panels of Figure \ref{fig:non_lin_g2_spec_astro}). 
As non-linear effects become more important (i.e., at higher tidal amplitudes), the spectrum gets extended to higher degrees $\ell$.

For the spectrum along order $m$, we see that the $m=2$ and the $m=0$ modes dominate, respectively, for the poloidal component and for the toroidal component (right panels of Figure \ref{fig:non_lin_g2_spec_astro}). 
We also see that odd orders $m$ are non-relevant for the rotating cases and only start to become relevant for the non-rotating case with the strongest non-linear effects while being below even modes.  
We see that the amplitude of the toroidal mode $m=0$ becomes more and more dominant as tidal amplitude gets higher, which is consistent with the ratio of axisymmetric energy in the time evolution. The $m=0$ mode of the poloidal component stays much lower than the $m=2$ mode.

\begin{figure}[ht]
    \centering
\begin{subfigure}[b]{0.47\textwidth}   \includegraphics[width=0.45\textwidth]{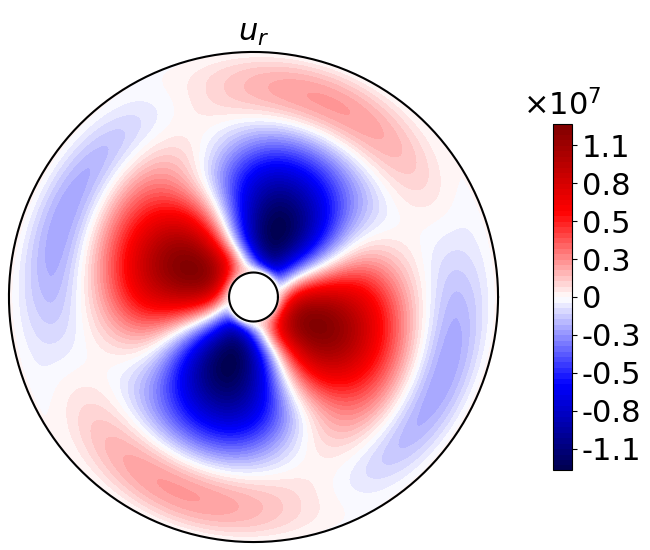}
    \includegraphics[width=0.45\textwidth]{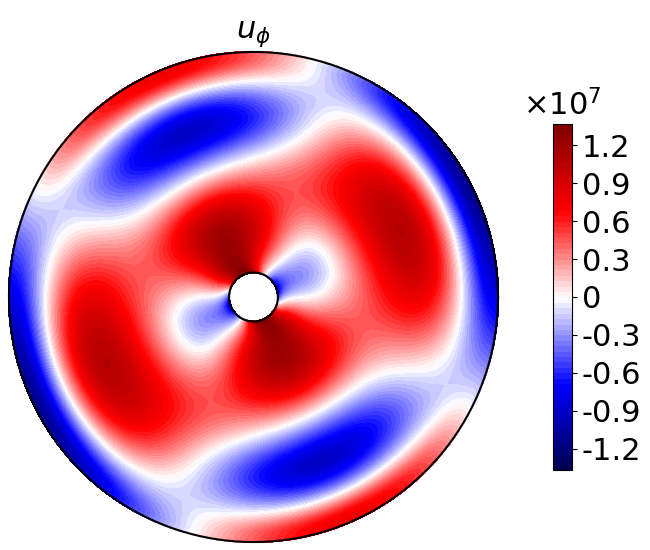}
    \caption{Equatorial snapshots of the $_2g$ gravito-inertial mode for $\Omega_s = 100/\pi$~Hz and $C_t = 9.3 \times 10^{-4}$.}
\end{subfigure}
\begin{subfigure}[b]{0.47\textwidth}
    \includegraphics[width=0.45\textwidth]{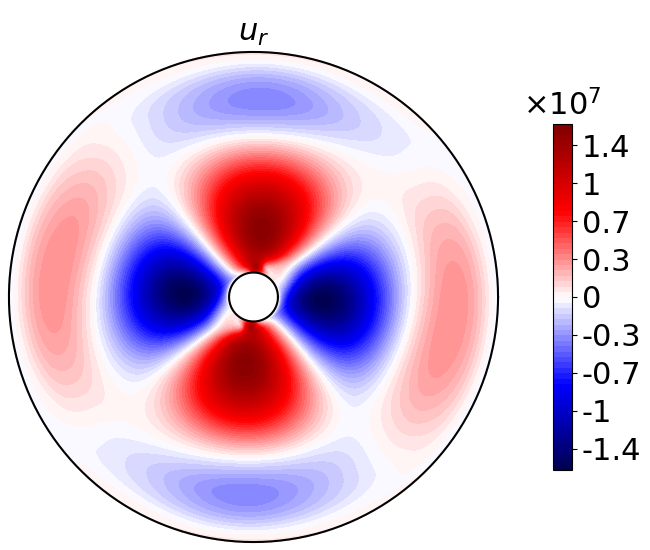}
    \includegraphics[width=0.45\textwidth]{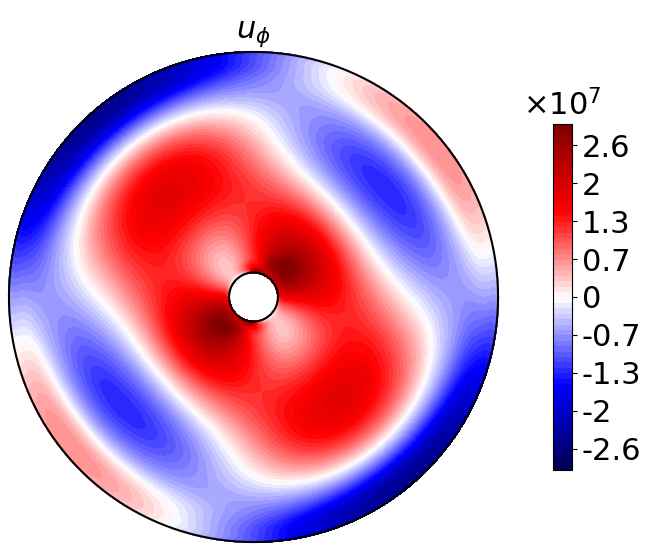}
    \caption{As above but for $\Omega_s = 100$ Hz and $C_t = 4.1 \times 10^{-3}$.}
\end{subfigure}
\begin{subfigure}[b]{0.47\textwidth}
    \includegraphics[width=0.45\textwidth]{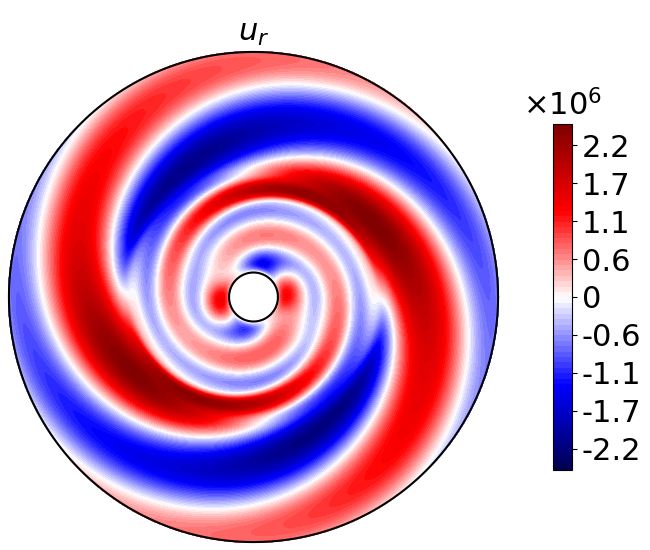}
    \includegraphics[width=0.45\textwidth]{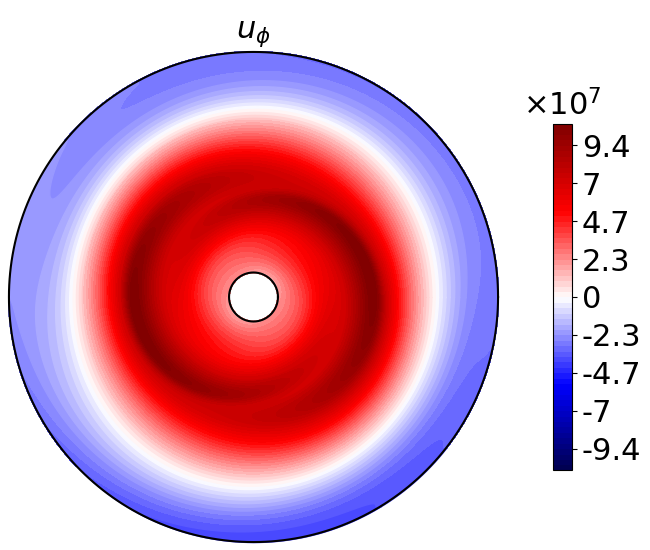}
    \caption{As above but for $\Omega_s=0$ with tidal amplitude $C_t = 4.1 \times 10^{-3}$.
    }
\end{subfigure}
    \caption{Velocity flows, $u_{r}$ (left) and $u_{\phi}$ (right), of the gravito-inertial $_
    2g$ mode in the equatorial plane for $\Omega_s = 100/\pi$~Hz, (top), $\Omega_s = 100$~Hz (middle), and $\Omega_s=0$ with an increased tidal amplitude (bottom).
    }
    \label{fig:u_non_lin_g2_sat_equat}
\end{figure}

To confirm the importance of the $m=0$ mode of the velocities, we show snapshots of $u_r$ and $u_\phi$ in the equatorial plane for the three different cases in Fig. \ref{fig:u_non_lin_g2_sat_equat}. 
In the rotating cases, we see that the the modes are quite similar: 
for $u_r$, the differences between the maximum amplitude of the dominant $m=2$ mode are within 30\%, and there is only a small difference in geometry close to the equatorial plane (not shown here). There are more differences for $u_\phi$ as its amplitude increases with rotation and tidal amplitude. The increase of the axisymmetric component with tidal amplitude can also be seen as the negative amplitude close to the inner core neighbors zero in the fast rotating case. In a similar fashion, the difference between positive and negative amplitudes of $u_\phi$ close to the outer boundary is lower in the fast rotating case. This means that the inner region is spun-up and the outer region is instead spun-down for this particular resonance.

The non-rotating case is both qualitatively and quantitatively different as compared to the rotating cases: for $u_r$, the $m=2$ mode is still dominant but the number of radial nodes is increased, especially in the inner region, and the amplitude is a factor $\sim 5$ lower. 
For $u_\phi$, the $m=0$ mode is dominant with the addition of ``spiral arms'' that looks like a superposition of $m=2$ and $m=1$ modes. 
The amplitude of $u_\phi$ is also $\sim 4$ times stronger compared to the rotating cases. This strong spin-up in the inner region might explain why different radial modes are excited for this non-rotating case.
Indeed, the spin-up leads to a rotation rate of $\Omega_{\rm spin-up}\approx 20$ Hz, which would modify the tidal forcing frequency from $\hat{\omega}= 0.127 N$ to $\hat{\omega}\approx0.038 N$. 
This would lead to the excitation of higher radial-number gravity-modes and could explain the splitting of the domain in two regions, where the excited modes appear more like $_3g$ ($_1g$) in the inner (outer) region.

\begin{figure}[ht]
    \centering
    \begin{subfigure}[b]{0.45\textwidth}
    \includegraphics[width=0.45\textwidth]{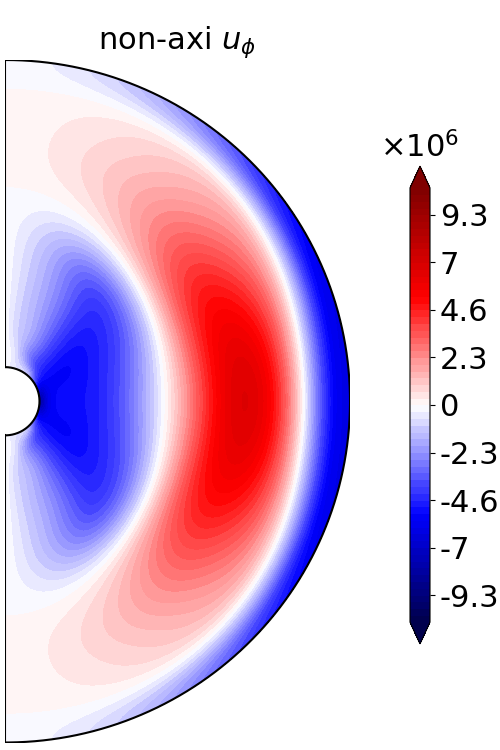}
    \includegraphics[width=0.45\textwidth]{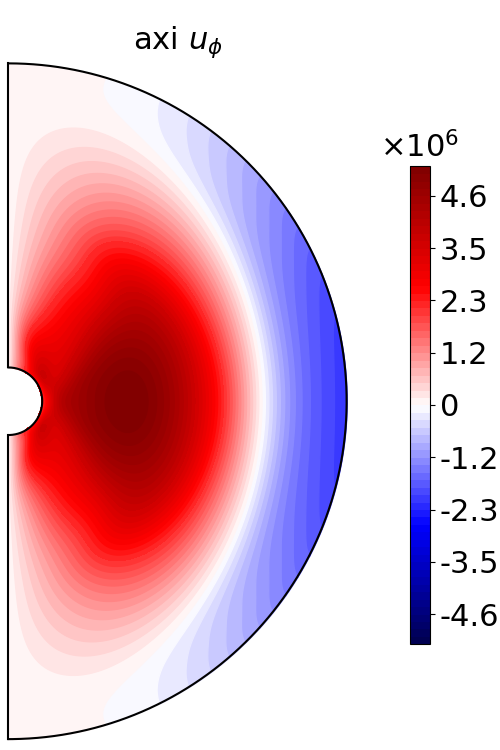}
    \caption{Non-axisymmetric (left) and axisymmetric (right) toroidal velocity $u_\phi$ of the $_2g$ mode for $\Omega_s = 100/\pi$ Hz and $C_t = 9.3 \times 10^{-4}$.}
\end{subfigure}
\begin{subfigure}[b]{0.45\textwidth}
    \includegraphics[width=0.45\textwidth]{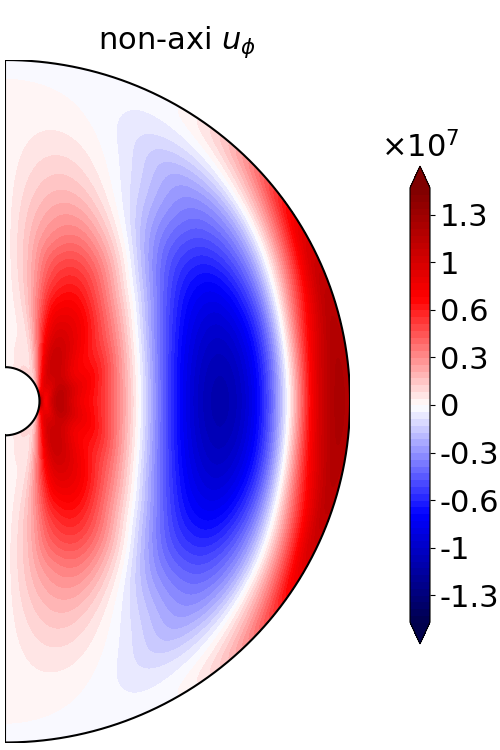}
    \includegraphics[width=0.45\textwidth]{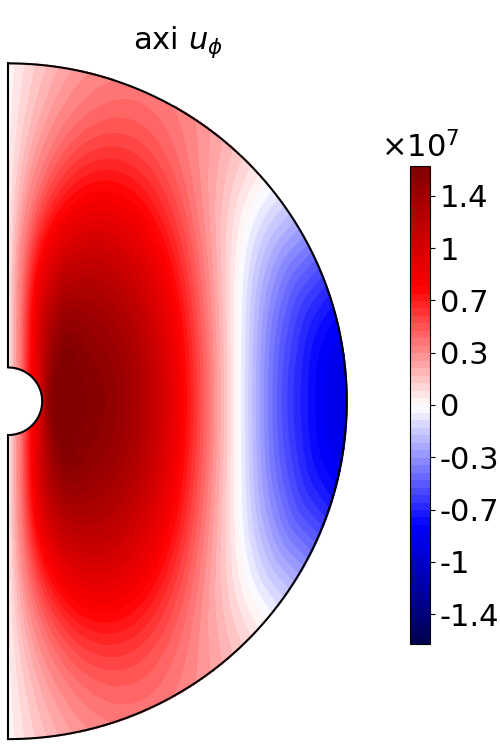}
    \caption{As above but for $\Omega_s = 100$ Hz and $C_t = 4.1 \times 10^{-3}$.}
\end{subfigure}
\begin{subfigure}[b]{0.45\textwidth}
    \includegraphics[width=0.45\textwidth]{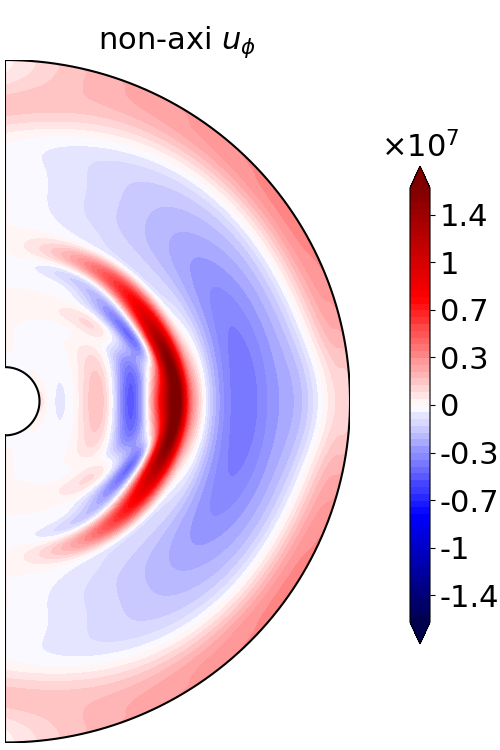}
    \includegraphics[width=0.45\textwidth]{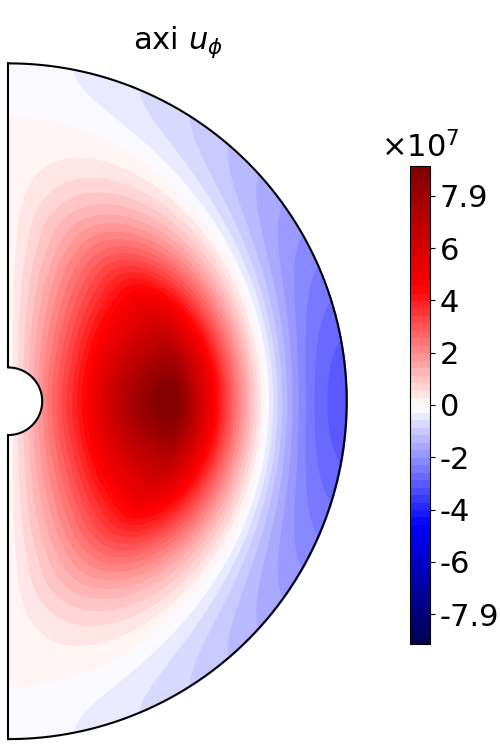}
    \caption{As above but for $\Omega_s = 0$ with tidal amplitude $C_t = 4.1 \times 10^{-3}$.}
\end{subfigure}
    \caption{Velocity flow of the gravito-inertial $_2g$ mode in the \emph{meridional} plane for $\Omega_s = 100/\pi$~Hz, (top), $\Omega_s = 100$~Hz (middle), and $\Omega_s=0$ with an increased tidal amplitude (bottom).}
    \label{fig:non_lin_g2_axi}
\end{figure}

We can confirm the overall spin-up in the domain for all three cases by looking at the toroidal velocity $u_\phi$ in the meridional plane (Figure \ref{fig:non_lin_g2_axi}). First, we see that the non-axisymmetric contribution is stronger than the axisymmetric contribution at a lower tidal amplitude as non-linear effects are less important. When modes become more non-linear, the axisymmetric part becomes of the same order as the non-axisymmetric component and even dominates for the non-rotating case with increased amplitude. 
By increasing rotation, the modes becomes more cylindrical as it can be seen in the $\Omega_s=100$ Hz case, while they are spherical when $\Omega_s=100/\pi$ Hz and $\Omega_s=0$.
In the non-rotating case, we also see the same radial geometry observed in the equatorial plane: the gravity $_2g$ mode seems to be split into a $_3g$ mode in the inner region and $_1g$ mode in the outer region.
We compute the vertical angular momentum following
\begin{gather}
    L_z=\int_V (r\ \mathrm{sin} \theta)^2 \Omega \,\mathrm{d}V,
\end{gather}
with $\Omega=u_\phi/(r\sin\theta)$, to estimate the spin up in the inner half of the domain $V$ from $r=0.1 R_o$ to $r=0.55 R_o$. 
We then estimate the solid body rotation at the end of the simulation $\Omega_{s,end}$ that would correspond to this angular momentum, viz.
 \begin{gather}
     \Omega_{s,end}= \frac{L_z}{\int_V (r \ \mathrm{sin} \theta)^2 \,\mathrm{d}V}.
     \label{eq:Omsend}
 \end{gather}
We find that $\Omega_{s,end} \in [21.58 \rm Hz, 1.0476 \Omega_s,1.042 \Omega_s]  $ for initial values $\Omega_s \in [0,100/\pi,100]$ Hz. 
With stronger non-linear effects (i.e, stronger tidal amplitude and/or reduced rotation), the absolute value of the spin-up is stronger as it goes from $\Omega_{s,end}-\Omega_s = 1.5$ Hz to $21.58 \rm Hz$.

\subsubsection{$_1g$ mode}
\begin{figure}[ht]
    \centering
    \includegraphics[width=0.49\textwidth]{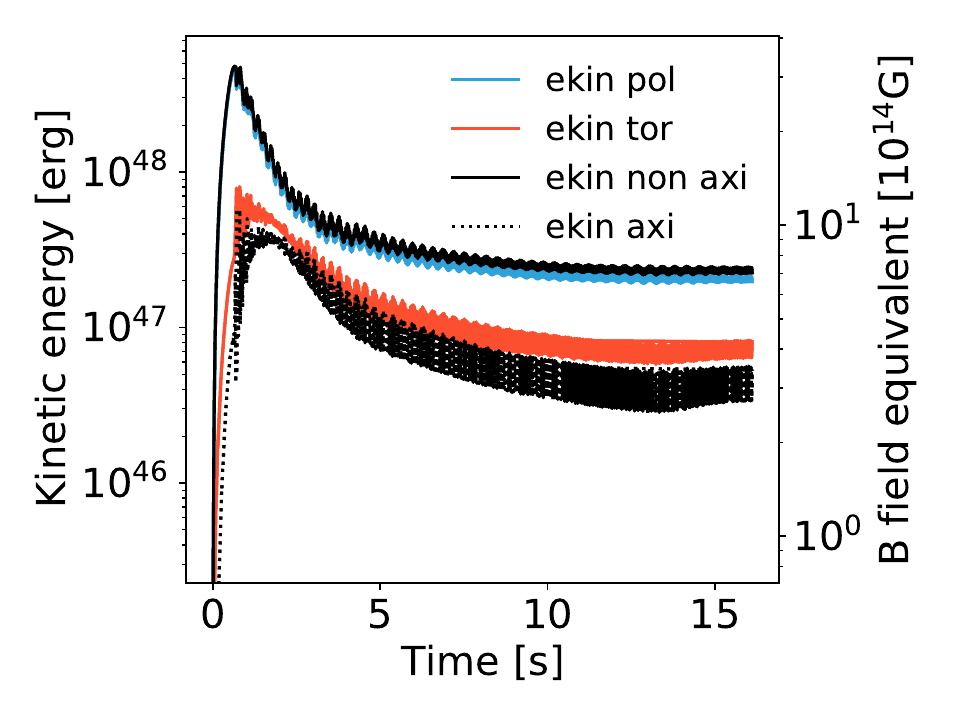}
    \includegraphics[width=0.49\textwidth]{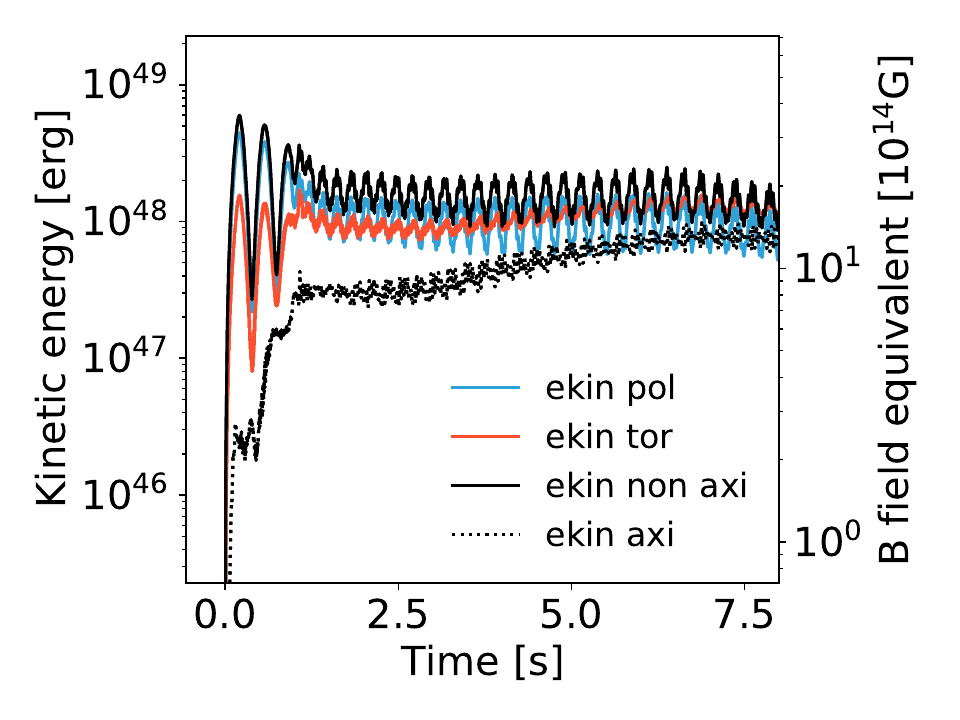}
    \includegraphics[width=0.49\textwidth]      {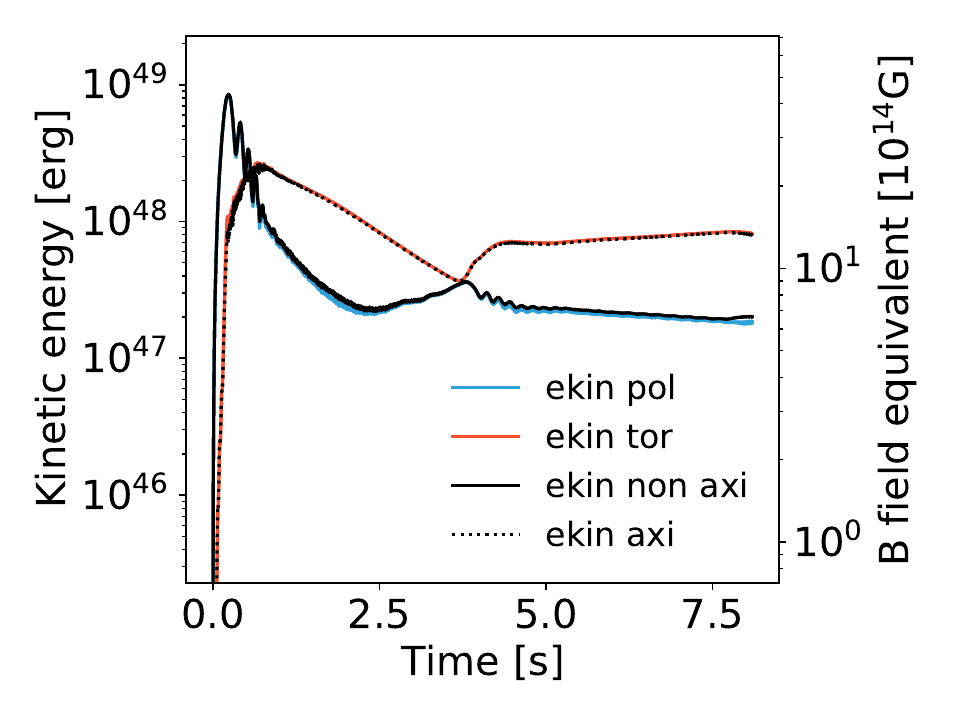}
\caption{Time evolution of the kinetic energy for the $_1g$ mode for $\Omega_s = 100/\pi$~Hz (top), $\Omega_s = 100$~Hz (middle), and $\Omega_s=0$ with an increased tidal amplitude $C_t = 5.3 \times 10^{-3}$ (bottom).
}
\label{fig:astro_g1_non_lin}
\end{figure}

Using the same strategy, we turn our attention to $_1g$ modes.
The tidal forcing frequencies of the $_1g$ mode are $\hat{\omega} \in [0.235 N, 0.21775 N, 0.13180 N]$ for $\Omega_s \in [0,100/\pi,100]$ Hz, and the tidal dimensionless amplitudes are $C_t = 1.6 \times 10^{-3}$ and $C_t = 5.3 \times 10^{-3}$ for the two rotating cases. 
As for the $_2g$ mode, we use the same tidal amplitude in the non-rotating case as the more-rapidly rotating case $C_t = 5.3 \times 10^{-3}$ instead of the realistic one $C_t = 1.36 \times 10^{-3}$. 
Figure \ref{fig:astro_g1_non_lin} shows time evolutions for the different kinetic energies, which for all simulations would be enough to generate a magnetic field stronger than %>
$ 10^{15}$~G assuming equipartition. The non-linearities become relevant for both modes after less than $1$ s after initialization. 

The evolution of the $_1g$ modes is quite different from the $_2g$ mode as after a first growth, some fast oscillations of the kinetic energy appear. These oscillations may be indicating that the $_1g$ mode is subject to parametric instabilities or corotation resonance instabilities. 
Before looking more into the instabilities, we note that similar results to those obtained for the $_2g$ mode are found for the importance of non-linear effects. Indeed, for $\Omega_s=100/\pi$~Hz, the poloidal kinetic energy dominates and the axisymmetric energy represents again around $\sim 20\%$ of the total kinetic energy. 
For $\Omega_s=100$~Hz, the non-linear effects are even stronger as axisymmetric energy represents $\approx 50\%$ of the budget. 
In addition, the toroidal and poloidal components contribute evenly to the total kinetic energy at the end of the simulation.
To understand the impact of rotation and tidal amplitude, we compare this evolution with that of the non-rotating case with the same tidal amplitude.
We find a critical difference for $\Omega_s=0$: the non-linear effects become important sooner in the evolution, and the axisymmetric toroidal kinetic energy
%velocity
becomes dominant after $1$ s. The time evolution for the non-rotating case is quite interesting as the first growth is higher than the other cases but it drops quicker, until stabilizing after $4$ s. The first decreasing part is due to the evolution of an instability, as we see small oscillations in the poloidal kinetic energy until the $m=2$ mode growths again and dominates the energy contribution for the second part after $4$ s. 

\begin{figure}[ht]
    \centering
\begin{subfigure}[b]{0.47\textwidth}
    \includegraphics[width=0.48\textwidth]{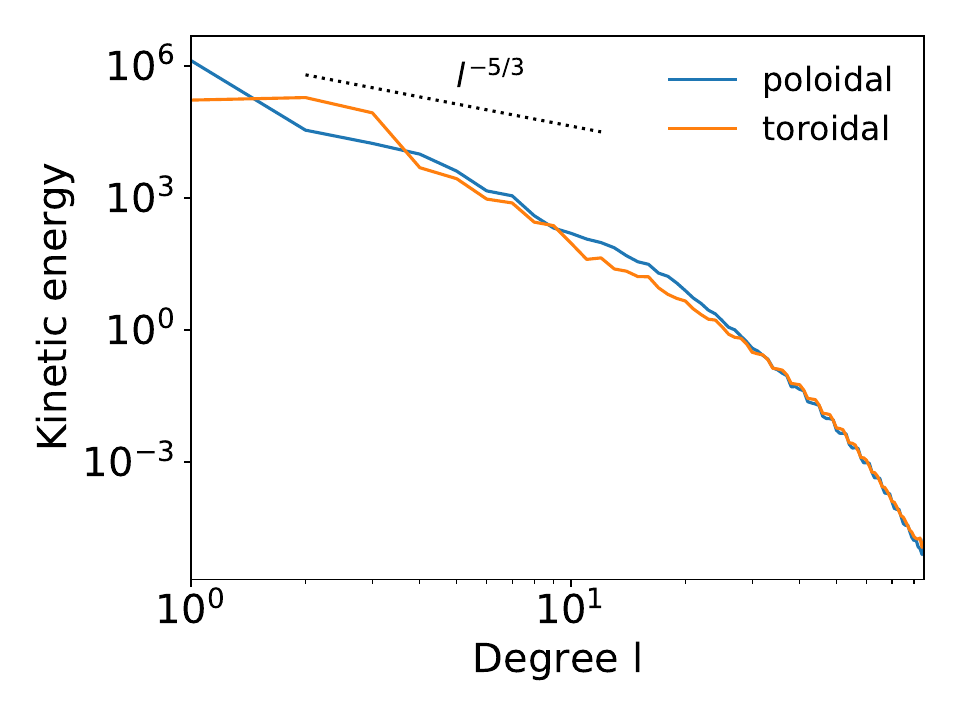}
    \includegraphics[width=0.48\textwidth]{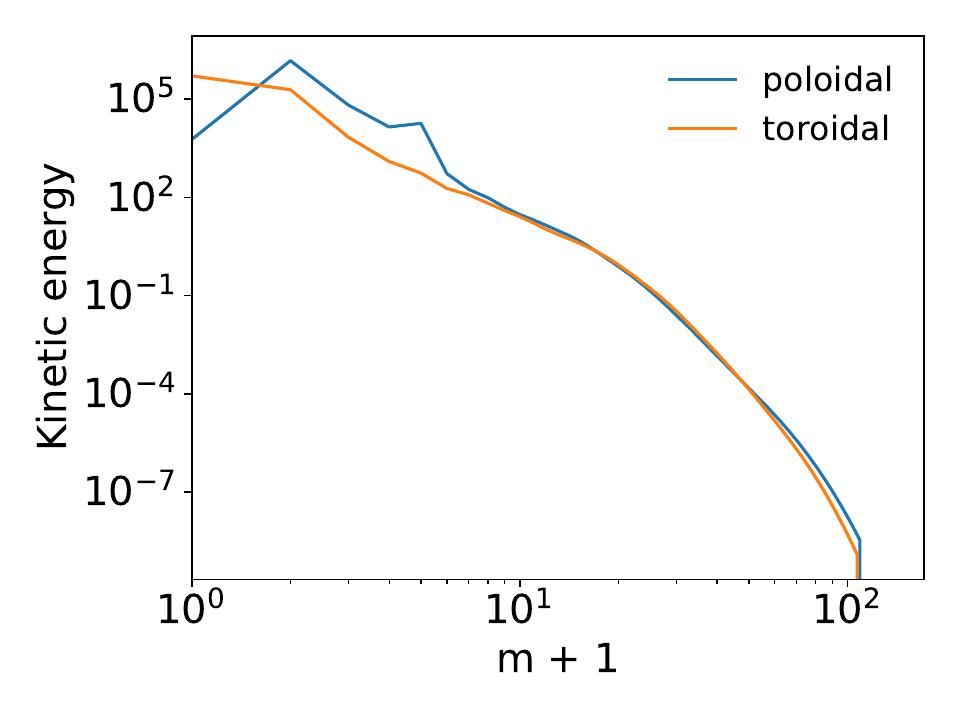}
    \caption{Kinetic energy spectra of the $_2g$ mode for $\Omega_s=100/\pi$ Hz and and $C_t = 1.36 \times 10^{-3}$.}
\end{subfigure}
\begin{subfigure}[b]{0.47\textwidth}
    \includegraphics[width=0.48\textwidth]{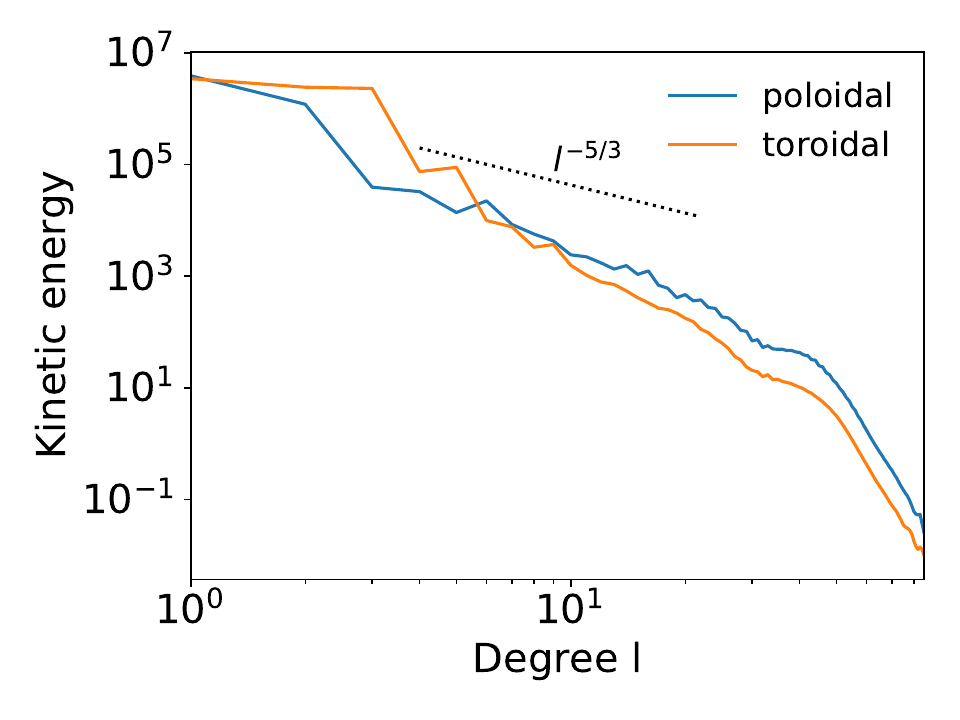}
    \includegraphics[width=0.48\textwidth]{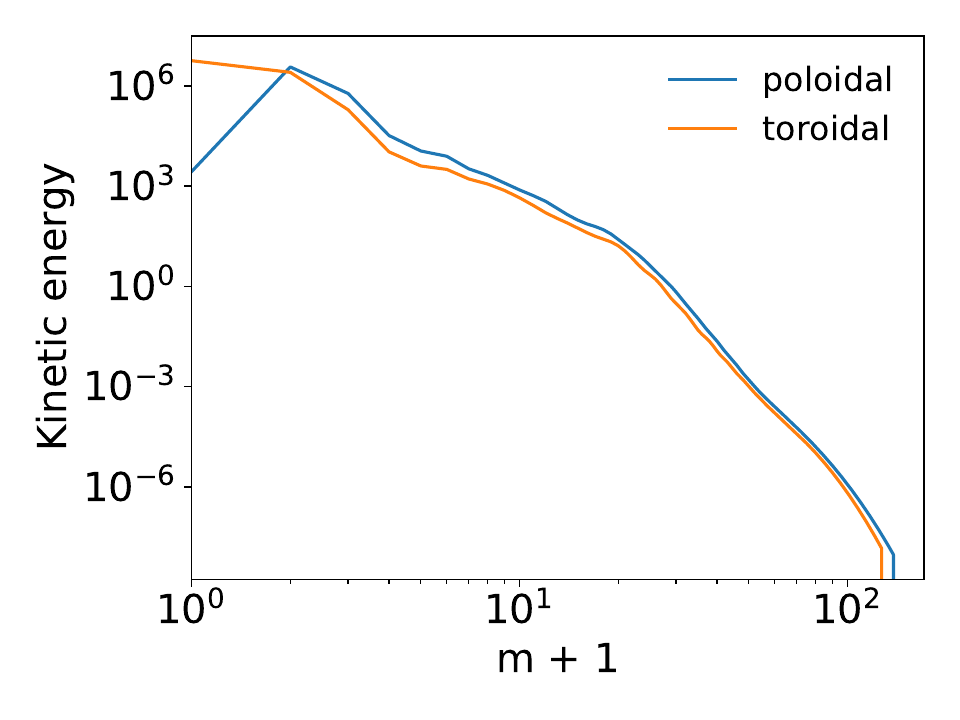}
    \caption{As above but for $\Omega_s=100$ Hz and $C_t = 5.3 \times 10^{-3}$.}
\end{subfigure}
\begin{subfigure}[b]{0.47\textwidth}
    \includegraphics[width=0.48\textwidth]{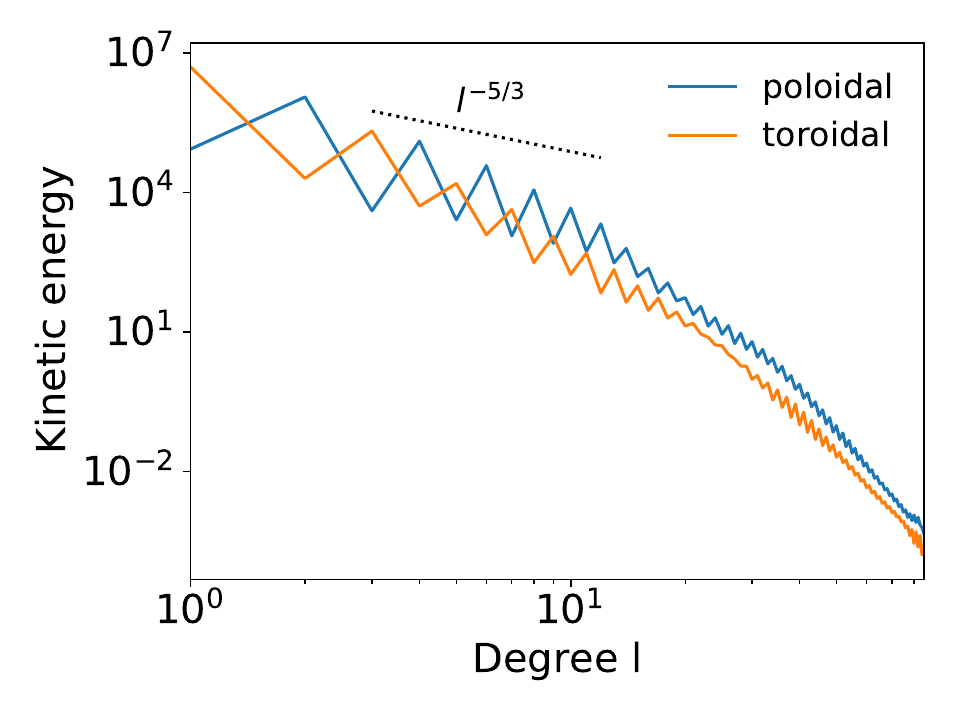}
    \includegraphics[width=0.48\textwidth]{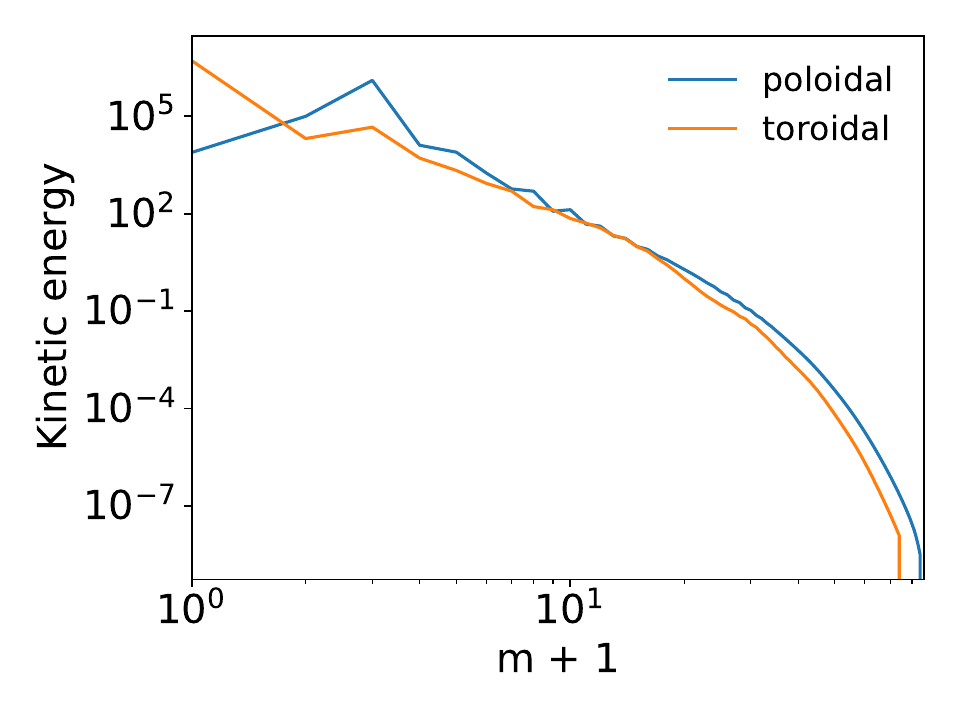}
    \caption{As above but for $\Omega_s=0$ with tidal amplitude $C_t = 5.3 \times 10^{-3}$.}
\end{subfigure}
    \caption{Similar to Fig.~\ref{fig:non_lin_g2_spec_astro} but for the $_1g$-mode. The dotted line corresponds to the slope $\ell^{-5/3}$.} \label{fig:non_lin_g1_spec_astro}
\end{figure}

To better understand these three simulations, we look at the kinetic energy spectra along degrees $\ell$ and orders $m$ at the end of the time evolution (Figure \ref{fig:non_lin_g1_spec_astro}). The difference with the $_2g$ mode is quite clear as there are no patterns between odd and even degrees. This means that equatorial symmetry is broken by an instability. 
For both rotating cases, the spectrum is continuous, meaning that non-linear effects are strong. This can be seen as well on the order spectrum, where, for the rotating cases, the dominant mode for the poloidal component is not the $m=2$ mode (of the initial tidally-forced waves) but the $m=1$ mode.
For the toroidal component, there is also a strong $m=1$ mode but the $m=0$ mode is dominant.
The emergence of this $m=1$ mode indicates that an instability appears in this simulation.
In the cases of fastest rotation and no rotation for which non-linear effects are the strongest, %case with greater spin,
the energy cascade from high to small spatial scales approaches a Kolmogorov spectrum, $\propto \ell^{-5/3}$, at moderate spherical harmonics degrees. In the low rotation case [panel (a) in Fig. \ref{fig:non_lin_g1_spec_astro}], %other cases,
the spectra are steeper than $\ell^{-5/3}$. 
The energy cascade towards different parity modes and smaller spatial scales may indicate the onset of wave turbulence, as found in other setup of unstratified hydrodynamical turbulence driven by triadic resonances of inertial waves \citep[see for instance][]{G2003,BT2018,BT2024}. The stronger tidal amplitude enhances the non-linear effects but, in the non-rotating case, the scaling $\ell^{-5/3}$ is not seen in the kinetic spectra even when the triadic resonance is dominant. 

\begin{figure}
    \centering    
    \includegraphics[width=0.44\textwidth]{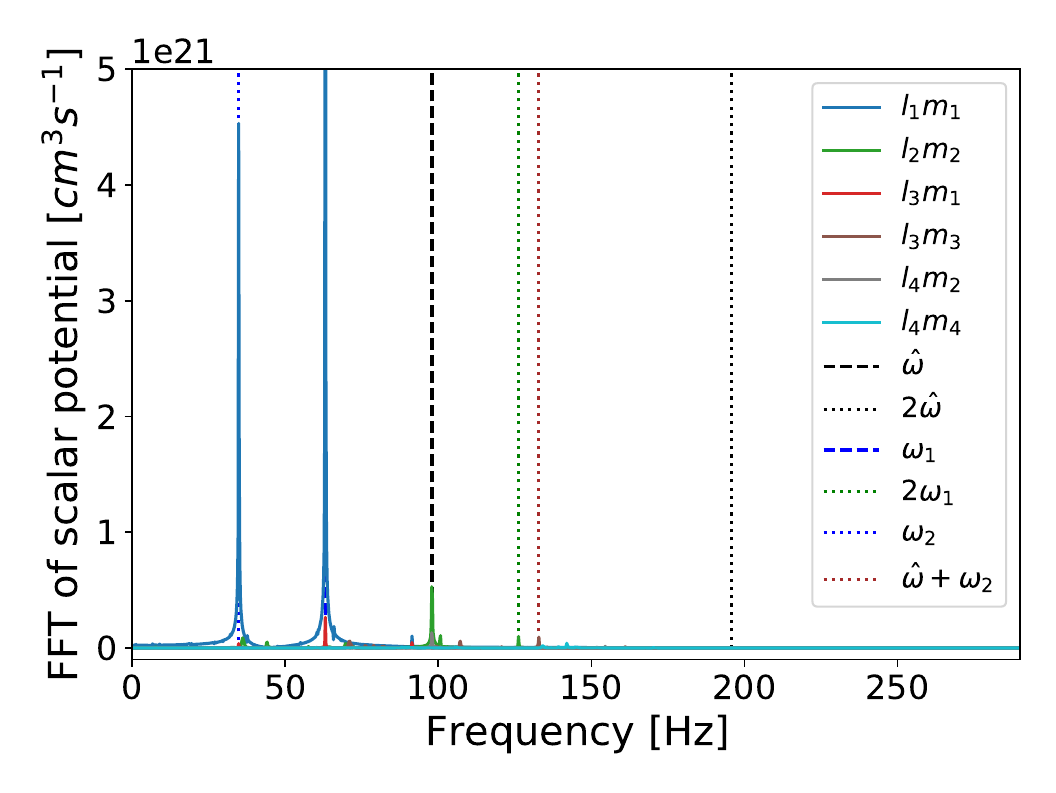}
    \includegraphics[width=0.44\textwidth]{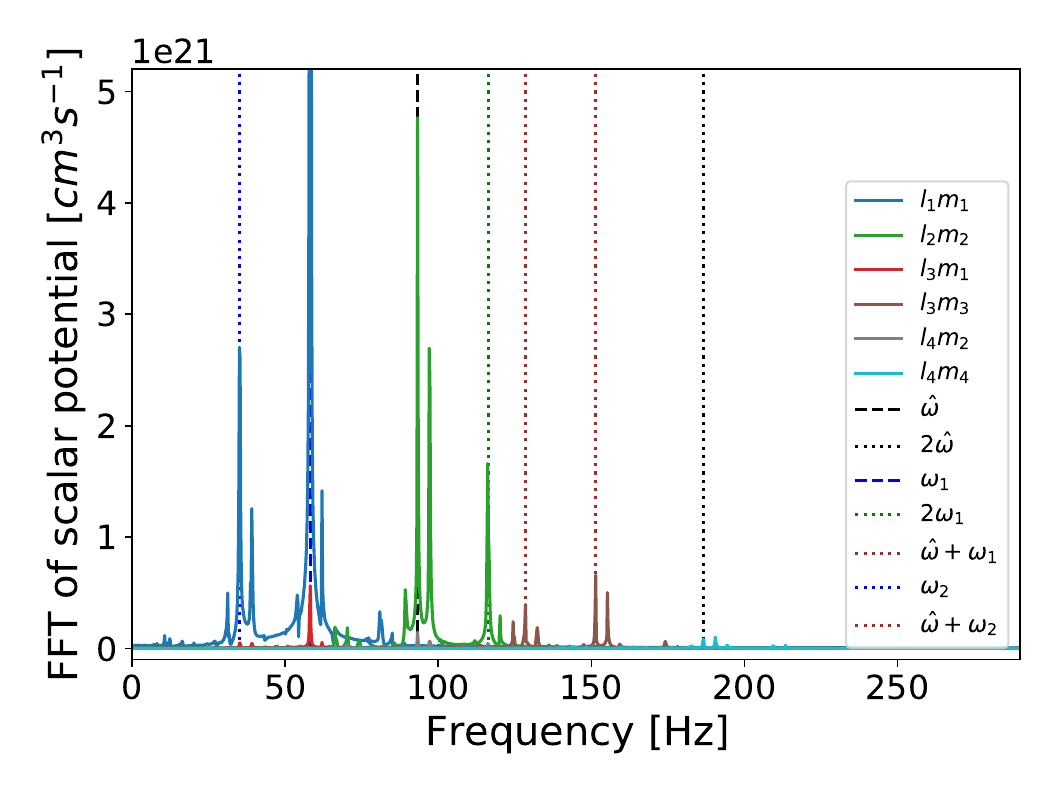}
    \includegraphics[width=0.44\textwidth]{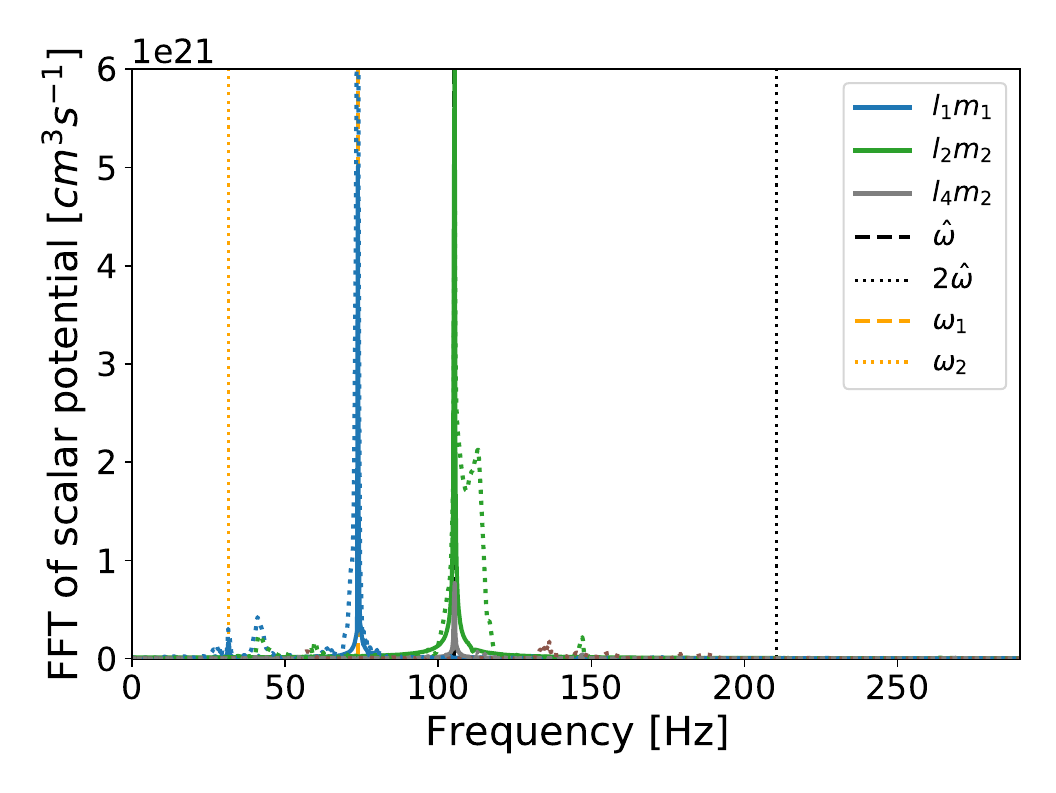}
    \caption{
    Fourier transform of the poloidal velocity potential $W$ showing the different main modes for $\Omega_s = 100/\pi$~Hz (top), %left)
    $\Omega_s = 100$~Hz (middle), and $\Omega_s=0$ (bottom) %right)
    . We have $\omega_1 \in[63.1,58.2,73.7]$ Hz and $\omega_2=\hat{\omega}-\omega_1$. In the non-rotating case, the FFT in the first phase $t\in[0,4]$ s is plotted in dotted line, while the second phase $t\in[4,10]$ s is plotted in solid lines. 
    }
    \label{fig:param_FFT_g1_astro}
\end{figure}

To characterize the instability and the oscillations that appears for the rotating simulations, we perform a Fourier transform of the poloidal scalar potential $W(l,m,r_{FFT})$ at mid-point of the domain $r_{FFT}=0.55$ for the different main modes for the slower and faster rotating cases and non-rotating case (Figure \ref{fig:param_FFT_g1_astro}). 
For both rotating cases, we find that the dominating component is the $(l,m)=(1,1)$ mode and the oscillations of this mode are divided in two frequencies $\omega_1$ and $\omega_2$. We call $\omega_1$ the main frequency at $\omega_1 =63.098$ Hz and $\omega_1=58.184$ Hz, for $\Omega_s=100/\pi$ Hz and $\Omega_s=100$ Hz, respectively. We also see that the second main mode is the expected $(l,m)=(2,2)$ modes but some further modes are also excited along with a number of even $\ell + m$ modes (due to symmetry; while odd $\ell+m$ being excited for the toroidal velocity potential), especially for the highest-rotation case where the tidal amplitude and non-linear effects are stronger.
The $(l,m)=(2,2)$ is excited at the tidal forcing frequency $\hat{\omega}$ and the $(l,m)=(4,4)$ is excited at its superharmonics $2\hat{\omega}$ in the fast rotating case only. 
The second tallest peak of the $(l,m)=(1,1)$ mode is at the frequency $\omega_2=\hat{\omega}-\omega_1$. 
Therefore, it seems that the $m=1$ mode is growing at the expense of the main $m=2$ forcing mode, which would explain why the latter is not dominant anymore (compared to simulations without the instability). 
All these observations 
provide clear evidence for triadic resonances \citep[theorized and observed for inertial waves in][]{K1999,BT2018,2022Astoul}
since all of the excited modes can be recovered by a linear combination of $\hat{\omega}$ and $\omega_1$ and their corresponding degrees and orders. In particular, we recover the modes $(l,m)=(3,3) = (2,2)+(1,1) = (4,4) -(1,1)$, with both excited frequencies   
$\omega_{3,3} \in [\hat{\omega} +\omega_1, 2\hat{\omega}-\omega_1] = \hat{\omega} + \omega_2$. 

In the fast rotating case, we can see the impact of rotation as main frequencies are split between a slight higher frequency and lower frequency. The new peaks appearing with fast rotation can also be recovered by linear combination of the $\hat{\omega}$ and $\omega_1$ and their corresponding split frequencies. This may be a sign that the rotation is increased in some part of the domain and decreased in the other part, which would shift, respectively, the frequency to lower frequencies or higher frequencies. 

\begin{figure}[ht]
    \centering
\begin{subfigure}[b]{0.47\textwidth}
    \includegraphics[width=0.45\textwidth]{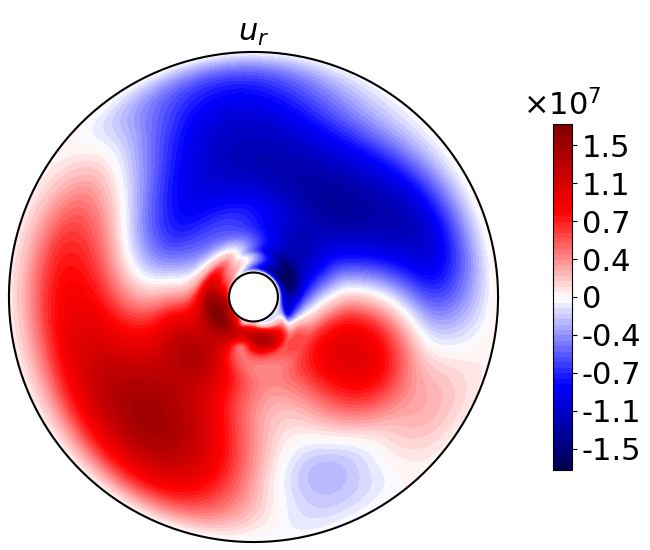}
    \includegraphics[width=0.45\textwidth]{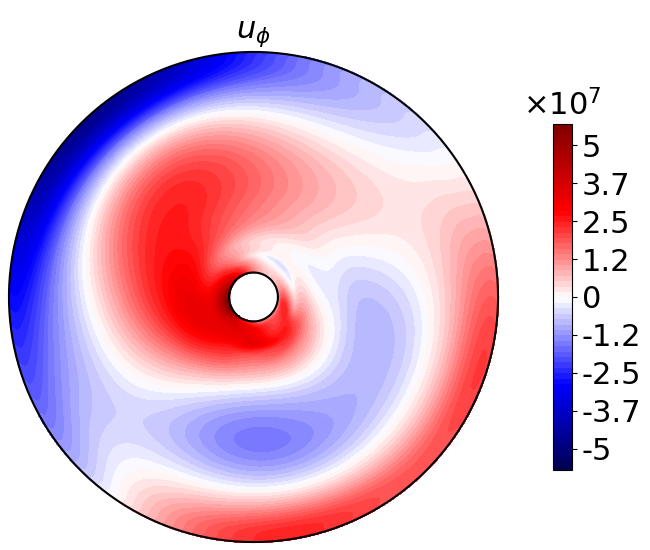}
    \caption{Equatorial snapshot of velocities associated with the $_1g$ gravito-inertial mode for $\Omega_s = 100/\pi$ Hz and $C_t = 1.36 \times 10^{-3}$.}
\end{subfigure}
\begin{subfigure}[b]{0.47\textwidth}
    \includegraphics[width=0.45\textwidth]{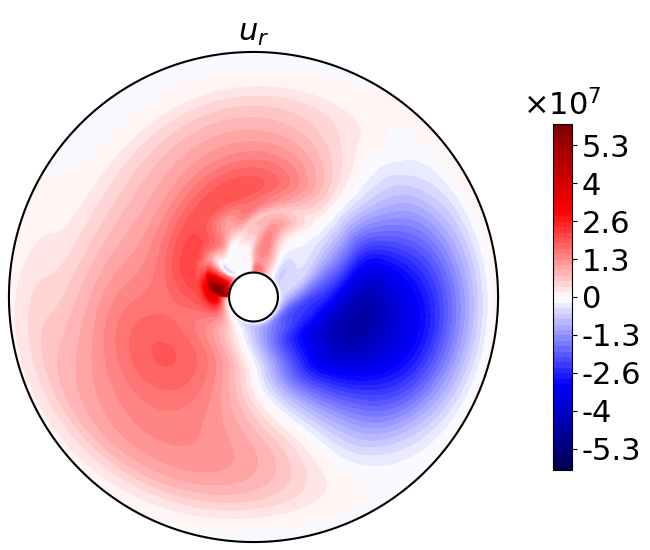}
    \includegraphics[width=0.45\textwidth]{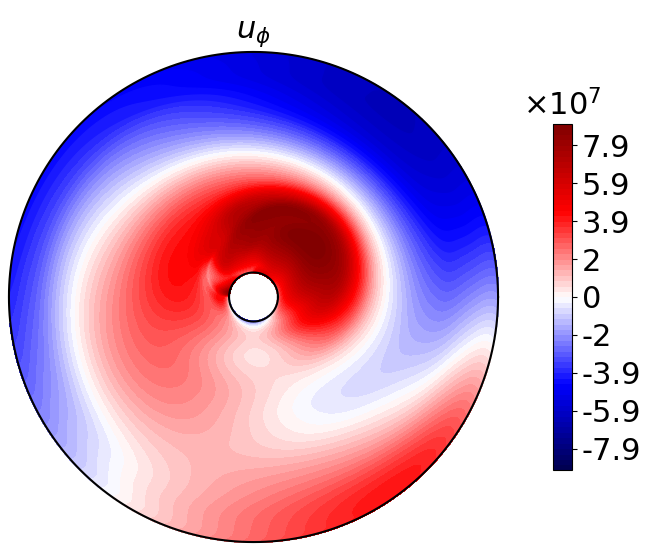}
    \caption{As above but for $\Omega_s = 100$ Hz and $C_t=5.3\times 10^{-3}$.}
\end{subfigure}
\begin{subfigure}[b]{0.47\textwidth}
    \includegraphics[width=0.45\textwidth]{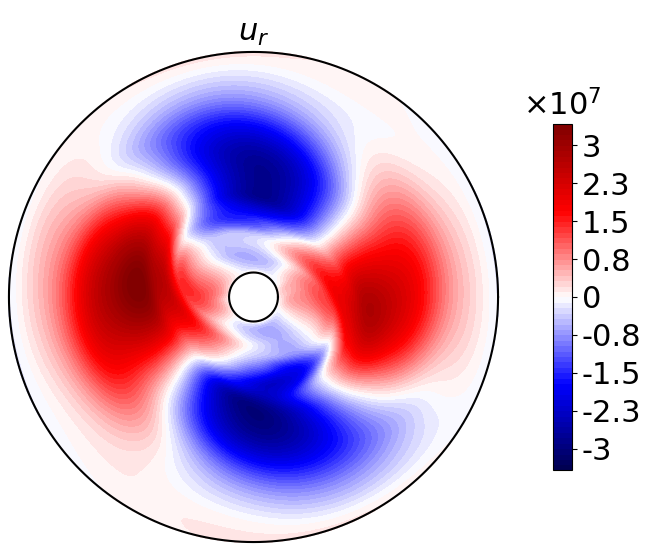}
    \includegraphics[width=0.45\textwidth]{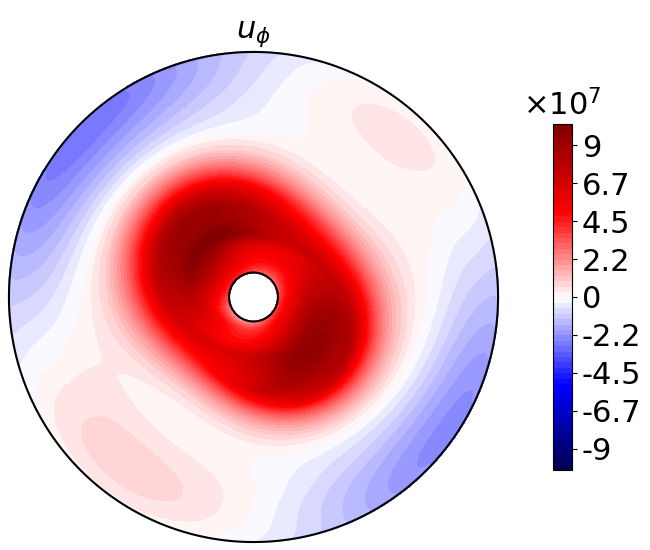}
    \caption{As above but for $\Omega_s=0$ with tidal amplitude $C_t=5.3\times 10^{-3}$.}
\end{subfigure}
    \caption{Similar to Fig.~\ref{fig:u_non_lin_g2_sat_equat} but for $_1g$ modes.}
    \label{fig:u_non_lin_g1_sat_equat}
\end{figure}

To spatially visualize the triadic resonances, we first look at snapshots of the radial $u_r$ and toroidal $u_\phi$ velocity in the equatorial plane (Figure \ref{fig:u_non_lin_g1_sat_equat}).
As expected from previous diagnostics, the $m=1$ mode is clearly dominant for both rotating cases. 
The velocity components $u_r$ and $u_\phi$ are overall stronger in the case with higher rotation.
The radial structure in the two cases is similar with the half-wavelength mode dominating for $u_r$ and full-wavelength one for $u_\phi$, as expected for the $_1g$ mode. Some small scale structure with strong amplitude is present close to the inner boundary. 
One could think that the inner boundary, and the development of a Stewartson layer at the inner tangent cylinder \citep{S1966,BT2018,BT2024}, is at the origin of this parametric instability, but we also observe a similar growing $\ell=m=1$ mode in a test simulation with a full sphere (not shown here).
The increase in non-linear effects with the fast rotation can also be seen in $u_\phi$ as the $m=1$ mode is less important compared to the $m=0$ mode.
In this case, the zonal flow looks more like the one for the $_2g$ mode, where the inner region is spun-up and the outer region is spun-down. 

\begin{figure}[ht]
    \centering
    \begin{subfigure}[b]{0.45\textwidth}
    \includegraphics[width=0.45\textwidth]{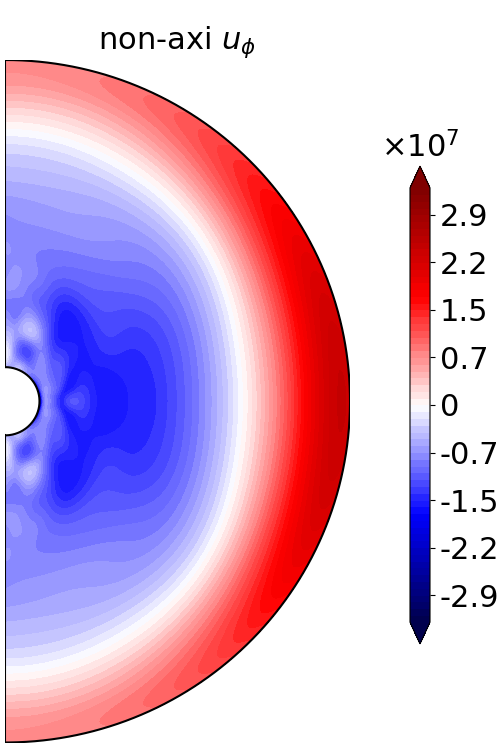}
    \includegraphics[width=0.45\textwidth]{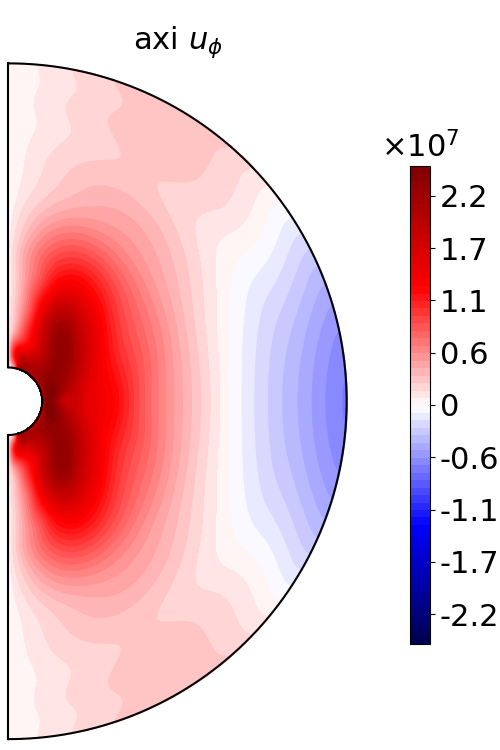}
    \caption{Non-axisymmetric (left) and axisymmetric (right) toroidal velocity $u_\phi$ of the $_1g$ mode for $\Omega_s = 100/\pi$ Hz and $C_t=1.36\times 10^{-3}$.}
\end{subfigure}
\begin{subfigure}[b]{0.45\textwidth}
    \includegraphics[width=0.45\textwidth]{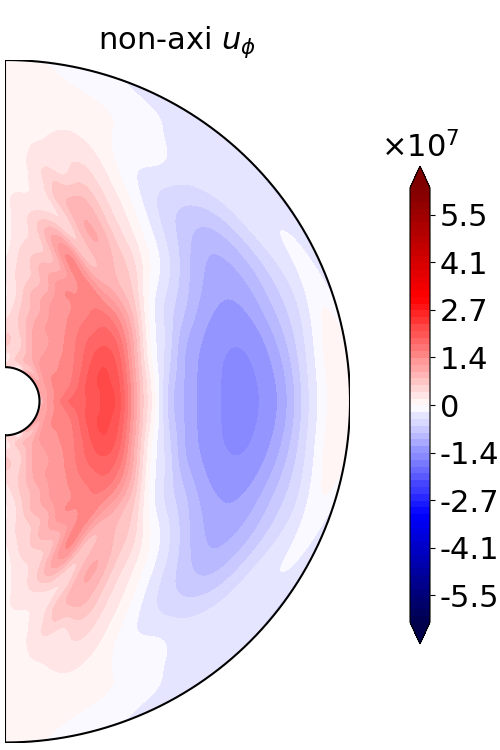}
    \includegraphics[width=0.45\textwidth]{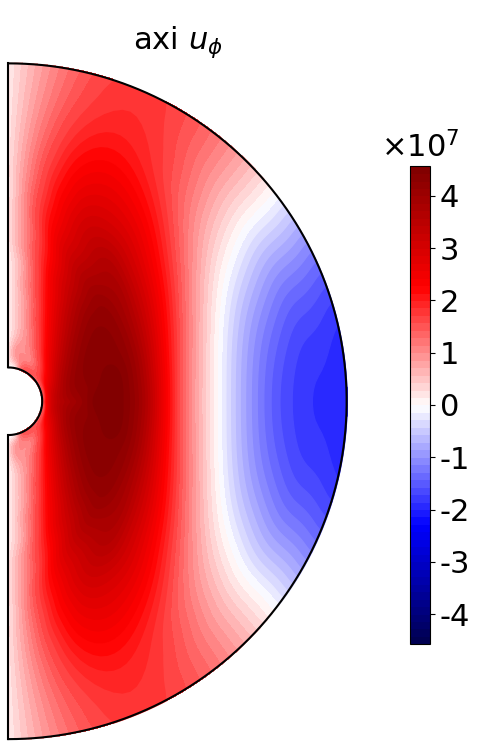}
    \caption{As above but for $\Omega_s = 100$ Hz and $C_t=5.3\times 10^{-3}$.}
\end{subfigure}
\begin{subfigure}[b]{0.45\textwidth}
    \includegraphics[width=0.45\textwidth]{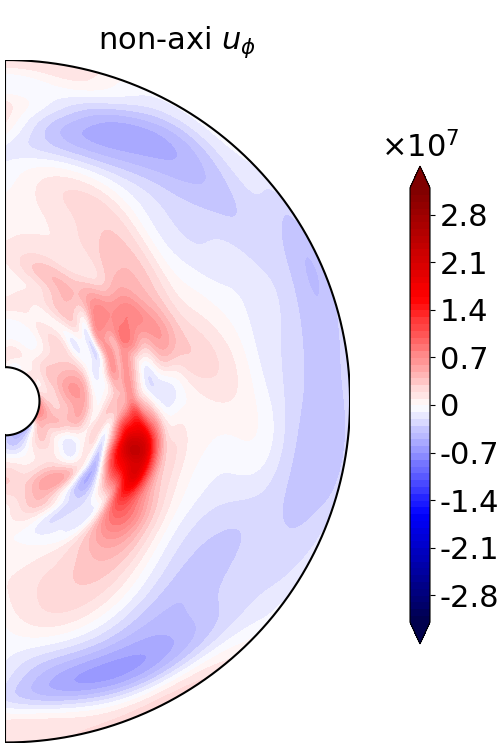}
    \includegraphics[width=0.45\textwidth]{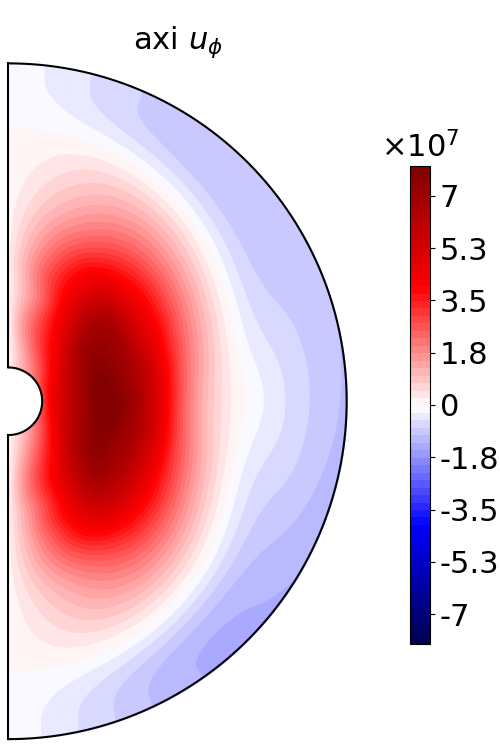}
    \caption{As above but for $\Omega_s = 0$ with tidal amplitude $C_t=5.3\times 10^{-3}$.}
\end{subfigure}
    \caption{Similar to Fig.~\ref{fig:non_lin_g2_axi} but for the $_1g$ mode.}
    \label{fig:non_lin_g1_axi}
\end{figure}

In the non-rotating case, we find similar results to the previous $_2g$ mode where instead $m=2$ ($m=0$) is dominant for the radial (toroidal) velocity. In terms of amplitude, the velocity components of the $_1g$ mode seem to have a similar amplitude as the faster rotating case (comparing panels (b) and (c) of Fig. \ref{fig:astro_g1_non_lin}), though it is difficult to compare directly as one is due to triadic resonances, while, at the end of the non-rotating simulation, the mode is a more typical tidally-excited $\ell=m=2$ mode for the poloidal component (panel (c) in Fig. \ref{fig:u_non_lin_g1_sat_equat}). This is confirmed by the comparison of the fast Fourier transform (FFT) of the non-rotating case before $4$ s (dotted lines) and after $4$ s (solid lines); see the bottom panel of Figure \ref{fig:param_FFT_g1_astro}).
In the non-rotating case, there seems to be a weaker mode close to the inner region and a stronger mode in the center of the domain. This may be due once again to the strong spin-up that we see in the inner region via $u_\phi$ in the panels (c) in Figures \ref{fig:u_non_lin_g1_sat_equat} and \ref{fig:non_lin_g1_axi}. 

We can confirm spin-up in the inner region 
for all three cases by looking at the axisymmetric toroidal velocity $u_\phi$ in the meridional plane (right panels of Figure \ref{fig:non_lin_g1_axi}). First, we see similar effects of rotation and increased tidal amplitude than for the $_2g$ mode: when modes become more non-linear (increased tidal amplitude/reduced rotation), the axisymmetric component becomes stronger compared to the non-axisymmetric component and clearly dominates for the non-rotating case with increased amplitude.
The impact of rotation is also clearly seen as the flow with $\Omega_s=100$ Hz is more cylindrical than the other cases, whereas the spherical geometry of gravity dominates in the two other cases.
The geometry for the non-axisymmetric components is different as there is only one radial wavelength and there are more structures close to the inner core. The axisymmetric component also seems to be shifted towards the interior compared to the $_2g$ mode (Figure \ref{fig:non_lin_g2_axi}).
By comparing the right panels of Figure \ref{fig:non_lin_g2_axi} and Figure \ref{fig:non_lin_g1_axi}, we see that the axisymmetric toroidal velocity is stronger for the $_1g$ mode in the rotating cases,  
which may be due to the increased tidal amplitude and also that the $_1g$ mode resonance is stronger even for a similar tidal amplitude (see Section \ref{sec:Nonlinear}).

To confirm that the inner spin-up is stronger for this simulation, we estimate the spin up in the first half of the shell as in Eq. (\ref{eq:Omsend}). We find that $\Omega_{s,end} \in [17.3 \ \rm Hz,1.1 \Omega_s, 1.18 \Omega_s] $ for an initial $\Omega_s \in [0,100/\pi,100]$ Hz. 
This is consistent with the fact that the axisymmetric azimuthal velocities are stronger for the $_1g$ mode in rotating cases.

\subsubsection*{Implications for dynamo activity and gravitational waves}
\label{sec:Astro_consequences}

We find that the kinetic energy of the tidally excited flow would be enough to generate a magnetic field of $>10^{14}$ G for the $_2g$ mode and $>10^{15}$ G for the $_1g$ mode. However, the bulk of this energy will likely be localized to small scales and it is unclear if the corresponding dipole growth would be enough to explain precursor luminosities.
Under the assumption of an incompressible fluid and the smooth BV frequency shown in Fig. \ref{fig:GR_N}, the modes are localized closer to the core than to the crust and would most likely generate the magnetic field there. However, from linear calculations with a realistic NS profile with the APR4 EOS, the mode velocity is faster in the crust, so we can expect (in principle) to have the differential rotation in the crust, where it would amplify the magnetic field that is most relevant for electromagnetic observables.
If we assume that the strength of the dipole is $5\%$ of the total magnetic field, 
as found in simulations of the magnetorotationnal instabilities \citep{2021ReboulSalze,2022ReboulSalze}, it may be enough to explain the luminosity of precursors \citep{skk24}.
This need to be further studied in simulations with magnetic transport included and compared with a more-realistic interior, which is out of the scope of this paper. Still, the results are promising. 

We also find that the star, for $_2g$ and $_1g$ modes and for all rotation rates, would get spin-up by the resonance. This implies that we can expect the tidal forcing frequency $\hat{\omega}=2(\Omega_o - \Omega_s) $ to remain in resonance window \emph{longer} as the orbital frequency $\Omega_o$ evolves due to the emission of GWs. This depends on the size of the window but roughly lies in the range $0.99\omega_{\rm resonnant}\leq \hat{\omega} \leq 1.01 \omega_{\rm resonnant}$, so that a spin-up of $10\%$ would increase the resonant window timescale by a factor $\sim 5$.

\section{Discussions}
\label{sec:Disc}

\subsection{Consequences for BH-NS mergers}

The calculations done in this study are practically independent of the particulars regarding the secondary of mass $M_2$. We could therefore extrapolate our results to cases relevant for a black-hole with $M_2$ >$M_1$. This means that the tidal amplitude parameter $\epsilon = \frac{M_2}{M_1}\left(\frac{R_o}{a}\right)^3$ would increase for a given semi-major axis $a$. 
This may lead to stronger tidal amplitudes; in particular, for a given orbital frequency $\Omega_o$ corresponding to a resonant frequency, the tidal amplitude parameter $\epsilon$ can be written as $\epsilon=\frac{M_2}{M_1+{M_2}}\left(\frac{\Omega_o}{\omega_d}\right)^2$ by using $a^3=\frac{G(M_1+M_2)}{\Omega_o^2}$ and $R_o^3 = \frac{G M_1}{\omega_d^2} $, which would increase by a maximum factor $\sim 2$. Stronger non-linear effects are therefore expected, provided the resonance occurs, most likely without changing qualitatively the results of this study.

\subsection{Interior structure}
\label{disc:interior}
Comparison between results from Secs.~\ref{sec:Sim} and \ref{sec:astrophysics} shows the importance of the NS interior model. Indeed, the BV frequency in section \ref{sec:astrophysics} is higher by a factor $\approx 2-5$ from the inner to the outer boundaries. 
This increase in the BV frequency lead to stronger non-linear effects. As the BV frequency is strongly dependent on the internal microphysics, a change in the NS model could impact the conclusions of this study.   
Additionally, the tidal amplitude parameter is dependent on the NS radius $R_o$ and an increase of $20\%$ in this quantity for a different EOS at a fixed mass would lead to an increase of $\approx 1.7$ in the amplitude of a given mode, which may then change the non-linear saturation \cite[see also][]{2024Suvorov}.
The density profile also changes with the EOS, which would further impact on the mode eigenfunctions. Density profiles can be taken into account with an anelastic model, the development of which is left to future studies owing to numerical challenges: the density gradient have to be taken into account in the computation of the equilibrium tides.
This would be especially important in studying the resonant modes in the crust, where proposed late-stage amplifications of near-surface magnetic fields have been invoked to explain the luminosity of GRB precursors as the Ohmic decay timescale is expected to be much shorter than the orbital lifetime \cite[e.g.][]{pv19}.

\subsection{Formulation issues}

As with most numerical studies of astrophysical phenomena, some limitations are inevitable as physical time- and length-scales typically span many orders of magnitude in such systems.
Indeed, the formulation presented in this article has a few limitations, which we highlight here and in the sections below for completeness. 

(i) The formulation cannot reproduce $f$-mode resonances or $g$-mode resonances too close to the f-mode frequency as this mode is `hard coded' in the static tide (the Love number in the equilibrium velocity). Moreover, a fixed surface is a poor approximation for f-mode oscillations \citep{2023Pontin}. 
(ii) In the non-resonant regime, the formulation cannot reproduce results with a free surface. This can be corrected be changing the stress-free boundary conditions to match the free surface case. However, its implementation with the scalar potentials in the non-linear code can be difficult or may lead to unexpected effects and the study of the regime, away from the BV frequency, is left to further studies.
(iii) The use of a spherical shell leads to the neglecting of non-linear interactions between the equilibrium and dynamical tides. It can be justified by some considerations \cite[see][]{2022Astoul} but when the resonant modes occupy the largest scales, these non-linear interactions might become relevant. 

\subsection{Cowling approximation}
\label{disc:self_gravity}

A simplifying hypothesis employed throughout is the Cowling approximation, the neglect of equilibrium and dynamical perturbations to the background gravitational potential $\nabla \Phi_g = \bm g$. 
These effects could have a strong impact on the amplitude of the modes however; the linear study of \citet{2017xuresonant} found a reduction down to 20\% of the initial amplitude in cases without perturbations to the self-gravity potential. 
Nonetheless, it is unknown how the dynamical perturbations to self-gravity $\Phi_g'$ in our setup, where we assume a constant density, would impact the modes and their non-linear saturation. 
In addition, neglecting the equilibrium perturbations to self-gravity $\Phi_g^{e}$ removes a term $\propto (1+k_2)$ in the amplitude of the equilibrium tide, where $k_2$ is the Love number associated with the $\ell=2$ tide. For NSs, realistic calculations of Love numbers give $k_2 \gtrsim 0.1$ \citep{flan08,hind08,hind10}, meaning we underestimate the amplitude by $\gtrsim 10\%$. Overall, we seem to be neglecting both positive and negative effects of the self-gravity and the impact of this assumption needs to be further studied as it could be exacerbated to some degree for a realistic stellar density profile.

\subsection{Dissipation processes}

The simulations presented in this work have been carried out with unrealistic viscosities and thermal/compositional diffusivities that could change the results. For example, in a NS, the microphysical viscosities are estimated to be at least $10$ orders of magnitude lower than that used here. For the thermal/composition diffusivities, the value depends highly on the NS model and evolutionary pathway \cite[e.g. whether there was a history of recycling; see][]{2024Suvorov}.
In linear studies, such as \cite{2023Pontin,2024Pontin}, it has been found that decreasing the viscosity/thermal diffusivity by $\sim 2$ orders of magnitude leads to a comparable increase to the total (frequency-dependent) tidal dissipation and mode kinetic energy. 
On a positive note, this suggests that the mode amplitudes found in this study can be seen as \emph{lower limits} and may be larger in astrophysical circumstances. This also means that non-linear effects are more relevant, especially with respect to how saturation depends on damping processes.
It is for this reason that we chose to study the non-rotating case with an increased tidal frequency.
However, triadic resonances, corotation resonances, or other secondary instabilities (like shear-flow instabilities) may lead to a different saturation mechanism that is largely insensitive to the viscosity and thermal diffusivity below some critical values. 
Studying the impact of viscosity and thermal diffusivity is required to understand the non-linear saturation and to explore a wider range of parameters; this will be the subject of future work.  

Moreover, the resonant peaks (see, e.g., Fig.~\ref{fig:diss_g_test}) may also become narrower and the relevant resonant window itself could become smaller. A study that takes into account the evolution of the tidal forcing frequency and tidal amplitude due to the emission of GWs would be required to see whether the modes are able to draw enough energy from the time-varying orbit and reach a similar amplitude as the modes presented in this study.

\section{Conclusions and perspectives}
\label{sec:Conclusion}

In this article, a new formulation, building on that of \cite{2022Astoul,2023Astoul} but applied here to a stably- instead of neutrally-stratified region, was derived in order to simulate the non-linear saturation of gravito-inertial modes in a self-consistent manner that separates equilibrium and dynamical tides. To ensure consistency with previous formulations that do not distinguish between dynamical/equilibrium tides, we first tested the linear limit of this scheme and found excellent agreement with previous studies for both pure and gravito-inertial modes provided the tidal forcing frequency is not so high so as to be close to the f- or acoustic-mode frequencies (Sec.~\ref{sec:Linear}), where nonlinear effects may be important in any case \citep{kuan24}.

The non-linear code was benchmarked against the linear results in Sec.~\ref{sec:Linear} with a uniform entropy/composition gradient, finding again quantitative agreement in the expected regimes of validity. With a homogeneous BV frequency of $N=0.045 \omega_d$ and a dimensionless tidal amplitude of $C_t \sim 10^{-3}$, such a comparison demonstrated the impact of non-linear effects for several modes in the form of a generated, axisymmetric kinetic energy representing $\gtrsim 5\%$ of the total kinetic energy. 
In particular, we found in Sec.~\ref{sec:Sim} that the strongest resonant modes are indeed able to generate an \emph{axisymmetric} kinetic energy representing of order $10\%$ of the total kinetic energy. 
This axisymmetric component is dominated by the toroidal velocity $u_\phi$, which is optimistic from the perspective of enticing premerger dynamo activity and the generation of magnetic fields, as first suggested by \cite{2024Suvorov}.

Although the formalism is general, it has been primarily applied in the context of neutron-star binaries in this work; astrophysical applications for gravito-inertial modes were considered in Sec.~\ref{sec:astrophysics}. A BV profile based on TOV calculations with the APR4+DH EOS, smoothed out in the crust, was considered in an effort to consider realistic profiles even though the formalism here is strictly Newtonian; such hybrid schemes are common in the literature, especially because a fully-relativistic treatment of tides is difficult \cite[see][for a discussion]{2024Suvorov}. Comparisons with the GR linear calculations demonstrate quantitative agreement for pure g-modes in the stellar core but not in the crust (see Fig.~\ref{fig:GR_profile}). This is primarily due to the impact of the radially-varying density profile and compositional discontinuities in the crust that we neglect in this study via the Boussinesq approximation. The linear code was then used to compute the resonant frequencies to study the non-linear saturation of the $_1g$ and $_2g$modes. We found that with a stronger BV frequency compared to Sec.~\ref{sec:Sim}, non-linear effects dominate even more in determining the saturation of the modes. Remarkably, the axisymmetric kinetic energy can constitute up to $95 \%$ of the total mode energy.
Overall, we argue that if the kinetic energy was efficiently converted into magnetic energy a magnetic field of $10^{14}-10^{15}$ G could be generated.
If this occurs via the magnetorotational instability, appealing to previous studies that quantified multipolar energy distributions \citep{2021ReboulSalze,2022ReboulSalze} suggests the anticipated strength of the dipole component would be between $5\times10^{12}-5\times10^{13}$ G. 
This would be enough to explain the luminosity of (at least most) precursors \cite[e.g.][]{tsang12,xiao24}.

We also find that the NS should get spun-up by resonances as $u_{\phi}$ is non-negligible (e.g. Fig.~\ref{fig:non_lin_g2_axi}). The rotation rate in the inner half of the domain may be increased by up to $\sim 10 \%$ for already-rotating cases and up to $\sim 20$ Hz in the non-rotating case. 
Such an increase would lead to the tidal forcing frequency $\hat{\omega}=2(\Omega_o - \Omega_s) $ remaining in the resonance window for longer as the orbital frequency $\Omega_o$ sweeps up due to the emission of GWs. Linear studies of gravito-inertial modes in NSs may therefore \emph{underestimate} the amplitude of these modes as this lengthening, and hence the duration of energy transfer, has not been considered. 
Stronger saturation should therefore be studied in more detail as it could lead to a greater degree of dephasing and makes a more optimistic case for the detectability of dynamical tides with next-generation GW interferometers \citep{2023Ho}.

\section*{Acknowledgements}
Linear simulations were run on the Yamazaki cluster at the Max-Planck Institute for Gravitational Physics, Potsdam. Non-linear simulations were run on the Sakura cluster at the Max Planck Computing and Data Facility (MPCDF) in Garching, Germany. 
AGS is grateful for support provided by the Conselleria d'Educaci{\'o}, Cultura, Universitats i Ocupaci{\'o} de la Generalitat Valenciana through Prometeo Project CIPROM/2022/13. AA has been funded by a Leverhulme Trust Early Career Fellowship (ECF-2022-362). The authors thank A. Barker for helpful discussions about the treatment of tides, and more specifically the validity of different dynamical/equilibrium tide decompositions.

\bibliographystyle{aa} % style aa.bst
\bibliography{biblio} 

\begin{thebibliography}{93}
\expandafter\ifx\csname natexlab\endcsname\relax\def\natexlab#1{#1}\fi

\bibitem[{{Aerts} \& {Tkachenko}(2024)}]{aerts23}
{Aerts}, C. \& {Tkachenko}, A. 2024, \aap, 692, R1

\bibitem[{{Akmal} {et~al.}(1998){Akmal}, {Pandharipande}, \& {Ravenhall}}]{1998APR4}
{Akmal}, A., {Pandharipande}, V.~R., \& {Ravenhall}, D.~G. 1998, \prc, 58, 1804

\bibitem[{{Alexander}(1987)}]{alex87}
{Alexander}, M.~E. 1987, \mnras, 227, 843

\bibitem[{{Andersson} \& {Pnigouras}(2020)}]{ande20}
{Andersson}, N. \& {Pnigouras}, P. 2020, \prd, 101, 083001

\bibitem[{{Astoul} \& {Barker}(2022)}]{2022Astoul}
{Astoul}, A. \& {Barker}, A.~J. 2022, \mnras, 516, 2913

\bibitem[{{Astoul} \& {Barker}(2023)}]{2023Astoul}
{Astoul}, A. \& {Barker}, A.~J. 2023, \apjl, 955, L23

\bibitem[{{Astoul} \& {Barker}(2025)}]{astoulbark25}
{Astoul}, A. \& {Barker}, A.~J. 2025, arXiv e-prints, arXiv:2501.08722

\bibitem[{{Baiko}(2024)}]{baiko24}
{Baiko}, D.~A. 2024, \mnras, 528, 408

\bibitem[{{Baiko} \& {Chugunov}(2018)}]{bc18}
{Baiko}, D.~A. \& {Chugunov}, A.~I. 2018, \mnras, 480, 5511

\bibitem[{{Balbus} \& {Hawley}(1998)}]{1998ReviewMRIBalbus}
{Balbus}, S.~A. \& {Hawley}, J.~F. 1998, Reviews of Modern Physics, 70, 1

\bibitem[{{Barik} {et~al.}(2018){Barik}, {Triana}, {Hoff}, \& {Wicht}}]{BT2018}
{Barik}, A., {Triana}, S.~A., {Hoff}, M., \& {Wicht}, J. 2018, Journal of Fluid Mechanics, 843, 211

\bibitem[{{Barik} {et~al.}(2024){Barik}, {Triana}, {Hoff}, \& {Wicht}}]{BT2024}
{Barik}, A., {Triana}, S.~A., {Hoff}, M., \& {Wicht}, J. 2024, Journal of Fluid Mechanics, 1001, A1

\bibitem[{{Barker}(2011)}]{B2011}
{Barker}, A.~J. 2011, \mnras, 414, 1365

\bibitem[{{Barker}(2016)}]{2016BarkerNLTides}
{Barker}, A.~J. 2016, \mnras, 459, 939

\bibitem[{{Barker}(2022)}]{barker22}
{Barker}, A.~J. 2022, \apjl, 927, L36

\bibitem[{{Barker} \& {Ogilvie}(2010)}]{BO2010}
{Barker}, A.~J. \& {Ogilvie}, G.~I. 2010, \mnras, 404, 1849

\bibitem[{{Bolmont} \& {Mathis}(2016)}]{BM2016}
{Bolmont}, E. \& {Mathis}, S. 2016, Celestial Mechanics and Dynamical Astronomy, 126, 275

\bibitem[{{Burns} {et~al.}(2020){Burns}, {Vasil}, {Oishi}, {Lecoanet}, \& {Brown}}]{burns20}
{Burns}, K.~J., {Vasil}, G.~M., {Oishi}, J.~S., {Lecoanet}, D., \& {Brown}, B.~P. 2020, Physical Review Research, 2, 023068

\bibitem[{Christensen \& Wicht(2015)}]{2015CHRISTENSEN245}
Christensen, U. \& Wicht, J. 2015, in Treatise on Geophysics (Second Edition), second edition edn., ed. G.~Schubert (Oxford: Elsevier), 245 -- 277

\bibitem[{{Ciolfi}(2020)}]{cio20}
{Ciolfi}, R. 2020, General Relativity and Gravitation, 52, 59

\bibitem[{{Coppin} {et~al.}(2020){Coppin}, {de Vries}, \& {van Eijndhoven}}]{cop20}
{Coppin}, P., {de Vries}, K.~D., \& {van Eijndhoven}, N. 2020, \prd, 102, 103014

\bibitem[{{Counselman}(1973)}]{counsel73}
{Counselman}, III, C.~C. 1973, \apj, 180, 307

\bibitem[{{Couston} {et~al.}(2018){Couston}, {Lecoanet}, {Favier}, \& {Le Bars}}]{2018CoustonDedalus}
{Couston}, L.-A., {Lecoanet}, D., {Favier}, B., \& {Le Bars}, M. 2018, \prl, 120, 244505

\bibitem[{{Damour} \& {Nagar}(2009)}]{dam09}
{Damour}, T. \& {Nagar}, A. 2009, \prd, 80, 084035

\bibitem[{{Dhouib} {et~al.}(2024){Dhouib}, {Baruteau}, {Mathis}, {Debras}, {Astoul}, \& {Rieutord}}]{2024DhouibIWinstratified}
{Dhouib}, H., {Baruteau}, C., {Mathis}, S., {et~al.} 2024, \aap, 682, A85

\bibitem[{{Dintrans} \& {Rieutord}(2000)}]{DR2000}
{Dintrans}, B. \& {Rieutord}, M. 2000, \aap, 354, 86

\bibitem[{{Dintrans} {et~al.}(1999){Dintrans}, {Rieutord}, \& {Valdettaro}}]{DR1999}
{Dintrans}, B., {Rieutord}, M., \& {Valdettaro}, L. 1999, Journal of Fluid Mechanics, 398, 271

\bibitem[{{Douchin} \& {Haensel}(2001)}]{2001DHcrust}
{Douchin}, F. \& {Haensel}, P. 2001, \aap, 380, 151

\bibitem[{{Duch{\^e}ne} \& {Kraus}(2013)}]{duch13}
{Duch{\^e}ne}, G. \& {Kraus}, A. 2013, \araa, 51, 269

\bibitem[{{El-Badry}(2024)}]{bad24}
{El-Badry}, K. 2024, \nar, 98, 101694

\bibitem[{{Favier} {et~al.}(2014){Favier}, {Barker}, {Baruteau}, \& {Ogilvie}}]{2014FavierNLIWtidal}
{Favier}, B., {Barker}, A.~J., {Baruteau}, C., \& {Ogilvie}, G.~I. 2014, \mnras, 439, 845

\bibitem[{{Flanagan} \& {Hinderer}(2008)}]{flan08}
{Flanagan}, {\'E}.~{\'E}. \& {Hinderer}, T. 2008, \prd, 77, 021502

\bibitem[{{Fuller} {et~al.}(2024){Fuller}, {Guillot}, {Mathis}, \& {Murray}}]{fuller24}
{Fuller}, J., {Guillot}, T., {Mathis}, S., \& {Murray}, C. 2024, \ssr, 220, 22

\bibitem[{{Galtier}(2003)}]{G2003}
{Galtier}, S. 2003, \pre, 68, 015301

\bibitem[{Gastine \& Wicht(2012)}]{2012GastineCode}
Gastine, T. \& Wicht, J. 2012, Icarus, 219, 428

\bibitem[{{Gilman} \& {Glatzmaier}(1981)}]{1981GilmanAnel}
{Gilman}, P.~A. \& {Glatzmaier}, G.~A. 1981, \apjs, 45, 335

\bibitem[{{Goldreich} \& {Nicholson}(1989)}]{G1989}
{Goldreich}, P. \& {Nicholson}, P.~D. 1989, \apj, 342, 1079

\bibitem[{{Guo} {et~al.}(2023){Guo}, {Ogilvie}, \& {Barker}}]{GO2023}
{Guo}, Z., {Ogilvie}, G.~I., \& {Barker}, A.~J. 2023, \mnras, 521, 1353

\bibitem[{{Hegade K.~R.} {et~al.}(2024){Hegade K.~R.}, {Ripley}, \& {Yunes}}]{hega24}
{Hegade K.~R.}, A., {Ripley}, J.~L., \& {Yunes}, N. 2024, \prd, 109, 104064

\bibitem[{{Hinderer}(2008)}]{hind08}
{Hinderer}, T. 2008, \apj, 677, 1216

\bibitem[{{Hinderer} {et~al.}(2010){Hinderer}, {Lackey}, {Lang}, \& {Read}}]{hind10}
{Hinderer}, T., {Lackey}, B.~D., {Lang}, R.~N., \& {Read}, J.~S. 2010, \prd, 81, 123016

\bibitem[{{Hinderer} {et~al.}(2016){Hinderer}, {Taracchini}, {Foucart}, {Buonanno}, {Steinhoff}, {Duez}, {Kidder}, {Pfeiffer}, {Scheel}, {Szilagyi}, {Hotokezaka}, {Kyutoku}, {Shibata}, \& {Carpenter}}]{hind16}
{Hinderer}, T., {Taracchini}, A., {Foucart}, F., {et~al.} 2016, \prl, 116, 181101

\bibitem[{{Ho} \& {Andersson}(2023)}]{2023Ho}
{Ho}, W. C.~G. \& {Andersson}, N. 2023, \prd, 108, 043003

\bibitem[{{Jackson} {et~al.}(2008){Jackson}, {Greenberg}, \& {Barnes}}]{jack08}
{Jackson}, B., {Greenberg}, R., \& {Barnes}, R. 2008, \apj, 678, 1396

\bibitem[{{Ji} {et~al.}(2023){Ji}, {Fuller}, \& {Lecoanet}}]{2023JiFullerTayler}
{Ji}, S., {Fuller}, J., \& {Lecoanet}, D. 2023, \mnras, 521, 5372

\bibitem[{{Kerswell}(1999)}]{K1999}
{Kerswell}, R.~R. 1999, Journal of Fluid Mechanics, 382, 283

\bibitem[{{Kokkotas} \& {Schafer}(1995)}]{ks95}
{Kokkotas}, K.~D. \& {Schafer}, G. 1995, \mnras, 275, 301

\bibitem[{{Kuan} {et~al.}(2024){Kuan}, {Kiuchi}, \& {Shibata}}]{kuan24}
{Kuan}, H.-J., {Kiuchi}, K., \& {Shibata}, M. 2024, arXiv e-prints, arXiv:2411.16850

\bibitem[{{Kuan} {et~al.}(2021{\natexlab{a}}){Kuan}, {Suvorov}, \& {Kokkotas}}]{kuan21a}
{Kuan}, H.-J., {Suvorov}, A.~G., \& {Kokkotas}, K.~D. 2021{\natexlab{a}}, \mnras, 506, 2985

\bibitem[{{Kuan} {et~al.}(2021{\natexlab{b}}){Kuan}, {Suvorov}, \& {Kokkotas}}]{kuan21b}
{Kuan}, H.-J., {Suvorov}, A.~G., \& {Kokkotas}, K.~D. 2021{\natexlab{b}}, \mnras, 508, 1732

\bibitem[{{Kuan} {et~al.}(2023){Kuan}, {Suvorov}, \& {Kokkotas}}]{kuan23b}
{Kuan}, H.-J., {Suvorov}, A.~G., \& {Kokkotas}, K.~D. 2023, \aap, 676, A59

\bibitem[{{Kwon} {et~al.}(2024){Kwon}, {Yu}, \& {Venumadhav}}]{2024Kwon}
{Kwon}, K.~J., {Yu}, H., \& {Venumadhav}, T. 2024, arXiv e-prints, arXiv:2410.03831

\bibitem[{{Kwon} {et~al.}(2025){Kwon}, {Yu}, \& {Venumadhav}}]{2025KwonNL}
{Kwon}, K.~J., {Yu}, H., \& {Venumadhav}, T. 2025, arXiv e-prints, arXiv:2503.11837

\bibitem[{{Lai} \& {Wu}(2006)}]{lw06}
{Lai}, D. \& {Wu}, Y. 2006, \prd, 74, 024007

\bibitem[{{Lazovik} {et~al.}(2024){Lazovik}, {Barker}, {de Vries}, \& {Astoul}}]{laz24}
{Lazovik}, Y.~A., {Barker}, A.~J., {de Vries}, N.~B., \& {Astoul}, A. 2024, \mnras, 527, 8245

\bibitem[{{Lecoanet} {et~al.}(2017){Lecoanet}, {Vasil}, {Fuller}, {Cantiello}, \& {Burns}}]{2017LecoanetgravityMHD}
{Lecoanet}, D., {Vasil}, G.~M., {Fuller}, J., {Cantiello}, M., \& {Burns}, K.~J. 2017, \mnras, 466, 2181

\bibitem[{{Lin} \& {Ogilvie}(2018)}]{LO2018}
{Lin}, Y. \& {Ogilvie}, G.~I. 2018, \mnras, 474, 1644

\bibitem[{{Mathis}(2009)}]{M2009}
{Mathis}, S. 2009, \aap, 506, 811

\bibitem[{{Ogilvie}(2009)}]{O2009}
{Ogilvie}, G.~I. 2009, \mnras, 396, 794

\bibitem[{{Ogilvie}(2013)}]{2013Ogilvie}
{Ogilvie}, G.~I. 2013, \mnras, 429, 613

\bibitem[{{Ogilvie}(2014)}]{og14}
{Ogilvie}, G.~I. 2014, \araa, 52, 171

\bibitem[{{Ogilvie} \& {Lin}(2004)}]{2004OgilvieL}
{Ogilvie}, G.~I. \& {Lin}, D.~N.~C. 2004, \apj, 610, 477

\bibitem[{{Passamonti} {et~al.}(2021){Passamonti}, {Andersson}, \& {Pnigouras}}]{pass21}
{Passamonti}, A., {Andersson}, N., \& {Pnigouras}, P. 2021, \mnras, 504, 1273

\bibitem[{{Peters}(1964)}]{peters64}
{Peters}, P.~C. 1964, Physical Review, 136, 1224

\bibitem[{{Pnigouras} \& {Kokkotas}(2015)}]{pk15}
{Pnigouras}, P. \& {Kokkotas}, K.~D. 2015, \prd, 92, 084018

\bibitem[{{Pons} \& {Vigan{\`o}}(2019)}]{pv19}
{Pons}, J.~A. \& {Vigan{\`o}}, D. 2019, Living Reviews in Computational Astrophysics, 5, 3

\bibitem[{{Pontin} {et~al.}(2023){Pontin}, {Barker}, \& {Hollerbach}}]{2023Pontin}
{Pontin}, C.~M., {Barker}, A.~J., \& {Hollerbach}, R. 2023, \apj, 950, 176

\bibitem[{{Pontin} {et~al.}(2024){Pontin}, {Barker}, \& {Hollerbach}}]{2024Pontin}
{Pontin}, C.~M., {Barker}, A.~J., \& {Hollerbach}, R. 2024, \apj, 960, 32

\bibitem[{{Press} \& {Teukolsky}(1977)}]{pres77}
{Press}, W.~H. \& {Teukolsky}, S.~A. 1977, \apj, 213, 183

\bibitem[{{Reboul-Salze} {et~al.}(2021){Reboul-Salze}, {Guilet}, {Raynaud}, \& {Bugli}}]{2021ReboulSalze}
{Reboul-Salze}, A., {Guilet}, J., {Raynaud}, R., \& {Bugli}, M. 2021, \aap, 645, A109

\bibitem[{{Reboul-Salze} {et~al.}(2022){Reboul-Salze}, {Guilet}, {Raynaud}, \& {Bugli}}]{2022ReboulSalze}
{Reboul-Salze}, A., {Guilet}, J., {Raynaud}, R., \& {Bugli}, M. 2022, \aap, 667, A94

\bibitem[{{Rieutord}(2009)}]{R2009}
{Rieutord}, M. 2009, in The Rotation of Sun and Stars, Vol. 765, 101--121

\bibitem[{{Schaeffer}(2013)}]{2013SHTNS}
{Schaeffer}, N. 2013, Geochem. Geophys. Geosyst., 14, 751

\bibitem[{{Semin} {et~al.}(2016){Semin}, {Facchini}, {P{\'e}tr{\'e}lis}, \& {Fauve}}]{SF2016}
{Semin}, B., {Facchini}, G., {P{\'e}tr{\'e}lis}, F., \& {Fauve}, S. 2016, Physics of Fluids, 28, 096601

\bibitem[{{Spiegel} \& {Veronis}(1960)}]{spi60}
{Spiegel}, E.~A. \& {Veronis}, G. 1960, \apj, 131, 442

\bibitem[{{Stewartson}(1966)}]{S1966}
{Stewartson}, K. 1966, Journal of Fluid Mechanics, 26, 131

\bibitem[{{Suvorov} \& {Kokkotas}(2020)}]{sk20}
{Suvorov}, A.~G. \& {Kokkotas}, K.~D. 2020, \prd, 101, 083002

\bibitem[{{Suvorov} {et~al.}(2022){Suvorov}, {Kuan}, \& {Kokkotas}}]{skk22}
{Suvorov}, A.~G., {Kuan}, H.~J., \& {Kokkotas}, K.~D. 2022, \aap, 664, A177

\bibitem[{{Suvorov} {et~al.}(2024{\natexlab{a}}){Suvorov}, {Kuan}, \& {Kokkotas}}]{skk24}
{Suvorov}, A.~G., {Kuan}, H.-J., \& {Kokkotas}, K.~D. 2024{\natexlab{a}}, Universe, 10, 441

\bibitem[{{Suvorov} {et~al.}(2024{\natexlab{b}}){Suvorov}, {Kuan}, {Reboul-Salze}, \& {Kokkotas}}]{2024Suvorov}
{Suvorov}, A.~G., {Kuan}, H.-J., {Reboul-Salze}, A., \& {Kokkotas}, K.~D. 2024{\natexlab{b}}, \prd, 109, 103023

\bibitem[{{Taniguchi} \& {Shibata}(2010)}]{tani10}
{Taniguchi}, K. \& {Shibata}, M. 2010, \apjs, 188, 187

\bibitem[{{Thompson} \& {Duncan}(1993)}]{1993ThompsonPNS}
{Thompson}, C. \& {Duncan}, R.~C. 1993, \apj, 408, 194

\bibitem[{{Tilgner} \& {Busse}(1997)}]{1997TilgnerJFM}
{Tilgner}, A. \& {Busse}, F.~H. 1997, J. Fluid Mech., 332, 359

\bibitem[{{Tsang} {et~al.}(2012){Tsang}, {Read}, {Hinderer}, {Piro}, \& {Bondarescu}}]{tsang12}
{Tsang}, D., {Read}, J.~S., {Hinderer}, T., {Piro}, A.~L., \& {Bondarescu}, R. 2012, \prl, 108, 011102

\bibitem[{{Wang} {et~al.}(2020){Wang}, {Peng}, {Zou}, {Zhang}, \& {Zhang}}]{wang20}
{Wang}, J.-S., {Peng}, Z.-K., {Zou}, J.-H., {Zhang}, B.-B., \& {Zhang}, B. 2020, \apjl, 902, L42

\bibitem[{{Weinberg} {et~al.}(2013){Weinberg}, {Arras}, \& {Burkart}}]{wein13}
{Weinberg}, N.~N., {Arras}, P., \& {Burkart}, J. 2013, \apj, 769, 121

\bibitem[{{Wicht}(2002)}]{2002WichtMagic}
{Wicht}, J. 2002, Phys. Earth Planet. Inter., 132, 281

\bibitem[{{Xiao} {et~al.}(2024){Xiao}, {Zhang}, {Zhu}, {Xiong}, {Gao}, {Xu}, {Zhang}, {Peng}, {Li}, {Zhang}, {Lu}, {Lin}, {Liu}, {Zhang}, {Ge}, {Tuo}, {Xue}, {Fu}, {Liu}, {Liu}, {Li}, {Wang}, {Zheng}, {Wang}, {Jiang}, {Li}, {Liu}, {Cao}, {Luo}, {Yang}, {Yi}, {Wang}, {Cai}, {Yi}, {Zhao}, {Xie}, {Li}, {Luo}, {Song}, {Zhang}, {Qu}, {Liu}, {Li}, {Xu}, \& {Li}}]{xiao24}
{Xiao}, S., {Zhang}, Y.-Q., {Zhu}, Z.-P., {et~al.} 2024, \apj, 970, 6

\bibitem[{Xu \& Lai(2017)}]{2017xuresonant}
Xu, W. \& Lai, D. 2017, Physical Review D, 96, 083005

\bibitem[{{Yu} {et~al.}(2024){Yu}, {Arras}, \& {Weinberg}}]{yu24}
{Yu}, H., {Arras}, P., \& {Weinberg}, N.~N. 2024, \prd, 110, 024039

\bibitem[{{Zahn}(1966)}]{zahn66}
{Zahn}, J.~P. 1966, Annales d'Astrophysique, 29, 489

\bibitem[{{Zahn}(2013)}]{zahn2013}
{Zahn}, J.-P. 2013, in Lecture Notes in Physics, Berlin Springer Verlag, ed. J.~{Souchay}, S.~{Mathis}, \& T.~{Tokieda}, Vol. 861, 301

\bibitem[{{Zhou} \& {Zhang}(2017)}]{zz17}
{Zhou}, Y. \& {Zhang}, F. 2017, \apj, 849, 114

\end{thebibliography}

\appendix

\end{document}